\documentclass[a4paper,11pt]{article}
\usepackage[utf8]{inputenc}
\pdfoutput=1 

\usepackage{jheppub} 

\usepackage[T1]{fontenc} 
\usepackage{graphicx}
\usepackage[makeroom]{cancel}
\usepackage{makecell}
\usepackage{bm}
\usepackage{amssymb}
\usepackage{amsmath}
\usepackage{epsfig}
\usepackage{hyperref}
\usepackage{hhline}
\usepackage{multirow}
\usepackage[normalem]{ulem}
\usepackage{subfig}
\usepackage{tikz}
\usepackage{epstopdf}
\usetikzlibrary{snakes}
\usepackage{slashed}
\usepackage{float}
\allowdisplaybreaks

\usepackage{framed}
\colorlet{shadecolor}{blue!10}

 \definecolor{jd}{rgb}{0.858, 0.188, 0.478}

\def\lapp{\mathrel{\rlap{\raise.5ex\hbox{$<$}}
                    {\lower.5ex\hbox{$\sim$}}}}
\def\gapp{\mathrel{\rlap{\raise.5ex\hbox{$>$}}
                    {\lower.5ex\hbox{$\sim$}}}}

{
{

\textwidth 15.55cm \textheight 22.5cm
\hoffset -1.6cm
\voffset -1cm

\newcommand{\bav}{\begin{array}{cccc}}

\newcommand{\be}{\begin{equation}}
\newcommand{\ee}{\end{equation}}
\newcommand{\bea}{\begin{eqnarray}}
\newcommand{\eea}{\end{eqnarray}}

\title{Singlet-Doublet Majorana Dark Matter and Neutrino Mass in a minimal Type-I Seesaw Scenario}
\author[a]{Manoranjan Dutta,}
\emailAdd{ph18resch11007@iith.ac.in}
\author[b]{Subhaditya Bhattacharya,}
\emailAdd{subhab@iitg.ac.in}
\author[c]{Purusottam Ghosh,}
\emailAdd{purusottamghosh@hri.res.in}
\author[a]{Narendra Sahu,}
\emailAdd{nsahu@phy.iith.ac.in}
  \affiliation[a]{Department of Physics, Indian Institute of Technology Hyderabad,\\ Kandi, Telangana-502285, India.}
  \affiliation[b]{Department of Physics, Indian Institute of Technology Guwahati,\\ North Guwahati, Assam-781039, India.}
  \affiliation[c]{Regional Centre for Accelerator-based Particle Physics, Harish-Chandra Research Institute, HBNI,\\ Chhatnag Road, Jhunsi, Allahabad - 211 019, India }
  
\abstract{In a bid to simultaneous explanation of dark matter (DM) and tiny but non-zero masses of left-handed 
neutrinos, we propose a minimal extension of the Standard Model (SM) by a vector-like fermion doublet and three 
right handed (RH) singlet neutrinos. The DM arises as a mixture of the neutral component of the fermion doublet 
and one of the RH neutrinos, both assumed to be odd under an additional $\mathcal{Z}_2$ symmetry. As a result, the 
DM emerges to be a dominantly Majorana particle and escapes from $Z$-mediated direct search constraints to mark a 
significant difference from singlet-doublet Dirac DM. The other two $\mathcal{Z}_2$ even heavy RH neutrinos give 
rise masses and mixing of light neutrinos via Type-I Seesaw mechanism. The particle content automatically allows 
us to extend the model by a gauged $U(1)_{B-L}$ symmetry, which is anomaly free and brings an additional portal 
between DM and SM particles. Relic density and direct search allowed parameter space for both the 
cases are investigated through detailed numerical scan, while collider search strategies are also indicated.}
\preprint{ HRI-RECAPP-2020-007}

\begin{document}

\maketitle
\flushbottom

\setcounter{footnote}{0}
\renewcommand*{\thefootnote}{\arabic{footnote}}
\section{Introduction}
\label{sec:intro}
Astrophysical observations like galaxy rotation curves, gravitational lensing, Cosmic Microwave Background (CMB) acoustic 
fluctuations etc. provide compelling evidences towards the existence of Dark Matter (DM)\cite{Bertone:2004pz, Jungman:1995df}, 
a form of matter that is electromagnetically inert and hence extremely difficult to detect, but can be inferred from its 
gravitational affects. In fact, satellite borne experiments like WMAP and PLANCK \cite{Hinshaw:2012aka, Ade:2013zuv}, which 
measure anisotropies in CMB, established that DM constitute almost $85\%$ of the total matter content and $26.8\%$ of the total 
energy budget of the universe. Even after this tantalising hint, we have no answer to the question what DM actually is. 
DM as a fundamental particle answers many puzzles together like structure formation, self interaction, rotation curve etc., 
hence studied elaborately. Since no SM particle resembles the properties of a DM particle expected to have, it is believed that 
DM is essentially one or more particles beyond the Standard Model (BSM) content. Several BSM scenarios have been formulated 
to explain the particle nature of the DM, with additional field content and stabilising symmetry. Amongst different class of 
possibilities, Weekly Interacting Massive Particles (WIMP) has been the most popular due to its phenomenological richness, where 
DM can be explained as the thermal relics of the universe \cite{Kolb:1990vq}. 

Another equally important puzzle in particle physics is the tiny neutrino mass which has been established by the solar 
and atmospheric neutrino oscillation experiments like T2K~\cite{Abe:2011sj, Abe:2013hdq}, Double Chooz~\cite{Abe:2011fz, DoubleChooz:2019qbj}, Daya Bay~\cite{An:2012eh, An:2017osx, Adey:2018zwh}, Reno~\cite{Ahn:2012nd} and MINOS~\cite{Adamson:2013whj, Adamson:2014vgd}. Besides, the nature of neutrinos, whether Dirac or Majorana, is also not known. Neutrinoless double beta decay experiments~\cite{Azzolini:2018dyb} perhaps will shed light onto it. Within the SM, neutrinos are assumed massless with no 
right handed (RH) neutrinos. Even if RH neutrinos are incorporated to the SM, the required Yukawa coupling to explain sub-eV 
neutrino mass through spontaneous symmetry breaking via Dirac mass term turns out to be as tiny as $10^{-12}$, almost six orders of 
magnitude smaller than the electron Yuwaka coupling and seems pretty unnatural. Assuming that the neutrinos are Majorana, which 
violates lepton number by two units, the tiny neutrino masses can be realised via the dimension five gauge invariant effective 
Weinberg operator $LLHH/\Lambda$, where $\Lambda$ denotes the scale of new physics and  $L$, $H$ are respectively the lepton 
and Higgs doublets of the SM~\cite{Weinberg:1979sa}. After electroweak symmetry breaking (EWSB), the SM neutrinos acquire 
sub-eV masses given by $M_\nu = \langle H\rangle^2/\Lambda$. One possibility of generating this operator is to assume the 
presence of heavy RH neutrinos in the early universe, where the scale of new physics ($\Lambda$) is decided by the mass of 
RH neutrinos. Thus it is straightforward to see that for tiny neutrino mass of the order of $M_\nu \sim 0.1 \rm eV$, the 
new physics scale requires to be very heavy ($\Lambda \sim 10^{14} \rm GeV$) when the involved couplings are of order one. 
This is usually referred as type-I seesaw mechanism~\cite{Minkowski:1977sc, Yanagida:1981xy, GellMann:1980vs, Mohapatra:1979ia}.

While the origin of DM and neutrino mass is hitherto unknown, it is highly appealing and economical to find a model having 
simultaneous solution of both. In fact, such models are expected to have constrained parameter space in comparison to their 
individual counterpart and hence can be probed at ongoing and future terrestrial experiments. Motivated by 
this, here we consider a simple extension of the SM to explain simultaneously the sub-eV masses of neutrinos and DM 
content of the universe.  

We consider a singlet-doublet WIMP like fermion DM~\cite{ArkaniHamed:2005yv,Freitas:2015hsa,Cynolter:2015sua, Calibbi:2015nha, Abe:2014gua, Cheung:2013dua, Cohen:2011ec, Enberg:2007rp, DEramo:2007anh, Mahbubani:2005pt, Banerjee:2016hsk, DuttaBanik:2018emv, Horiuchi:2016tqw, Restrepo:2015ura, Badziak:2017the,Betancur:2020fdl,Abe:2019wku, Abe:2017glm,Bhattacharya:2017sml,Barman:2019tuo,Bhattacharya:2018cgx,Bhattacharya:2016rqj,Bhattacharya:2015qpa,
Bhattacharya:2018fus,Konar:2020vuu}. The motivation of considering a singlet-doublet fermion DM has already been established; this is because a purely singlet case requires a higher dimensional 
effective operator for DM-SM interaction, which is mostly ruled out from direct search bound excepting for the Higgs resonance region, while the pure doublet case is also ruled out from relic density and direct search bound upto several TeVs of DM 
mass making the model inaccessible to probe.
Our model consists of a vector-like fermion doublet $\Psi^T=(\psi^0, \psi^-)$ and three right 
handed neutrinos ($N_{R_i},i=1,2,3)$. A $\mathcal{Z}_2$ symmetry is imposed under which the doublet $\Psi$ and one 
of the right handed neutrinos, say $N_{R_1}$ are odd, while other particles are even. As a result there is mixing between 
the neutral component of the doublet and the singlet through the Yukawa interaction and DM emerges out to be a 
mixed state of the doublet $\psi^0$ and $N_{R_1}$ after EWSB. Due to Majorana mass of the RH singlet $N_{R_1}$, the DM 
is dominantly a Majorana particle. As a result it escapes the $Z$-mediated vector current direct search interaction 
and provide a distinction from the earlier vector like singlet-doublet DM~\cite{Bhattacharya:2017sml, Barman:2019tuo,Bhattacharya:2018cgx, Bhattacharya:2016rqj,Bhattacharya:2015qpa,Bhattacharya:2018fus}. 
The field content permits us to extend the model easily to a gauged $U(1)_{B-L}$ scenario, which allows an additional gauge 
mediated interaction for DM. We find the relic density and direct search allowed parameter space for both the 
cases and also indicate possible collider search strategies. The neutrino mass arises from the Yukawa interaction of 
$\mathcal{Z}_2$ even RH neutrinos together with Majorana mass term in a minimal Type-I Seesaw framework. Since two RH 
neutrinos take part in the seesaw, one of the light neutrino mass is exactly zero. The masses of RH neutrinos, including 
the one which constitutes DM, originate from the $U(1)_{\rm B-L}$ symmetry breaking scale. We assume their masses to be 
of same order and derive constraints from lepton flavour violating processes like $\mu \to e\gamma$ .

The paper has been arranged as follows. In section-\ref{Model}, we explain the details of the model, followed by a summary of 
different theoretical and experimental constraints. We discuss the relic abundance of dark matter in section-\ref{relics} and 
direct detection in section-\ref{direct_detection}. Then we discuss the gauged $U(1)_{B-L}$ extension of the model in section-\ref{gauged_model}. 
We briefly summarise collider search strategy for both the cases in section-\ref{collider}. 
In section-\ref{neutrino}, we discuss the light neutrino mass and finally conclude in section-\ref{conclusion}.
\section{The Model for singlet-doublet Majorana DM}\label{Model}

In this work the SM has been extended by one vector-like fermion doublet (VLFd) $\Psi^T=(\psi^0, \psi^-)$ (with hypercharge 
$Y=-1$, where we use $Q=T_3+Y/2$) and three heavy right handed neutrinos (RHN) ${N}_{R_i} (i=1,2,3)$, which are singlets under 
the SM gauge group. All the newly added particles are also singlet under $SU(3)_C$, i.e. colour neutral. An additional 
$\mathcal{Z}_2$ symmetry is imposed under which $\Psi$ and $N_{R_1}$ are odd, while all other fields are even. 
It is well known that the stability of DM is ensured by some additional symmetry and $\mathcal{Z}_2$ serves as the minimal one.
The quantum numbers of the BSM fields under $SU(3)_c\times SU(2)_L\times U(1)_Y \times \mathcal{Z}_2$ are listed in Table \ref{tab:tab1}. 
\begin{table}[h]
\resizebox{\linewidth}{!}{
 \begin{tabular}{|c|c|c|c|}
\hline \multicolumn{2}{|c}{Fields}&  \multicolumn{1}{|c|}{ $\underbrace{ SU(3)_C \otimes SU(2)_L \otimes U(1)_Y}$ $\otimes   \mathcal{Z}_2 $} \\ \hline
\multirow{2}{*} 
{VLFd} & $\Psi=\left(\begin{matrix} \psi^0 \\ \psi^- \end{matrix}\right)$& ~~~1 ~~~~~~~~~~~2~~~~~~~~~~~-1~~~~~~~~~ - \\
\hline
{RHNs} &  ${N}_{R_1}$& ~~~1 ~~~~~~~~~~~1~~~~~~~~~~~~0~~~~~~~~~ - \\ [0.5em] \cline{2-3}
       &  ${N}_{R_2}$&  ~~~1 ~~~~~~~~~~~1~~~~~~~~~~~~0~~~~~~~~ + \\ [0.5em] \cline{2-3}
       &  ${N}_{R_3}$&  ~~~1 ~~~~~~~~~~~1~~~~~~~~~~~~0~~~~~~~~ + \\
\hline
\hline
Higgs doublet & $H=\left(\begin{matrix} w^+ \\ \frac{h+v+iz}{\sqrt{2}} \end{matrix}\right)$ & ~~~1 ~~~~~~~~~~~2~~~~~~~~~~~~1~~~~~~~~~+ \\
\hline
\end{tabular}
}
\caption{\footnotesize{Charge assignment of BSM fields with SM Higgs doublet under the gauge group $\mathcal{G} \equiv \mathcal{G}_{\rm SM} \otimes \mathcal{Z}_2$  where $\mathcal{G}_{\rm SM}\equiv SU(3)_C \otimes SU(2)_L \otimes U(1)_Y$.}}
    \label{tab:tab1}
\end{table}
The Lagrangian of the model (as guided by Table \ref{tab:tab1}) is given by:
\begin{equation}
\label{model_Lagrangian}
 \mathcal{L} = \mathcal{L}_{SM} + \overline{\Psi} \left( i\gamma^\mu D_\mu - M \right) \Psi +\overline{{N}_{R_i}} i\gamma^\mu\partial_\mu {N}_{R_i}- (\frac{1}{2}M_{R_i} \overline{{N}_{R_i}} \left({N}_{R_i}\right)^c + h.c) + \mathcal{L}_{yuk}.    
\end{equation}
Apart from kinetic pieces, it is straightforward to note that since $\Psi$ is a vector-like Dirac fermion, it possesses a 
bare Dirac mass term $M$, while all the three right handed neutrinos have Majorana mass $M_{R_i}$. Also worthy to note 
that $D_\mu$ denotes the covariant derivative involving the $SU(2)_L$ gauge boson triplet $W_\mu^a$ ~$(a=1,2,3)$ and $U(1)_Y$ gauge boson $B_\mu$ given by:
 \begin{equation}
 D_\mu = \partial_\mu - i\frac{g}{2}\tau_a.W_\mu^a - ig'\frac{Y}{2}B_\mu;
 \end{equation}
where $\tau_a$ are Pauli spin matrices (generators of $SU(2)$), $g$ and $g^{'}$ denote $SU(2)$ and $U(1)$ coupling strength 
respectively. This ensures that $\Psi$ has $SU(2)$ gauge interaction with the SM.

We note that the Yukawa interaction plays the key role in this model and can be written as:
\begin{equation}
    -\mathcal{L}_{yuk} =\left[ \frac{Y_1}{\sqrt{2}}\overline{\Psi}\Tilde{H}\big(N_{R_1}+(N_{R_1})^c\big) +h.c\right]+ \left(Y_{j \alpha }\overline{N_{R_j}} \Tilde{H^\dagger} L_{\alpha} + h.c.\right).
    \label{eq:yukawa}
\end{equation}
 where $\Tilde{H}=i\tau_2 H^{*}$  and $L$ denotes SM lepton doublet with indices $j=2,3$ and $\alpha = e,\mu, \tau$. 
 ${N}_{R_1}$ being odd under $\mathcal{Z}_2$ has Yukawa coupling to fermion doublet $\Psi$ and determines the DM of the model after spontaneous symmetry breaking (SSB), as elaborated below. $N_{R_2}$ and ${N}_{R_3}$ being $\mathcal{Z}_2$ even, do not couple to 
$\Psi$, but couple to the SM lepton doublets and hence generate Dirac masses for SM neutrinos after SSB, which will be 
discussed in details later.

\subsection{Masses and mixing of dark sector particles}
\label{dark_matter}

Thanks to the Yukawa interaction given in \ref{eq:yukawa}, the electromagnetic charge neutral component of 
$\Psi$ {\it viz.} $\psi^0$ and $N_{R_1}$ mixes after the SM Higgs acquires vacuum expectation value (vev): 
$\langle H\rangle=\frac{1}{\sqrt{2}}\left(\begin{matrix} 0 \\ v \end{matrix}\right)$. The mass terms for these fields can then 
be written together as:

\begin{equation}
  -\mathcal{L}_{mass} = M\overline{\psi^0_L}\psi^0_R + \frac{1}{2}M_{R_1}\overline{N}_{R_1}(N_{R_1})^c + \frac{m_D}{\sqrt{2}} (\overline{\psi^0_L}N_{R_1}+\overline{\psi^0_R}(N_{R_1})^c) + h.c. 
  \label{l_mass}
\end{equation}

where $m_D=\frac{{Y_1 v }}{\sqrt{2}}$, where $ v = 246$ GeV. Writing these mass terms in the basis 
$ ((\psi^0_R)^c, \psi^0_L, (N_{R_1})^c)^T$, we get the following mass matrix:
\begin{equation}\label{dark-sector-mass}
\mathcal{M}=
\left(
\begin{array}{ccc}
0 &M &\frac{m_D}{\sqrt{2}}\\
M &0 &\frac{m_D}{\sqrt{2}}\\
\frac{m_D}{\sqrt{2}} &\frac{m_D}{\sqrt{2}} &M_{R_1}\\
\end{array}
\right)\,.
\end{equation}
In the above equation, assuming $\mathcal{M}$ is symmetric, 

it can be diagonalised by a single unitary matrix 
$\mathcal{U (\theta)}=U_{13}(\theta_{13}=\theta).U_{23}(\theta_{23}=0).U_{12}(\theta_{12}=\frac{\pi}{4})$, which is essentially characterised by a single angle $\theta_{13}=\theta$. So we diagonalize the mass matrix $\mathcal{M}$ by $\mathcal{U}.\mathcal{M}.\mathcal{U}^T = \mathcal{M}_{Diag.}$, 
where the unitary matrix $\mathcal{U}$ is given by:
\begin{equation}
\label{diagonalizing_matrix}
\mathcal{U}= \left(
\begin{array}{ccc}
1 & 0 & 0\\
0 & e^{i\pi/2} & 0\\
0 & 0 & 1\\
\end{array}
\right)
\left(
\begin{array}{ccc}
\frac{1}{\sqrt{2}}\cos\theta &\frac{1}{\sqrt{2}}\cos\theta &\sin\theta\\
-\frac{1}{\sqrt{2}} &\frac{1}{\sqrt{2}} &0\\
-\frac{1}{\sqrt{2}}\sin\theta &-\frac{1}{\sqrt{2}}\sin\theta &\cos\theta\\
\end{array}
\right)\\,
\end{equation}
where the extra phase matrix is multiplied to make sure all the eigenvales are positive.

The diagonalisation of 
the mass matrix \ref{dark-sector-mass} requires:
\begin{equation}
\tan2\theta = \frac{2 m_D}{M-M_{R_1}} .
\end{equation}
 
The physical states that emerge are defined as $\chi_{_i}=\frac{\chi_{_{iL}}+(\chi_{_{iL}})^c}{\sqrt{2}}~(i=1,2,3)$ and are 
related to the unphysical states as:
\begin{equation}
\begin{aligned}
\chi_{_{1L}} & = \frac{\cos\theta}{\sqrt{2}}( \psi^0_L+(\psi^0_R)^c  )+\sin\theta (N_{R_1})^c,
\\
\chi_{_{2L}} & =  \frac{i}{\sqrt{2}}(\psi^0_L - (\psi^0_R)^c), 
\\
\chi_{_{3L}} & =  -\frac{\sin\theta}{\sqrt{2}}(\psi^0_L + (\psi^0_R)^c ) +\cos\theta(N_{R_1})^c \,.
\end{aligned}
\end{equation}

All the three physical states $\chi_{_1}, \chi_{_2} ~{\rm and} ~\chi_{_3}$ are therefore of Majorana nature and their 
mass eigenvalues can be expressed respectively as,
\begin{equation}
\begin{aligned}
m_{\chi_{_1}} & = M \cos^2\theta + M_{R_1} \sin^2\theta + m_D\sin2\theta,
\\
m_{\chi_{_2}} & = M,
\\
m_{\chi_{_3}} & = M_{R_1} \cos^2\theta + M\sin^2\theta - m_D\sin2\theta\,.
\\
\end{aligned}
\end{equation}

In the small mixing limit ($\theta\to 0$), the eigenvalues can be further simplified as,
\begin{equation}
\begin{aligned}
m_{\chi_{_1}} & \approx M + \frac{ m^2_D}{M - M_{R_1}},
\\
m_{\chi_{_2}} & = M,
\\
m_{\chi_{_3}} & \approx M_{R_1} - \frac{m^2_D}{M - M_{R_1}}.
\\
\end{aligned}
\end{equation}
where we have assumed $m_D << M, M_{R_1}$. Hence it is clear that $m_{\chi_{_1}}>m_{\chi_{_2}}>m_{\chi_{_3}}$ 
and $\chi_{_3}$ becomes the stable DM candidate. We may note here that the analysis taken up before in~\cite{Barman:2019tuo,Bhattacharya:2015qpa,Bhattacharya:2018cgx,Bhattacharya:2016rqj,Bhattacharya:2017sml,Bhattacharya:2018fus}, where the $\mathcal{Z}_2$ odd doublet $\Psi$ mixes with a {\it vector like singlet}, providing a Dirac DM state with
one heavy electromagnetically charged Dirac state as opposed to two heavy Majorana states here.

Using the relation $\mathcal{U}.\mathcal{M}.\mathcal{U}^T = \mathcal{M}_{Diag.}$, one can express $Y_1$, $M$ and $M_{R_1}$ in 
terms of the physical masses and the mixing angle as,
\begin{equation}
\begin{aligned}
Y_1 & = \frac{\sqrt{2}~\Delta M ~\sin2\theta}{v},
\\
M & = m_{\chi_{_1}} \cos^2\theta +m_{\chi_{_3}} \sin^2\theta, 
\\
M_{R_1} & = m_{\chi_{_3}} \cos^2\theta +  m_{\chi_{_1}}\sin^2\theta; 
\\
\end{aligned}
\label{y1}
\end{equation}

where $\Delta M=(m_{\chi_{_1}}-m_{\chi_{_3}})$. We can also see that in the limit of $m_D<<M$, $m_{\chi_{_1}}\simeq m_{\chi_{_2}}=M$, 
where $M$ is the mass of electrically charged components $\psi^{\pm}$ of vector-like fermion doublet $\Psi$. The phenomenology of dark sector is therefore governed mainly by the three independent parameters, DM mass, splitting with the heavier neutral component, and doublet-singlet mixing :

\begin{equation}
 \textrm{Dark~Parameters}: ~~~~\{~~ m_{\chi_{_3}},\Delta M=(m_{\chi_{_1}}-m_{\chi_{_3}})\approx(m_{\chi_{_2}}-m_{\chi_{_3}}),~ \sin\theta~~ \}.
 \label{eq:parameters1}
\end{equation}

\subsection{Theoretical and Experimental constraints }
\label{sec:constraints}
\noindent $\bullet$\textbf{ Perturbativity:} 
In order to maintain perturbativity of the model, Yukawa couplings should satisfy the following limits:
\begin{align}
 |Y_1| < \sqrt{4\pi}, ~~~~~|Y_{\alpha j}| < \sqrt{4\pi}~~~ .
\end{align}
\noindent $\bullet$\textbf{LEP limits:} 
 LEP exclusion bound on charged fermion mass $m_{\psi^\pm}=M >102.7$ GeV~\cite{Abdallah:2003xe}. The bound from LHC applies to 
 a typical case of type III seesaw model, for which  $m_{\psi^\pm}=M \gtrsim 800$ GeV~\cite{Sirunyan:2017qkz, Sirunyan:2019bgz}. Note that we do not abide by the bound from LHC as 
 the decay channels are widely different.
\\
\noindent $\bullet$\textbf{Relic Density and Direct Search of Dark Matter:}
The observed number density of DM is constrained by the combined WMAP~\cite{Hinshaw:2012aka} and PLANCK~\cite{Ade:2013zuv} data as: 
\bea
0.1166 \leq \Omega_{DM} h^2\leq 0.1206.
\eea

For direct search, we have used the current stringent bounds from non -observation of DM at XENON-1T~\cite{Aprile:2018dbl} ($\sim 10^{-47}~{\rm cm^2}$). We also note that the fluctuation recently observed at XENON 1T at $\sim$ KeV scale~\cite{Aprile:2020tmw} do not apply to our case. 

\section{Relic Abundance of singlet-doublet Majorana Dark Matter}
\label{relics}
\subsection{Annihilation/Coannihilation processes and Boltzmann Equations}

\begin{figure}[!hb]
    \centering
    \subfloat{{\includegraphics[width=4.6cm]{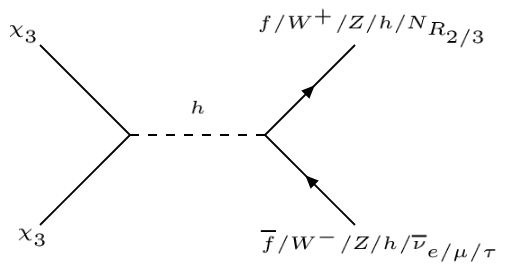} }}%
    \qquad
    \subfloat{{\includegraphics[width=2.1cm]{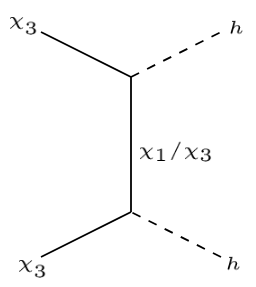} }}%
    \qquad
    \subfloat{{\includegraphics[width=2.6cm]{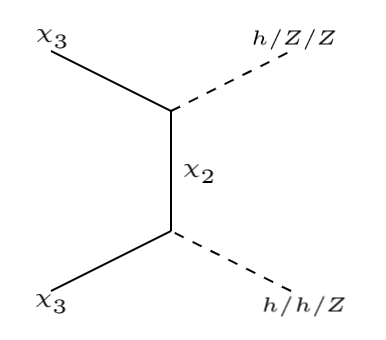} }}%
    \qquad
     \subfloat{{\includegraphics[width=2.5cm]{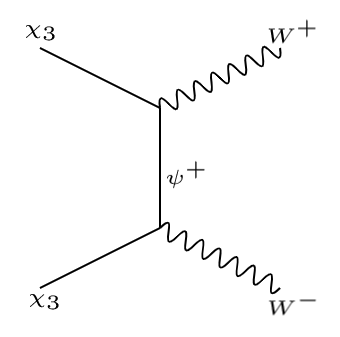} }}%
    \caption{\footnotesize{Annihilation channels to the SM through which the DM ($\chi_{_3}$) density depletes.}}%
    \label{annihilation}%
\end{figure}

\begin{figure}[!ht]
    \centering
    \subfloat{{\includegraphics[width=3cm]{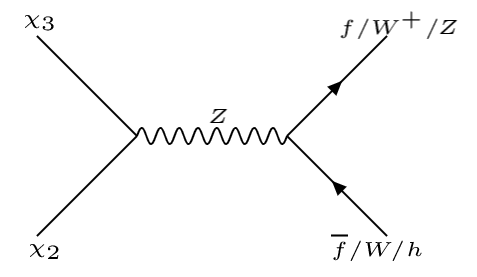} }}%
    \qquad
    \subfloat{{\includegraphics[width=3.2cm]{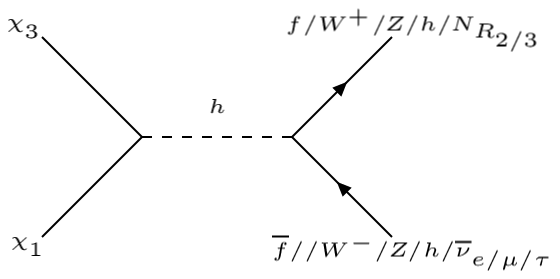} }}%
    \qquad
    \subfloat{{\includegraphics[width=3cm]{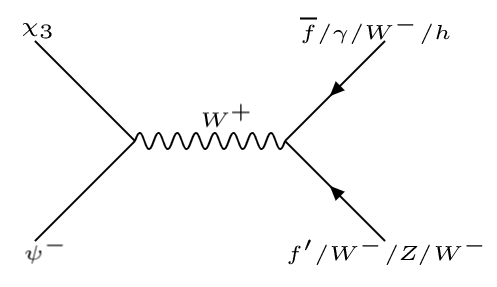} }}%
    \qquad\\
    \subfloat{{\includegraphics[width=2.6cm]{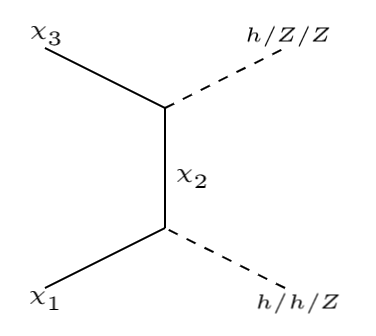} }}%
    \qquad
    \subfloat{{\includegraphics[width=2.5cm]{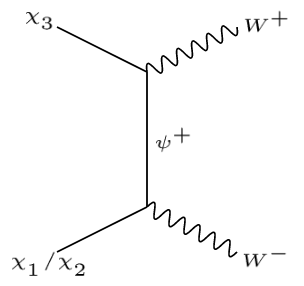} }}%
    \qquad
    \subfloat{{\includegraphics[width=2.6cm]{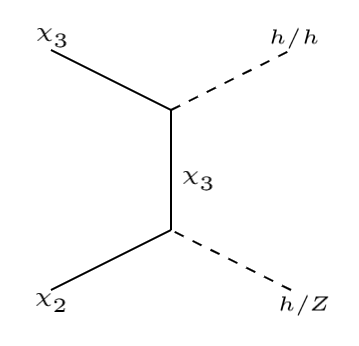} }}%
    \\
    \qquad
    \subfloat{{\includegraphics[width=2.6cm]{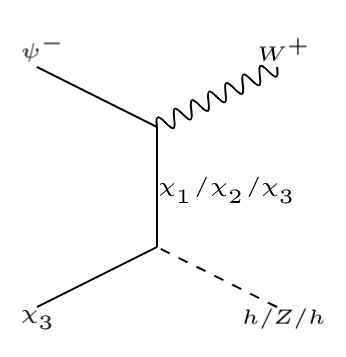} }}%
    \qquad
    \subfloat{{\includegraphics[width=2.6cm]{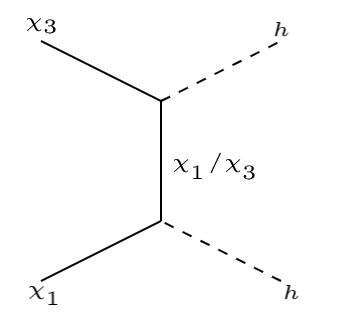} }}%
    \qquad
    \subfloat{{\includegraphics[width=2.6cm]{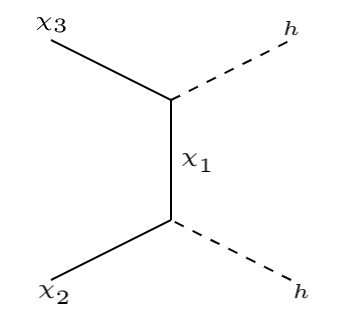} }}%
    \caption{\footnotesize{Coannihilation channels of DM ($\chi_{_3}$) with $\chi_{_1}$, $\chi_{_2}$ and $\psi^\pm$. }}%
    \label{coanni}
\end{figure}

\begin{figure}[!ht]
    \centering
    \subfloat{{\includegraphics[width=2.5cm]{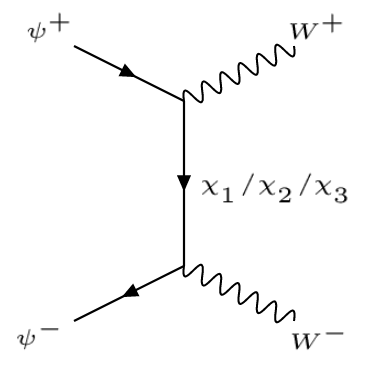} }}%
    \qquad
    \subfloat{{\includegraphics[width=2.7cm]{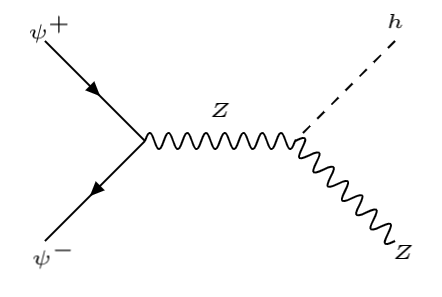} }}%
    \qquad
    \subfloat{{\includegraphics[width=2.7cm]{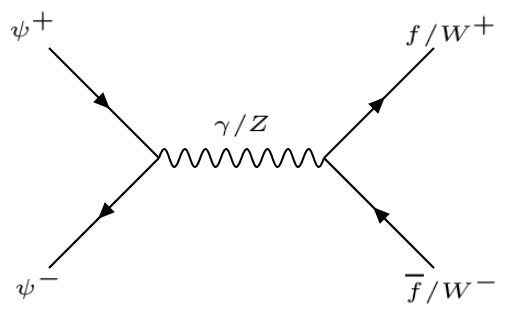} }}%
    \qquad
    \subfloat{{\includegraphics[width=2.5cm]{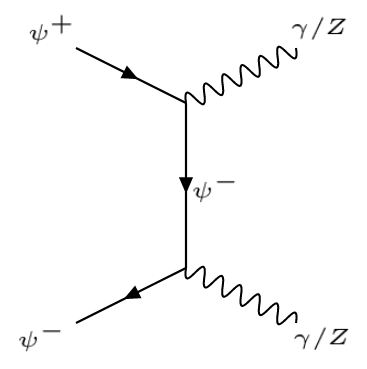} }}%
    \caption{\footnotesize{Annihilation channels of $\psi^+$ and $\psi^-$ that contribute to coannihilation of DM ($\chi_{_3}$).}}%
    \label{chargedcoanni}
\end{figure}

The basic assumption for calculation of relic density of the DM here is to assume that DM is in equilibrium with 
thermal bath due to its non-negligible interaction with the SM particles in the early universe. It thereafter 
`freezes out' from the hot soup of the SM particles via the number changing processes through which DM number density 
depletes as the universe expands to provide correct relic density. The dark sector consists of DM $\chi_{_3}$ as well 
as heavy neutral components $\chi_{_1}$, $\chi_{_2}$ and charged components $\psi^\pm$ (all odd under the dark symmetry 
$\mathcal{Z}_2$). The number density of DM ($\chi_{_3}$) is therefore governed by its annihilation as well as coannihilations 
with other dark sector particles ($\chi_{_1}$, $\chi_{_2}$ and $\psi^\pm$) into SM final states. Feynman diagrams of relevant 
annihilation and coannihilation processes are shown in Fig.~\ref{annihilation}, Fig.~\ref{coanni} and Fig.~\ref{chargedcoanni}. 
The DM-SM interaction terms which essentially contribute to the relic density has been detailed in Appendix~\ref{dm_sm_int}. 


The relic density of DM in this scenario can be estimated by solving the Boltzmann equation in the following form:
\begin{equation}
\frac{dn}{dt} + 3Hn = -\langle\sigma v\rangle_{eff}\Big(n^2 - n^2_{eq}\Big),
\end{equation}
where $n$ denotes number density of DM, i.e. $n\sim n_{\chi_3}$ and $n_{eq}= g(\frac{mT}{2\pi})^{3/2}exp(-m/T)$ denotes 
equilibrium distribution, which DM is initially subjected to. Then it freezes out depending on $\langle\sigma v\rangle_{eff}$, 
which takes into account all number changing process listed in Fig.~\ref{annihilation}, Fig.~\ref{coanni} and Fig.~\ref{chargedcoanni} 
as well as those annihilations involving $\chi_{_{1,2}}$ (although the contribution is very small) and can be estimated as follows:

\begin{equation}
\begin{aligned}
\langle\sigma v\rangle_{eff} &= \frac{g^2_3}{g^2_{eff}}\langle\sigma v\rangle_{\overline{\chi_{_3}}\chi_{_3}}+ \frac{2g_3 g_2}{g^2_{eff}}\langle\sigma v\rangle_{\overline{\chi_{_3}}\chi_{_2}}\Big(1+\frac{\Delta M}{m_{\chi_{_3}}}\Big)^{\frac{3}{2}}\exp(-x \frac{\Delta M}{m_{\chi_{_3}}})\\
& + \frac{2g_3 g_1}{g^2_{eff}}\langle\sigma v\rangle_{\overline{\chi_{_3}}\chi_{_1}}\Big(1+\frac{\Delta M}{m_{\chi_{_3}}}\Big)^{\frac{3}{2}}\exp(-x \frac{\Delta M}{m_{\chi_{_3}}})\\
& + \frac{2g_3 g_4}{g^2_{eff}}\langle\sigma v\rangle_{\overline{\chi_{_3}}\psi^\pm}\Big(1+\frac{\Delta M}{m_{\chi_{_3}}}\Big)^{\frac{3}{2}}\exp(-x \frac{\Delta M}{m_{\chi_{_3}}})\\
& + \frac{2g_2 g_4}{g^2_{eff}}\langle\sigma v\rangle_{\overline{\chi_{_2}}\psi^\pm}\Big(1+\frac{\Delta M}{m_{\chi_{_3}}}\Big)^3 \exp(-2x \frac{\Delta M}{m_{\chi_{_3}}})\\
& + \frac{2g_1 g_4}{g^2_{eff}}\langle\sigma v\rangle_{\overline{\chi_{_1}}\psi^\pm}\Big(1+\frac{\Delta M}{m_{\chi_{_3}}}\Big)^3 \exp(-2x \frac{\Delta M}{m_{\chi_{_3}}})\\
& + \frac{g^2_2}{g^2_{eff}}\langle\sigma v\rangle_{\overline{\chi_{_2}}\chi_{_2}}\Big(1+\frac{\Delta M}{m_{\chi_{_3}}}\Big)^3 \exp(-2x \frac{\Delta M}{m_{\chi_{_3}}})\\
& + \frac{g_1g_2}{g^2_{eff}}\langle\sigma v\rangle_{\overline{\chi_{_1}}\chi_{_2}}\Big(1+\frac{\Delta M}{m_{\chi_{_3}}}\Big)^3 \exp(-2x \frac{\Delta M}{m_{\chi_{_3}}})\\
& + \frac{g^2_1}{g^2_{eff}}\langle\sigma v\rangle_{\overline{\chi_{_1}}\chi_{_1}}\Big(1+\frac{\Delta M}{m_{\chi_{_3}}}\Big)^3 \exp(-2x \frac{\Delta M}{m_{\chi_{_3}}})\\
& + \frac{g^2_4}{g^2_{eff}}\langle\sigma v\rangle_{\psi^+\psi^-}\Big(1+\frac{\Delta M}{m_{\chi_{_3}}}\Big)^3 \exp(-2x\frac{\Delta M}{m_{\chi_{_3}}}),\\
\end{aligned}\\
\label{Boltzmann}
\end{equation}
where $\Delta M=m_i - m_{\chi_3}$ and $m_i$ denotes the mass of $\chi_{_1}$, $\chi_{_2}$ and $\psi^\pm$. 
Here we have defined $g_{eff}$ as the effective degrees of freedom given by,
\begin{equation}
\begin{aligned}
g_{eff} &= g_3 + g_2 \Big(1+\frac{\Delta M}{m_{\chi_{_3}}}\Big)^{\frac{3}{2}}\exp(-x \frac{\Delta M}{m_{\chi_{_3}}})\\
&+ g_1 \Big(1+\frac{\Delta M}{m_{\chi_{_3}}}\Big)^{\frac{3}{2}}\exp(-x \frac{\Delta M}{m_{\chi_{_3}}})  + g_4 \Big(1+\frac{\Delta M}{m_{\chi_{_3}}}\Big)^{\frac{3}{2}}\exp(-x\frac{\Delta M}{m_{\chi_{_3}}}),
\end{aligned}
\end{equation}
where $g_3$, $g_2$, $g_1$ and $g_4$ are the internal  degrees of freedom of $\chi_{_3}$, $\chi_{_2}$, $\chi_{_1}$ and $\psi^\pm$ respectively. 
The dimensionless parameter $x$ is defined as $x=\frac{m_{\chi_{_3}}}{T}$. We also note that the contributions from processes which do not directly involve DM, like $\psi^+\psi^-$ in effective annihilation 
$\langle\sigma v\rangle_{eff}$ is further Boltzmann suppressed by $\exp(-2x\frac{\Delta M}{m_{\chi_{_3}}})$.
The relic density of the DM ($\chi_{_3}$) then 
can be given by {\cite{Griest:1990kh},\cite{Chatterjee:2014vua},\cite{Patra:2014sua}:
\begin{equation}
\Omega_{\chi_{_3}}h^2 = \frac{1.09\times 10^9 GeV^{-1}}{g^{1/2}_* M_{Pl}}\frac{1}{J(x_f)}
\end{equation}
where $g_*=106.7$ and $J(x_f)$ is given by 
\begin{equation}
J(x_f)=\int_{x_f}^{\infty} \frac{\langle\sigma v\rangle_{eff}}{x^2} dx  ~~ .
\end{equation}
Here $x_f = \frac{m_{\chi_{_3}}}{T_f}$, where $T_f$ denotes the freeze-out temperature of the DM. We may note here that 
for correct relic $x_f \simeq 20$.

It is worthy to mention here that we have adopted a numerical way of computing annihilation cross-section and relic 
density by inserting the model into the package {\tt MicrOmegas}~\cite{Belanger:2008sj}, where the model files are generated 
using another package {\tt FeynRule}~\cite{Christensen:2008py, Alloul:2013bka}.

\subsection{Parameter Space Scan}

In order to understand the DM relic density, let us first study the dependence on important relevant parameters: the 
mass of DM ($m_{\chi_{3}}$), the mass splitting ($\Delta M$) between the DM $\chi_{_3}$ and the next-to-lightest stable 
particle (NLSP) $\chi_{_2}$ and the mixing angle $\sin \theta$. Note that the charged components of $\Psi$ namely 
$\psi^\pm$ which contribute dominantly to the coannihilation channels has the same mass as that of $\chi_{_2}$, 
{\it i.e.,} $m_{\chi_2}=m_{\psi^\pm}$. Variation of relic density of DM $\chi_{_3}$ is shown in Fig.~\ref{fig:relicsplot1} 
as a function of its mass for different choices of $\Delta M$ = 1-10 GeV, 10-30 GeV, 30-50 GeV, 50-100 GeV shown by 
different colour shades as in the inset of the figure and for different choices of $\sin\theta = 0.01, 0.1, 0.3, 0.5$ 
in the top left, top right, bottom left and bottom right panels respectively.

\begin{figure}[h]
$$
 \includegraphics[height=5.0cm]{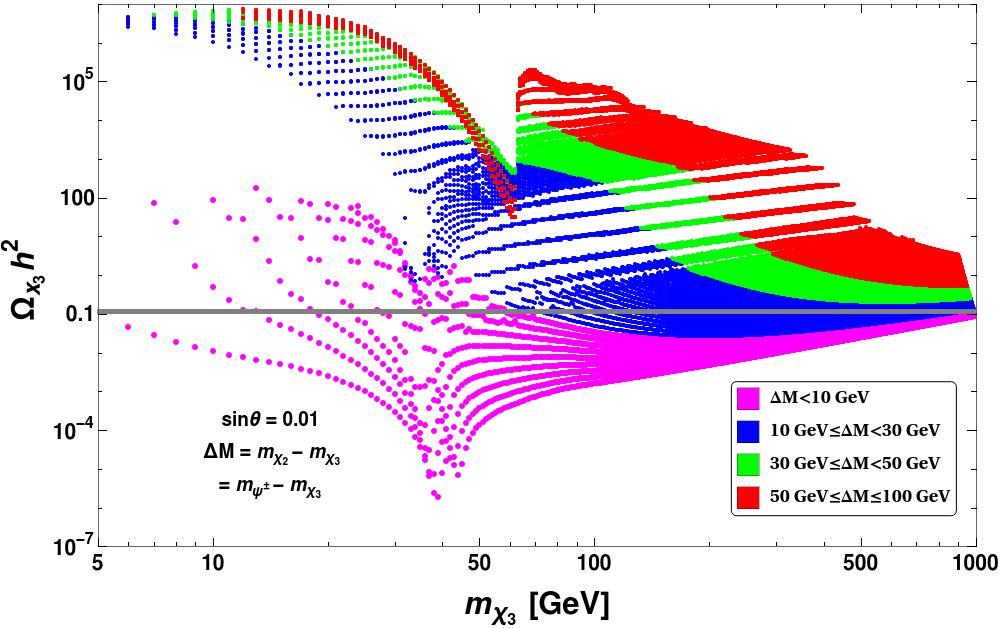} 
 \includegraphics[height=5.0cm]{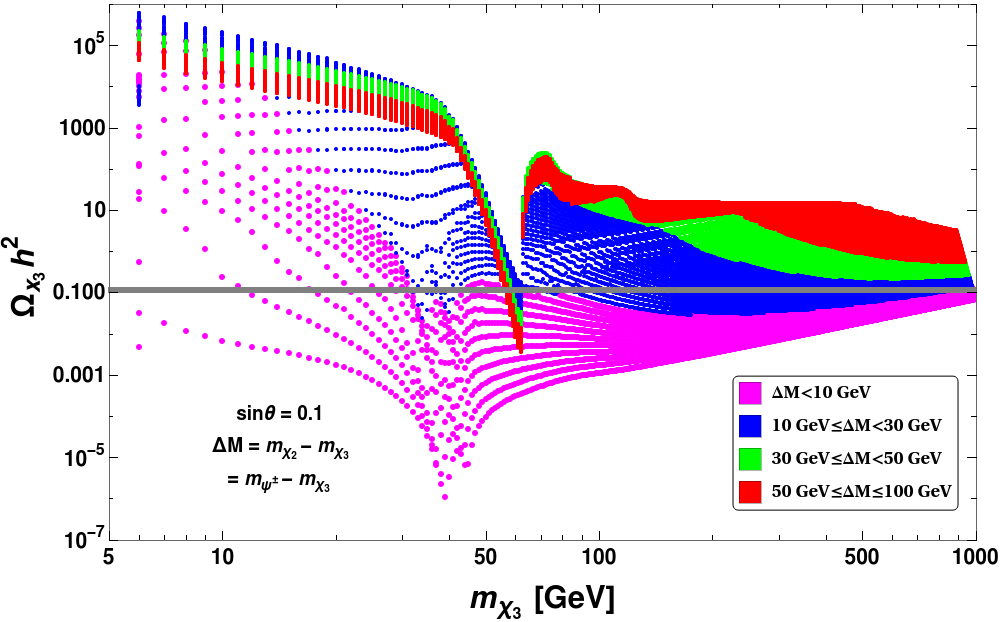}
 $$
 $$
 \includegraphics[height=5.0cm]{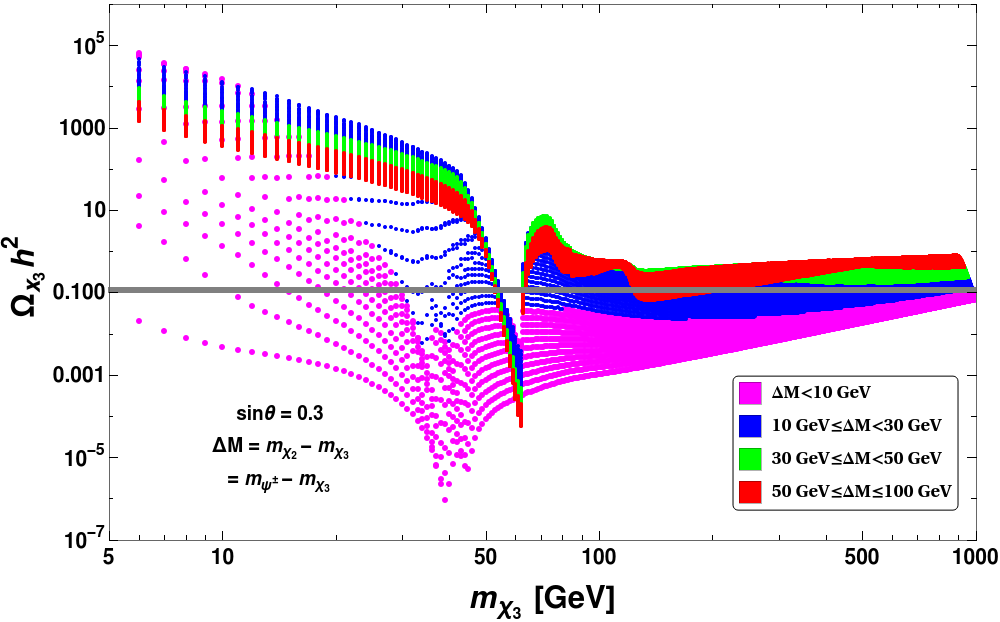} 
 \includegraphics[height=5.0cm]{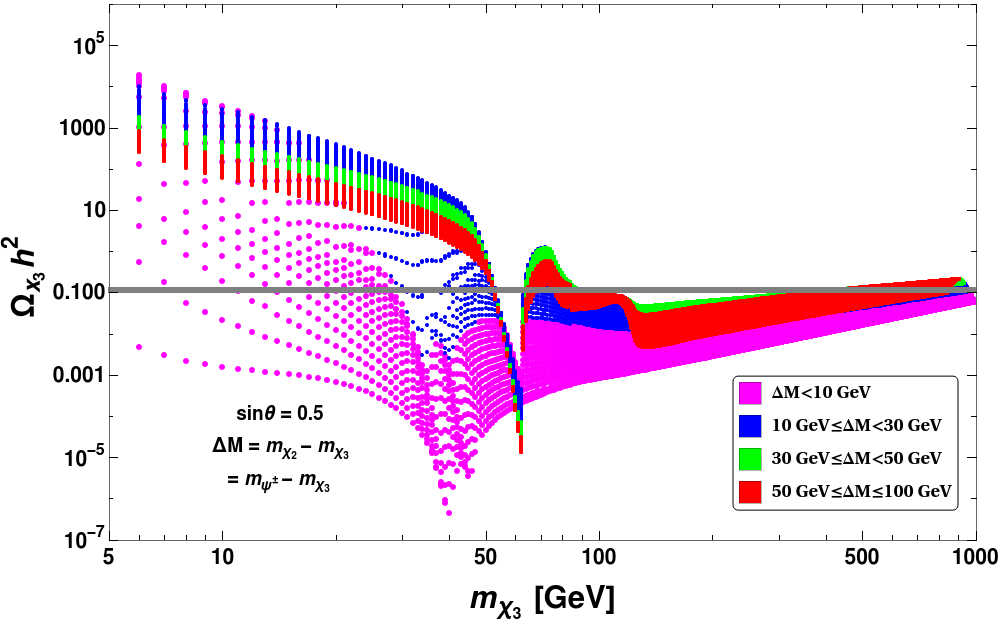}
$$ 
 \caption{\footnotesize{DM relic density as a function of DM mass ($m_{\chi_3}$) for different mass splitting $\Delta M$ between the DM and the NLSP (as mentioned in figure inset in GeV) 
 for $\sin\theta=0.01$ (top left panel), $\sin\theta=0.1$ (top right panel), $\sin\theta=0.3$ (bottom left panel) and  $\sin\theta=0.5$ (bottom right panel). 
 Correct relic density region from PLANCK data ($0.1166\leq\Omega_{DM} h^2 \leq 0.1206$) is indicated by the silver horizontal line.}}
 \label{fig:relicsplot1}
\end{figure}

 As seen from Fig.~\ref{fig:relicsplot1}, when $\Delta M$ is small, relic density is smaller due to large coannihilation 
contribution from flavour changing  $Z$-mediated processes as well as $W^\pm$ mediated processes (less Boltzmann suppression 
followed from Eq.~\ref{Boltzmann}). The resonance drops at $m_{Z}/2$ is seen due to s-channel off-diagonal $Z$ mediated 
coannihilation interactions. As none of these neutral current interactions are diagonal, we observe the resonance to be 
somewhat flattened rather than a sharp spike that would have been expected if the interactions were diagonal. These coannihilation 
channels dominantly contribute towards the relic density as long as the mass splitting between the DM and NLSP is small, {\it e.g.,} 
for $\Delta M=10 ~{\rm GeV}$. As $\Delta M$ increases, these coannihilations become less and less effective, and Higgs mediated processes 
takes over. For $\Delta M=30~ {\rm GeV}$, both contributions are present comparable while for $\Delta M> 30~{\rm GeV}$, the contributions from 
vector current (coannihilation) interactions are practically negligible and the the Higgs mediated channel dominates. Consequently, 
we see a resonance drop at $m_{h}/2$, while the drop at $m_{Z}/2$ disappears. We have also observed that as long as $\Delta M$ is 
small and the coannihilation channels dominate, the effect of $\sin\theta$ on relic density is quite negligible. For smaller 
$\sin \theta$, the annihilation cross-section due to Higgs portal (see Eqn.~\ref{DMHiggs}) is small leading to larger relic 
abundance, while for large $\sin \theta$, the effective annihilation cross-section is large leading to small relic abundance. 
However, this can only be observed when $\Delta M$ is sufficiently large enough and coannihilation processes are negligible. 
In Fig.~\ref{fig:relicsplot1}, we also show the correct relic density by the silver horizontal line. In Fig.~\ref{relic3}, the 
correct relic density allowed parameter space has been shown in the plane of $\Delta M$ vs $m_{\chi_{_3}}$ for wide range of 
mixing angle $\{\sin\theta= 0.001-0.01, 0.01-0.1, 0.1-0.2, 0.2-0.4,0.4-0.6\}$, indicted by different colours. 
We can see that in Fig.~\ref{relic3}, there is a bifurcation around $\Delta M \sim 50$ GeV, so the allowed plane of $m_{\chi_{_3}}-\Delta M$ are separated in two regions: 
(i) the bottom portion with small $\Delta M$, where $\Delta M$ decreases with larger DM mass ($m_{\chi_{_3}}$) and (ii) the top portion of the figure with large $\Delta M$, 
where $\Delta M$ increases slowly with larger $m_{\chi_{_3}}$. 

\begin{figure}[h]
\centering
\includegraphics[width = 99 mm]{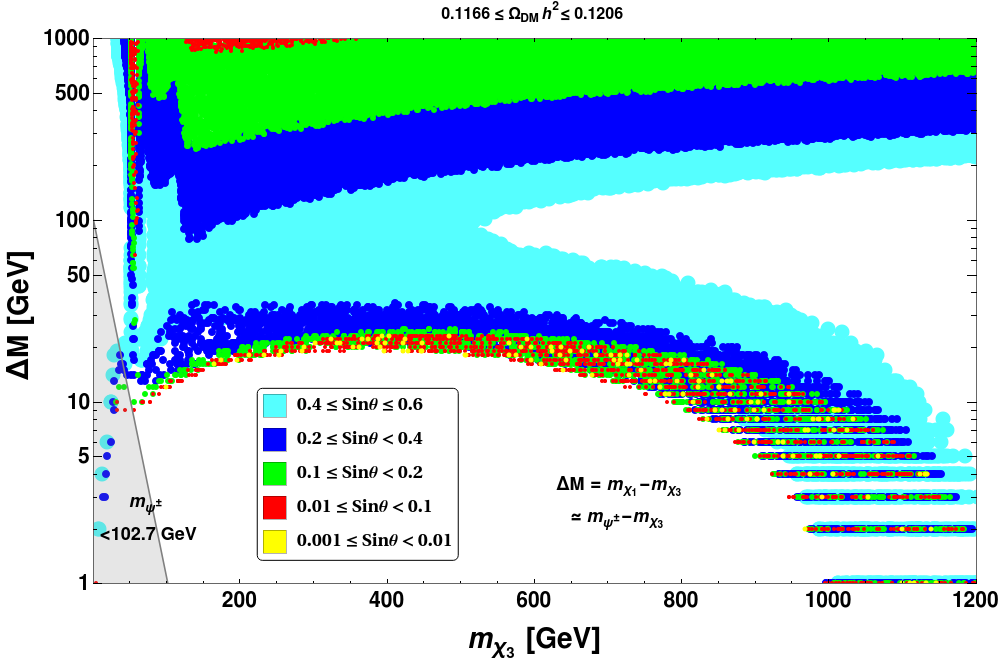}
\caption{\footnotesize{DM relic density ($0.1166\leq\Omega_{DM} h^2\leq 0.1206$) allowed parameter space in the plane of $\Delta M $ vs $m_{\chi_3}$ for different 
ranges of $\sin\theta$ as mentioned in the figure inset. The shaded region in the bottom left corner is ruled out by LEP exclusion bound on charged fermion mass, $m_{\psi^\pm}=M >102.7$ GeV.}}
\label{relic3}
\end{figure}

In region (i), given a specific range of $\sin\theta$, the annihilation cross-section decreases with larger DM mass 
$m_{\chi_{_3}}$ (from annihilation diagrams) and hence more co-annihilation contribution is required to get correct 
relic density, resulting $\Delta M$ to decrease. This also implies that the region below the each coloured 
zone is under-abundant (small $\Delta M$ implying large co-annihilation for a given $m_{\chi_{_3}}$), while the region 
above is over-abundant by the same logic. In this region the Yukawa coupling $Y_1$ which governs the annihilation cross-section is comparatively small since 
$Y_1\propto \Delta M \sin\theta$ and $\Delta M$ is small. Also the annihilation cross-section decreases with increase in DM mass. 
Therefore, when DM mass is sufficiently heavy ($m_{\chi_3} > 1.2$ TeV), annihilation becomes too weak to be compensated by the coannihilation even when $\Delta M \rightarrow 0$, producing over abundance.  
Hence, for small $\Delta M$, the allowed region has a maximum DM mass, as the region beyond $m_{\chi_3} \sim 1.2$ TeV is overabundant.

In region (ii), we note that, the co-annihilation contribution is much smaller 
due to large $\Delta M$, so the annihilation processes effectively contribute to the relic density. Annihilation processes are essentially gauge or Higgs mediated. We already noted that Higgs Yukawa 
coupling is proportional to both $\sin\theta$ and $\Delta M$ as $Y_1 \propto \Delta M \sin2\theta$. Hence, for a given $\sin\theta$, larger $\Delta M$ leads to larger $Y_1$ and hence 
larger annihilation cross-section to yield under abundance, which can only be tamed down to correct relic density by having a larger 
DM mass.  Also larger $\sin\theta$ requires smaller $\Delta M$ for the same reason. Therefore, the region above each coloured zone 
(allowed by relic density for a specific range of $\sin\theta$) is under abundant, while the region below each coloured zone is over abundant.

Let us come back to region 
(i) again and note that allowed parameter space indicates larger DM mass requires smaller and smaller $\Delta M$ and we reach a maximum DM mass ($\sim$ 1 TeV) 
for $\Delta M \to 0$. However, with $\Delta M \to 0$, the charged companions $\psi^{\pm}$ are degenerate to DM and are stable. 
This is not acceptable as DM won't be dark then. Hence, $\Delta M$ can not be arbitrarily small. 
We can put a lower bound on $\Delta M$ by requiring the charged partners $\psi^{\pm}$ of the 
DM to decay before the onset of Big Bang Nucleosynthesis ($ \tau_{\rm BBN} \sim 1$ sec.). The decay rate for the processes  $\psi^\pm \rightarrow\chi_{_3} l^{\pm} \nu_{_l}$ in the limit of small $\Delta M$ is given by~\footnote{Semi-leptonic processes {\it e.g.}  $\psi^{\pm}\rightarrow \chi_{_3} \pi^{\pm}$ are also possible, see for  example~\cite{Jana:2019tdm}} :

\begin{equation}
\Gamma_{\psi^\pm} = \frac{1}{15 (2\pi)^3}\frac{e^4 \sin^2\theta}{\sin^4\theta_w}\frac{(\Delta M)^5}{M^4_W},
\label{eq:decay}
\end{equation}

\begin{figure}[ht]
\centering
\includegraphics[width = 80 mm]{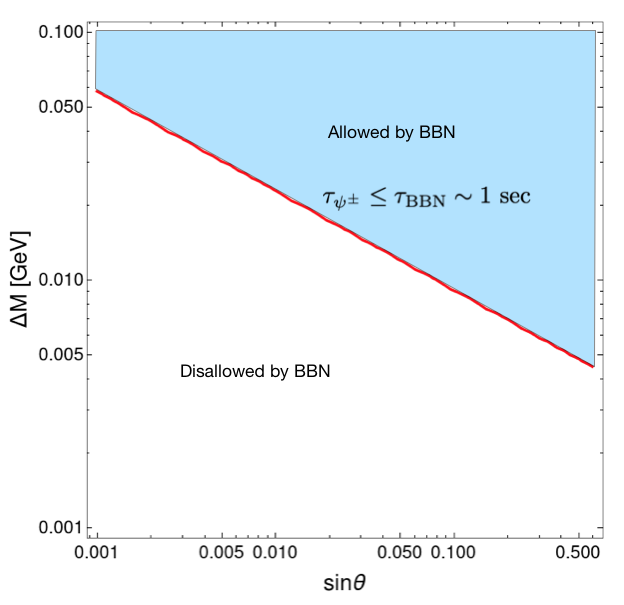}
\caption{\footnotesize{Lower bound on $\Delta M$ as a function of $\sin\theta$ from Big-Bang Nucleosynthesis (BBN). The shaded region is allowed.}}
\label{DWlog}
\end{figure}

By requiring that the charged fermions should decay before the onset of BBN, we can get a lower bound on $\Delta M $ as,
\begin{equation}
\tau_{\psi^\pm}=\frac{1}{\Gamma_{\psi^\pm}} \leq \tau_{\rm BBN}\sim1~ {\rm sec} \implies \Big(\frac{\Delta M}{\rm GeV}\Big)^5 \geq \frac{6.4\times10^{-13}}{\sin^2\theta}.
\end{equation}

In Fig.~\ref{DWlog}, we show the lower bound on $\Delta M$ for the range of $\sin\theta$ we used in our work. The region above the red line is allowed by the constraint. It is obvious that the bound 
is more stringent for smaller $\sin\theta$. }

\section{Direct Detection of singlet-doublet Majorana Dark Matter}
\label{direct_detection}
\begin{figure}[!hb]
\centering
\includegraphics[width = 40 mm]{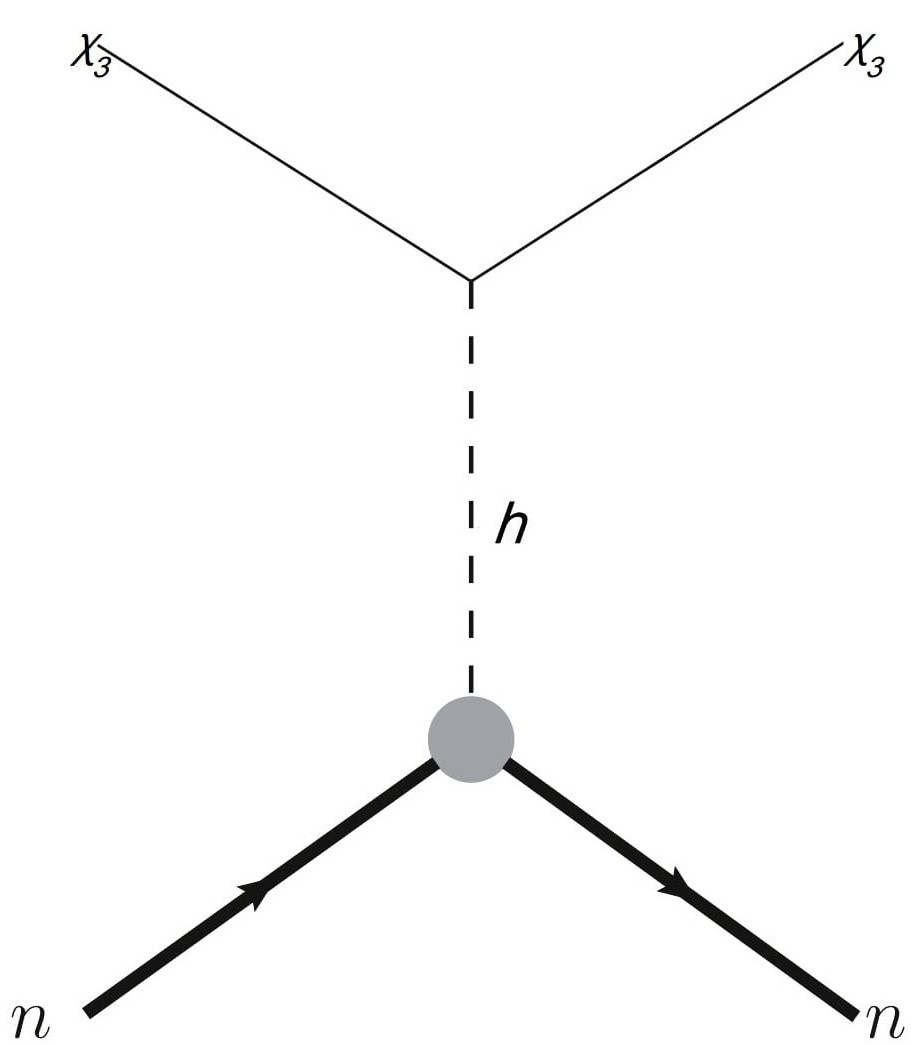}
\caption{\footnotesize{Feynman diagram for elastic scattering of DM off nuclei at terrestrial laboratory.}}
\label{DD1}
\end{figure}

Among different possibilities of detecting DM, one major experimental procedure is direct DM search.
Direct detection of the DM ($\chi_3$) at a terrestrial laboratory is possible through elastic scattering of the DM off nuclei via 
Higgs-mediated interaction represented by the Feynman diagram shown in Fig.~\ref{DD1}.
The presence of only Higgs mediated diagram for direct search makes this model crucially segregated from that of a vector like singlet-doublet DM as 
elaborated in~\cite{Bhattacharya:2017sml,Barman:2019tuo,Bhattacharya:2018cgx,Bhattacharya:2016rqj,Bhattacharya:2015qpa,Bhattacharya:2018fus}. 
Here, the DM being a Majorana fermion only has off diagonal Z-coupling and therefore do not contribute to direct search as it is very difficult to produce 
a heavier particle in the low energy scattering as in direct search experiment. The absence of $Z$ mediation 
crucially alters the available parameter space of the model as we describe below. The corresponding vertex of $\chi_{_3} \chi_{_3} h$ 
can be obtained from the Lagrangian $\mathcal{L}_{DM-Higgs}$ given by Eq. \ref{DMHiggs}. The cross section per nucleon for the spin-independent 
(SI) DM-nucleon interaction is then given by:
\begin{equation}\label{da}
    \sigma_{SI} = \frac{1}{\pi A^2}\mu^2_r|\mathcal{M}|^2,
\end{equation}
where A is the mass number of the target nucleus, $\mu_r$ is the reduced mass of the DM-nucleon system and ${\mathcal M}$ 
is the amplitude for the DM-nucleon interaction, which can be written as:
\begin{equation}
\label{dd}
    \mathcal{M}=\Big[Z f_p +(A-Z)f_n\Big],
\end{equation}
where $f_{p}$ and $f_{n}$ denote effective interaction strengths of DM with proton and neutron of the nuclei used for the experiment with $A$ 
being mass number and $Z$ being atomic number. The effective interaction strength can then further be decomposed in terms of interaction with parton as:
\begin{equation}
\label{dda}
    f_{p,n}=\sum_{q=u,d,s}f^{p,n}_{Tq}\alpha_{q}\frac{m_{(p,n)}}{m_q} + \frac{2}{27}f^{p,n}_{TG}\sum_{q=c,b,t}\alpha_q \frac{m_{(p,n)}}{m_{q}};
\end{equation}  
with 
\begin{equation}
\label{dda2}
\alpha_q =\frac{Y_1 \sin2\theta}{M^2_h}\frac{m_q}{v}=\frac{~\Delta M ~\sin^22\theta m_q}{v^2 M^2_h}; 
\end{equation}
coming from DM interaction with SM via Higgs portal coupling. Further, in Eq.\ref{dda}, the different coupling strengths between DM and light quarks are given by 
Bertone et al \cite{Bertone:2004pz,Alarcon:2012nr} as $f^p_{Tu} = 0.020 \pm 0.004, f^p_{Td} = 0.026 \pm 0.005, f^p_{Ts} = 0.014 \pm 0.062$, $f^n_{Tu} = 0.020 \pm 0.004, 
f^n_{Td} = 0.036 \pm 0.005, f^n_{Ts} = 0.118 \pm 0.062$. The coupling of DM with the gluons in target nuclei is parameterised by:
\begin{equation*}
f^{(p,n)}_{TG} = 1- \sum_{q=u,d,s}f^{p,n}_{Tq}.
\end{equation*}
Using Eqs. \ref{da}, \ref{dd}, \ref{dda} and \ref{dda2}, the spin-independent DM-nucleon cross-section is given by:
\begin{equation}
\label{ddaf}
\begin{aligned}
 \sigma_{SI} &= \frac{4}{\pi A^2}\mu^2_r\frac{Y^2 \sin^2 2\theta}{M^4_h}\Big[\frac{m_p}{v}\Big(f^{p}_{Tu} + f^{p}_{Td} + f^{p}_{Ts} + \frac{2}{9}f^{p}_{TG}\\
 &+\frac{m_n}{v}\Big(f^{n}_{Tu} + f^{n}_{Td} + f^{n}_{Ts} + \frac{2}{9}f^{n}_{TG}\Big)\Big]^2
\end{aligned}
\end{equation}
In the above equation for DM-nucleon direct search cross-section, two parameters from model that enter are the Higgs-DM Yukawa coupling ($Y_1$) and singlet-doublet mixing parameter ($\sin2\theta$), which can be constrained by requiring that $\sigma_{SI}$ is less than the current DM-nucleon cross-sections dictated by non-observation of DM 
in current direct search data. Recently, there has been a signal like event from electron recoil reported in XENON-1T data~\cite{Aprile:2020tmw} observed at sub-GeV DM mass, 
which remains out of our scan. 

\begin{figure}[htb!]
$$
\includegraphics[height=5.0cm]{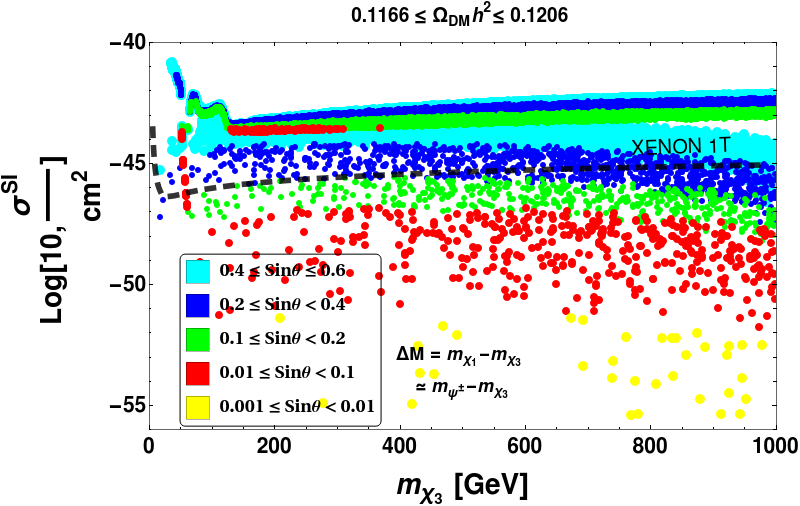}
 \includegraphics[height=5.0cm]{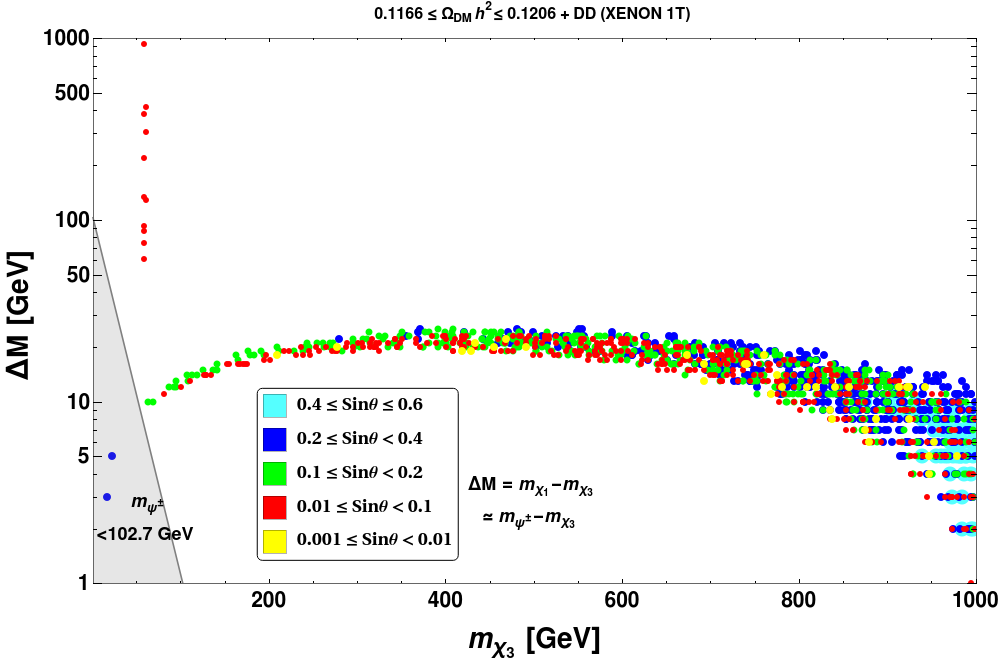}  
$$ 
 \caption{\footnotesize{[Left]: Direct detection cross section for the DM ($\chi_3$) confronted with bounds on spin-independent elastic scattering cross section by XENON-1T~\cite{Aprile:2018dbl} over and above relic density constraint from PLANCK; 
 [Right]: Correct DM relic density in $\Delta M-m_{\chi_{_3}}$ plane constrained by XENON-1T bound. Different coloured points indicate different ranges of $\sin\theta$ as mentioned in figure inset. The shaded region in the bottom left corner of 
 right panel plot is ruled out by LEP exclusion bound on charged fermion mass, $m_{\psi^\pm}=M >102.7$ GeV. }}
 \label{directdetection}
\end{figure}

In the left panel of Fig.~\ref{directdetection}, we confront the direct detection cross section obtained for the model as 
a function of DM mass, with bounds on spin-independent elastic scattering cross section from XENON-1T~\cite{Aprile:2018dbl}, 
shown by black dashed curve. It is worth mentioning that all points shown in left panel of Fig.~\ref{directdetection} also 
satisfies relic density constraints from PLANCK. Different coloured patches indicate different ranges of mixing angle ($\sin\theta$) 
as indicated in figure panel. Obviously those regions that appear below the XENON-1T line can be allowed by the bound. It is 
obvious that $Y_1$ being proportional to $\sin\theta$ (see Eqn.~\ref{y1}) and due to the explicit presence of $\sin 2\theta$ in 
the direct search cross section as in Eq. \ref{dda2}, parameter space with smaller $\sin\theta$ survive the cut. This is what is shown  
in $\Delta M-m_{\chi_{_3}}$ plane in the right hand side (RHS) of Fig.~\ref{directdetection}, where we plot those points 
which simultaneously satisfy relic density~\cite{Ade:2013zuv} and direct search XENON-1T bound~\cite{Aprile:2018dbl} together. It 
is seen that null observation from direct search crucially tames down the relic density allowed parameter space, which is 
evident when we compare the RHS of Fig.~\ref{directdetection} with that of Fig.~\ref{relic3}, where only relic density allowed 
parameter space is depicted. It is seen in RHS of Fig.~\ref{directdetection}, that $\sin\theta$ is correlated to DM mass 
and $\Delta M$. For example, $\sin\theta$ is very small for smaller DM mass with moderate $\Delta M$ ($\sin\theta 
\lesssim 0.2$ for $m_{DM} \sim$ 500 GeV with $\Delta M \sim$ 20 GeV shown by red and green points); 
while larger $\sin\theta \sim 0.6$ is allowed at higher DM mass $\sim 1000$ GeV, with very small $\Delta M \lesssim 2$ GeV (Cyan points). 
This is simply because, the direct search cross-section is proportional to $\sim Y_1\sin 2\theta\sim \Delta M \sin^22\theta$, therefore larger 
$\sin\theta$ requires $\Delta M$ to be smaller to remain within correct direct search limit. However, due to larger coannihilation contribution with small $\Delta M$, 
the relic density drops below the PLANCK bound, unless we restore it to the correct ballpark by having larger DM mass (annihilation cross-section is inversely proportional to 
DM mass). This feature crucially distinguishes the model at hand from vector like singlet-doublet scenario with Dirac dark matter, where the presence of $Z$
mediated direct search graph tames $\sin\theta$ to much smaller values like $\sim 0.2$ (for details see~\cite{Barman:2019tuo,Bhattacharya:2015qpa,Bhattacharya:2018cgx,Bhattacharya:2016rqj,Bhattacharya:2017sml,Bhattacharya:2018fus}). Higgs resonance $m_{\chi_3}\sim m_h/2$ is seen to satisfy
both relic density and direct search bound, where $\Delta M$ can be very large having very small $\sin\theta \sim 0.2$.

\section{Singlet-doublet Majorana DM in gauged $U(1)_{B-L}$ Extension of the SM}\label{gauged_model}
\subsection{The Model}
Due to the presence of three right handed neutrinos $N_{R_i}$ and the fermion doublet $\Psi$ being vector-like, the model is 
automatically $U(1)_{B-L}$ anomaly free if we assign one unit of B-L charge to each of these fields. This is because of the 
fact that in a gauged B-L theory with only SM fermion content, non-zero anomalies are associated with the following two 
triangular diagrams:
\begin{equation}
\begin{aligned}
\mathcal{A}_1[U(1)^3_{B-L}]&=\mathcal{A}^{SM}_1[U(1)^3_{B-L}]&=-3\,,\\
\mathcal{A}_2[(Gravity)^2 \times U(1)_{B-L}]&=\mathcal{A}^{SM}_2[(Gravity)^2 \times U(1)_{B-L}]&=-3\,,
\end{aligned}
\end{equation}
which are exactly cancelled by anomalies from three additional right handed neutrinos since,
\begin{equation}
\begin{aligned}
\mathcal{A}^{RHN}_1[U(1)^3_{B-L}]&=3\\
\mathcal{A}^{RHN}_2[(Gravity)^2 \times U(1)_{B-L}]&=3\,.
\end{aligned} 
\end{equation}

Motivated by this fact, we extend the gauge group of the model to $SU(3)_C \times SU(2)_L \times U(1)_Y \times U(1)_{B-L} 
\otimes \mathcal{Z}_2$. Besides, one new complex scalar singlet $\Phi_{BL}$ is added with lepton number $-2$. The particle content 
and the corresponding quantum numbers under the symmetry of the model are listed in the Table \ref{tab:tab2}. Since two 
of the right handed neutrinos, say $N_{R_2},N_{R_3}$ are chosen to be even under the imposed $\mathcal{Z}_2$ symmetry, they can couple to 
the SM lepton and Higgs doublets to explain non-zero masses and mixing of light neutrinos. On the other hand, the vectorlike 
fermion doublet $\Psi$ and $N_{R_1}$ are chosen to be odd under the imposed $\mathcal{Z}_2$ symmetry. As a result the DM emerges as a 
mixture of the neutral component of the doublet $\Psi$ {\it viz.} $\psi^0$ and $N_{R_1}$, similar to section \ref{Model}. 
However, we notice certain differences in the mass matrix of dark sector neutral fermions in comparison to Eq. \ref{dark-sector-mass} 
due to the conservatoin of $B-L$ charge. In the following we discuss in details the corresponding phenomenology. 

\vspace{1cm}
%
\begin{table}[ht]
\resizebox{\linewidth}{!}{
 \begin{tabular}{|c|c|c|c|}
\hline \multicolumn{2}{|c}{Fields}&  \multicolumn{1}{|c|}{ $\underbrace{ SU(3)_C \otimes SU(2)_L \otimes U(1)_Y}$ $\otimes U(1)_{B-L} \otimes \mathcal{Z}_2 $} \\ \hline
\multirow{2}{*} 
{VLFd} & $\Psi=\left(\begin{matrix} \psi^0 \\ \psi^- \end{matrix}\right)$&   ~~1 ~~~~~~~~~~~2~~~~~~~~~-1~~~~~~~~~~-1~~~~~~~~~~  - \\
\hline
{RHNs} &  ${N}_{R_1}$&   ~~1 ~~~~~~~~~~~1~~~~~~~~~~0~~~~~~~~~~-1~~~~~~~~~~ - \\ [0.5em] \cline{2-3}
       &  ${N}_{R_2}$&  ~~~1 ~~~~~~~~~~~1~~~~~~~~~~0~~~~~~~~~~-1~~~~~~~~~~ + \\ [0.5em] \cline{2-3}
       &  ${N}_{R_3}$& ~~~1 ~~~~~~~~~~~1~~~~~~~~~~0~~~~~~~~~~-1~~~~~~~~~~ + \\
\hline
\hline
Higgs doublet & $H=\left(\begin{matrix} w^+ \\ \frac{h+v+iz}{\sqrt{2}} \end{matrix}\right)$ & ~~~1 ~~~~~~~~~~~2~~~~~~~~~~1~~~~~~~~~~~0~~~~~~~~~~~+ \\
\hline
Scalar Singlet& $\Phi_{BL}=\frac{\phi+v_{BL} + i z_\phi }{\sqrt{2}}$ & ~~~1 ~~~~~~~~~~~1~~~~~~~~~~0~~~~~~~~~~-2~~~~~~~~~~~+ \\
\hline
\end{tabular}
}
\caption{\footnotesize{Charge assignment of BSM fields along with the SM Higgs doublet under the gauge group $\mathcal{G} \equiv \mathcal{G}_{\rm SM} \otimes U(1)_{B-L} \otimes \mathcal{Z}_2$, where $\mathcal{G}_{\rm SM}\equiv SU(3)_C \otimes SU(2)_L \otimes U(1)_Y$ . }}
    \label{tab:tab2}
\end{table}

Owing to the symmetry and charge assignments of the particles given in Tab.~\ref{tab:tab2}, the Lagrangian of the Model can be 
given as:
\begin{equation}
\label{mdl_Lag_BL}
 \mathcal{L} = \overline{\Psi}(i\slashed{D}-M)\Psi + \overline{N}_{R_i}i\slashed{\Tilde{D}}N_{R_i} + \mathcal{L}_{yuk} + \mathcal{L}_{Gauge}+\mathcal{L}_{\rm{scalar}}+\mathcal{L}_{SM} ;  
\end{equation}
where the covariant derivatives  $D_\mu$ and $\Tilde{D}_\mu$ are given by:
 \begin{equation}
 \label{coderv}
 \begin{aligned}
     D_\mu &= \partial_\mu - i\frac{g}{2}\tau.W_\mu - ig'\frac{Y}{2}B_\mu -ig_{BL} Y_{BL} Z_{BL},\\
     \Tilde{D}_\mu &= \partial_\mu - ig_{BL} Y_{BL} (Z_{BL})_\mu.\\
     \end{aligned}
 \end{equation}
 In the covariant derivative of $\Psi$, there is an additional term due to the lepton number assignment, i.e. its transformation 
under $U(1)_{B-L}$; $g_{BL}$ stands for $U(1)_{B-L}$ gauge coupling, which serves as a additional free parameter of the model. Note 
that $Y_{BL}$ can simply be replaced by the lepton number assignment as given in Tab.~\ref{tab:tab2}.
 
The Yukawa interaction of the model is given by:
\begin{equation}
-\mathcal{L}_{yuk} =\left[ Y_1\overline{\Psi}\Tilde{H}N_{R_1} +h.c\right]+ \left(Y_{j \alpha }\overline{N_{R_j}} \Tilde{H^\dagger} L_{\alpha} + h.c.\right)+\left[\frac{y'_i}{2}\Phi_{BL} \overline{{N}_{R_i}} \left({N}_{R_i}\right)^c+h.c.\right];
\label{eq:Yukawa2}
\end{equation}
 where $\alpha = e,\mu, \tau$, $j=2,3$ and $i=1,2,3$. Due to $B-L$ conservation, the Yukawa interaction term $\overline{\Psi}\tilde{H}(N_{R_1})^c$ that was allowed in the earlier case (see Eqn.~\ref{eq:yukawa}) is no longer allowed. For the 
same reason, the bare Majorana mass terms of right-handed neutrinos are also not allowed. The masses of right-handed neutrinos 
as well as the neutral gauge boson $Z_{BL}$ are generated from the vev of $\Phi_{BL}$. Thus the gauge sector is augmented by a 
new gauge boson $Z_{BL}$. The new gauge kinetic terms that appear in the Lagrangian constitute of,
\begin{equation}
    \mathcal{L}_{Gauge} = -\frac{1}{4} (Z_{BL})_{\mu\nu}Z^{\mu \nu}_{BL} - \frac{\epsilon}{2} (Z_{BL})_{\mu\nu} B^{\mu \nu};
\end{equation}
where $Z^{\mu \nu}_{BL}$ represents the field strength of the $U(1)_{B-L}$ gauge boson and is defined as:
\begin{equation}
Z^{\mu \nu}_{BL} = \partial^\mu(Z_{BL})^\nu - \partial^\nu(Z_{BL})^\mu\,.
\end{equation}
In the second term, $\epsilon$ parametrises the kinetic mixing between the $U(1)_{B-L}$ and $U(1)_Y$ gauge sectors. Such a mixing 
term can be generated through quantum corrections and approximated at one loop as $\epsilon \approx \frac{g' g_{BL}}{16\pi^2}$~\cite{Gherghetta:2019coi, Mambrini:2011dw}. Since $g_{BL}$ has tight upper bound from ATLAS, such one loop mixing is very very small compared to other relevant parameters of the model and the same has been neglected in rest of our analysis.

The Lagrangian of scalar sector is given by:
\bea
\mathcal{L}_{\rm{scalar}}&=& |\mathcal{D}_\mu{H}|^2+|\mathfrak{D}_\mu{\Phi}_{BL}|^2 - V(H,\Phi_{BL})
\eea
where $\mathcal{D}_\mu$ and $\mathfrak{D}_\mu$ are given as follows:
\begin{equation}
\begin{aligned}
\mathcal{D}_\mu&=\partial_\mu  - i\frac{g}{2}\tau.W_\mu - ig'\frac{Y}{2}B_\mu\\
\mathfrak{D}_\mu&=\partial_\mu - ig_{BL} Y_{BL} (Z_{BL})_\mu\,.
\end{aligned}
\end{equation}

The scalar potential is given by
\begin{equation}
\begin{aligned}
    V(H, \Phi_{BL}) &= -{\mu_H}^2 \left(H^\dagger H \right) + \lambda_H \left(H^\dagger H \right)^2 \nonumber \\
    & -{\mu_{\Phi}}^2 \left({\Phi_{BL}}^\dagger \Phi_{BL} \right) + \lambda_{\Phi} \left({\Phi_{BL}}^\dagger \Phi_{BL} \right)^2 \nonumber  + \lambda_{H \Phi} (H^\dagger H)\left({\Phi_{BL}}^\dagger \Phi_{BL} \right)\,.
    \end{aligned}
\end{equation}
We note here that $H$ do not have any transformation under the extended symmetry, while $\Phi_{BL}$ is a singlet under SM, 
the only gauge invariant terms that one can cook up are $H^\dagger H$ and $\Phi_{BL}^\dagger\Phi_{BL}$, resulting a simple scalar potential, where the only interaction term that can be written is $(H^\dagger H)\left({\Phi_{BL}}^\dagger \Phi_{BL} \right)$. 
$\lambda_{H \Phi}$ turns out to be an important additional parameter that contributes to the phenomenology. We also note that 
for both $H$ and $\Phi_{BL}$ to acquire non-zero vevs, we need both ${\mu_H}$ and ${\mu_{\Phi}}$ to be positive. 

We analyse the model as follows: scalar mixing in subsection-\ref{sca_mixing}, masses and mixing of dark sector particles in 
subsection-\ref{DM-mass-mixing}, theoretical and experimental constraints in subsection-\ref{sec:constraintsBL}, relic 
abundance of DM in subsection-\ref{blrelics}, direct detection in subsection-\ref{bldirect} and finally show the allowed 
parameter space in the light of ATLAS bound on $g_{\rm BL}$ versus $M_{Z_{BL}}$ in subsection \ref{gbl_versus_mzbl}.

\subsection{Spontaneous symmetry breaking and physical scalars}
\label{sca_mixing}
At TeV scales $\Phi_{BL}$ acquires a non-zero vev and breaks $U(1)_{\rm B-L}$ to identity. The non-zero vevs which 
spontaneously breaks $\mathcal{G}_{\rm SM} \otimes U(1)_{B-L} \otimes \mathcal{Z}_2$ down to $U(1)_Q \otimes \mathcal{Z}_2$ are given as:
\bea
 \langle\Phi_{BL}\rangle=\frac{v_{BL}}{\sqrt{2}},
~~~~~~~
\langle H\rangle= \left(\begin{matrix}  0  \\
   \frac{v}{\sqrt{2}}
   \end{matrix}
   \right).
\eea
The minimization conditions around the vev's are given by : 
\bea
\frac{\partial V}{\partial H}\bigg|_{v}=0:~\mu^2_{H}& =& \lambda_{H} v^2 + \frac{\lambda_{H\Phi} v^2_{BL}}{2}, \nonumber \\
\frac{\partial V}{\partial \Phi_{BL}}\bigg|_{v_{BL}}=0:~\mu^2_{\Phi} &=& \lambda_{\Phi} v^2_{BL} + \frac{\lambda_{H\Phi} v^2}{2}.
\eea

Due to presence of $(H^\dagger H)\left({\Phi_{BL}}^\dagger \Phi_{BL} \right)$ interaction in the scalar sector, both weak states $h$ and $\phi$ mix with each other. Using above minimization conditions, the mass terms of the scalar sector can be expressed as:
\bea
\mathcal{L}_{\rm scalar}^{\rm mass} &=& \frac{1}{2} \left(\begin{matrix} h && \phi \end{matrix} \right)
                                \left( \begin{matrix} 2 \lambda_H v^2 && \lambda_{H\Phi} v v_{BL}  \\
                                        \lambda_{H\Phi} v v_{BL} && 2 \lambda_{\Phi} v_{BL}^2 
                                       \end{matrix} \right) 
                                       \left(\begin{matrix} h \\ \phi \end{matrix} \right),  \nonumber \\
                                    &=&  \frac{1}{2} \left(\begin{matrix} h_1 && h_2 \end{matrix} \right)
                                \left( \begin{matrix} {m^2_{h_1}} && 0  \\
                                       0 && {m^2_{h_2}}
                                       \end{matrix} \right) 
                                       \left(\begin{matrix} h_1 \\ h_2 \end{matrix} \right). 
\eea

In order to obtain the mass eigenvalues, the flavor eigenstates are rotated by an orthogonal matrix as follows :
\begin{equation}
\begin{pmatrix}
h_{1}  & \\
h_{2}      & \\
\end{pmatrix}=\begin{pmatrix}
\cos\beta&	\sin\beta& \\
-\sin\beta	&	\cos\beta & \\
\end{pmatrix}
\begin{pmatrix}
h & \\
 \phi    & \\
\end{pmatrix};
\label{higgs_mixing}
\end{equation} 
where $h_1$ and $h_2$ are the physical mass eigenstates. We identify $h_1$ to be the physical Higgs discovered in 2012 at LHC 
with mass $m_{h_1}=125$ GeV and $m_{h_2}$ remains a scalar beyond the SM. How heavy $h_2$ requires to be is constrained from 
LHC data which we discuss in a moment. The CP odd states also mix with each other, but turns out to be massless states known 
as Goldstone Bosons. In unitary gauge they are accounted as the longitudinal modes of massive vector Bosons and do not enter 
into phenomenology explicitly. The scalar sector therefore accounts for three free parameters:
 \bea
 \{ m_{h_2},~ v_{BL}, ~\sin\beta\};
 \eea
 which are constrained from Higgs data at Collider. We will discuss them in the next subsection. 
 Other quartic couplings $\lambda_H,~\lambda_{\Phi}$ and $\lambda_{H \Phi}$ can be expressed in terms of the 
 physical parameters as:
 \bea
 \lambda_{H} &=& \frac{m^2_{h_1} \cos^2\beta + m^2_{h_2} \sin^2\beta }{2 v^2}, \nonumber \\
  \lambda_{\Phi} &=& \frac{m^2_{h_1} \sin^2\beta + m^2_{h_2} \cos^2\beta }{2 v^2}, \nonumber \\
  \lambda_{H\Phi} &=& \frac{\left(m^2_{h_2}-m^2_{h_1}\right)\sin2\beta}{2 v v_{BL}} ~~.
 \eea
 
The broken $U(1)_{B-L}$ gauge symmetry yields mass for $Z_{BL}$ as:
\bea
M_{Z_{BL}} = 2 g_{BL} v_{BL}
\label{mzbl}
\eea
$M_{Z_{BL}}$ and $g_{BL}$ are constrained from both LEP and LHC which we shall address later. 
So it follows from Eqn.~\ref{mzbl} that $v_{BL}$ is no longer a free parameter. Instead, in the combined gauged and scalar sector, the free parameters involved are:
\bea
 \{ m_{h_2},~M_{Z_{BL}}, ~g_{BL}, ~\sin\beta\};
 \eea
As we will see in the later sections, these parameters play a crucial role in DM phenomenology in the $U(1)_{B-L}$ extension of the 
SM model.

\subsection{Masses and mixing of dark sector particles}\label{DM-mass-mixing}
After electroweak symmetry breaking the mass term of the neutral dark sector particles can be written as,
\begin{equation}
  -\mathcal{L}_{mass} = M\overline{\psi^0_L}\psi^0_R + \frac{1}{2}M_{R_1}\overline{N}_{R_1}(N_{R_1})^c + 
m_D \overline{\psi^0_L}N_{R_1} + h.c.\,,
  \label{bl_mass} 
\end{equation}
where $m_D=\frac{Y_1\langle v \rangle}{\sqrt{2}}$ with $\langle v \rangle=246$ GeV being the vacuum expectation value (vev) 
of the SM Higgs $H$ and $M_{R_i}=\frac{y'_i v_{BL}}{\sqrt{2}}$, where $v_{BL}$ is the vev of new scalar $\Phi_{BL}$. Writing these mass terms in the basis $ ((\psi^0_R)^c, \psi^0_L, (N_{R_1})^c)^T$, we get the mass 
matrix:
\begin{equation}
\label{dark-mass-matrix}
\mathcal{M}=
\left(
\begin{array}{ccc}
0 &M &0\\
M &0 &m_D \\
0 &m_D &M_{R_1}\\
\end{array}
\right) \, .
\end{equation}
The above mass matrix of neutral dark sector particles can be diagonalized by using an orthogonal transformation: $\mathcal{M}_{\rm diag}=U.\mathcal{M}.U^T$, where $U=U_{13}(\theta_{13}).U_{23}(\theta_{23}).U_{12}(\theta_{12})$ and $U_{13}(\theta_{13})$, $U_{23}(\theta_{23})$ and $U_{12}(\theta_{12})$ are taken as three Euler rotation 
matrices. Assuming $m_D << M, M_{R_1}$, the mass eigenvalues are given by~\footnote{Similar to Eqn.~\ref{diagonalizing_matrix}, the mass matrix \ref{dark-mass-matrix} can be further rotated by a phase matrix $U_{\rm ph}$ to make sure all the eigenvalues are positive.}:
\begin{equation}\label{dark-eigenvalues}
\begin{aligned}
m_{\chi_{_1}} & \approx  M + \frac{m^2_D}{2(M - M_{R_1})},
\\
m_{\chi_{_2}} & \approx -\Big(M + \frac{m^2_D}{2(M + M_{R_1})}\Big), 
\\
m_{\chi_{_3}} & \approx  M_{R_1}\Big(1 - \frac{m^2_D}{M^2 - M^2_{R_1}}\Big). 
\\
\end{aligned}
\end{equation}
From Eqs. (\ref{dark-mass-matrix}) and (\ref{dark-eigenvalues}) we see that ${\rm Tr}{\mathcal M}= M_{R_1}=
\sum_{i=1}^3 m_{\chi_i}$. Note that the above diagonalization is upto $\mathcal{O}(\frac{m^2_D}{M+M_{R_1}})$. The 
corresponding physical eigenstates can be given in terms of flavour eigenstates as:
\begin{equation}
\begin{aligned}
\chi_{_{1L}} & = (c_{13}c_{12} + s_{13}s_{23}s_{12})(\psi^0_R)^c + (c_{13}s_{12} - s_{13}s_{23}c_{12})\psi^0_L + (s_{13}c_{23})N^c_{R_1},\\
\chi_{_{2L}} & =  (-c_{23}s_{12})(\psi^0_R)^c + (c_{23}c_{12})\psi^0_L + s_{23} N^c_{R_1},\\
\chi_{_{3L}} & =  (-s_{13}c_{12} + c_{13}s_{23}s_{12})(\psi^0_R)^c + (-s_{13}s_{12} - s_{23}c_{12}c_{13})\psi^0_L + (c_{13}c_{23})N^c_{R_1}.
\end{aligned}
\end{equation}
where we abbreviated $\cos\theta_{ij} = c_{ij}$ and $\sin\theta_{ij} = s_{ij}$, with $\{ij:12,13,23\}$. The diagonalisation of 
the mass matrix requires:
\begin{equation}
\begin{aligned}
\theta_{12} & = \frac{\pi}{4}\, ,
\\
\tan 2\theta_{23} & =  \frac{-\sqrt{2}m_D}{M+M_{R_1}}\, ,
\\
\tan 2\theta_{13} & =  \Big(\frac{\sqrt{2}m_D}{ M -M_{R_1} -\frac{m^2_D}{2(M+M_{R_1})}}\Big)\cos\theta_{23}\,.
\end{aligned}
\label{theta1}
\end{equation}
Thus in the effective theory the dark sector comprises of three phyiscal Majorana fermions $\chi_{_1}, \chi_{_2}, \chi_{_3}$ defined 
as $\chi_{_i}=\frac{\chi_{_{iL}}+ (\chi_{_{iL}})^c}{\sqrt{2}}~(i=1,2,3)$. We assume $m_{\chi_{_1}} > m_{\chi_{_2}} > m_{\chi_{_3}}$, so 
that  $\chi_{_3}$ serves as a stable dark matter candidate. In the limit $m_D << M,M_{R_1}$, from Eq.~\ref{theta1}, we can further write,
\begin{equation}
Y_{1}  \approx  \frac{\Delta M \sin2\theta_{13}}{v}\,,
\label{theta2}
\end{equation}
where $\Delta M =|m_{\chi_{_1}}|- |m_{\chi_{_3}}| \approx |m_{\chi_{_2}}|- |m_{\chi_{_3}}|$. The mixing angle $\theta_{23}$ 
can be obtained using values of $m_D$ in the definition of $\theta_{13}$. Therefore the phenomenology of dark sector is 
governed mainly by the following three independent parameters: DM mass $m_{\chi_{_3}}$, splitting with the heavier neutral 
components $\Delta M$ and mixing angle $\theta_{13}$. Thus the ultimate free parameters in the dark sector are: 
\begin{equation}
 \textrm{Dark~Parameters}: ~~~~\{~ m_{\chi_{_3}},~\Delta M,~ \sin\theta_{13}\},~{\rm or} ~\{~ M_{R_1}, ~M,~ \sin\theta_{13}\}.
 \label{eq:parameters2}
\end{equation}
%
\subsection{Theoretical and Experimental constraints }\label{sec:constraintsBL}
\noindent $\bullet$\textbf{ Stability of potential:}
In order to maintain stable vaccum, the quartic terms of the scalar potential should obey following co-positivity conditions~\cite{Kannike:2012pe,Chakrabortty:2013mha}:
\begin{align}
 \lambda_{H} \geq 0,~~~~\lambda_{\Phi} \geq 0 ~~~{\rm and}~~~\lambda_{H\Phi} + 2\sqrt{\lambda_H \lambda_\Phi} \geq 0 .
\end{align}

\noindent $\bullet$\textbf{ Perturbativity:} 
In order to maintain perturbativity of the model, Yukawa couplings should satisfy the following limits:
\begin{align}
 |\lambda_H| < 4 \pi,~~~~~~|\lambda_{\Phi}| < 4 \pi, ~~~~|\lambda_{H\Phi}| < 4\pi ~~~; \nonumber \\
 |Y_1| < \sqrt{4\pi}, ~~~~~|Y_{\alpha j}| < \sqrt{4\pi},~~~ |g_{BL}| < \sqrt{4\pi} ~~.
\end{align}
\noindent $\bullet$\textbf{ LEP limits:} 
 LEP exclusion bound on charged fermion mass, $m_{\psi^\pm}=M > 102.7$ GeV~\cite{Abdallah:2003xe}. Again, we note that the bound from LHC has been evaluated for 
 a typical case of type III seesaw model, $m_{\psi^\pm}=M \gtrsim 800$ GeV~~\cite{Sirunyan:2017qkz, Sirunyan:2019bgz}, which is not strictly applicable to our case.

\noindent $\bullet$\textbf{ Constraints on $M_{Z_{BL}}$:}
LEP II data puts lower bound on $M_{Z_{BL}} / g_{BL} \geq 7$ TeV~\cite{Cacciapaglia:2006pk}. Corresponding bound from ATLAS and CMS at LHC Run 2 is 
more severe than LEP II, $M_{Z_{BL}} $ > 4.3 TeV for $g_{BL}$ of the same order as that of SM coupling~\cite{Aaboud:2017buh, Sirunyan:2018xlo, Okada:2016gsh}. 
However, this constraint can be relaxed for lower value of $g_{BL}$. For $M_{Z_{BL}}=\mathcal{O}$ (1TeV), the upper bound on $g_{BL}$ can be as small as 0.009~\cite{Bhattacharya:2019tqq}.
\\
\noindent $\bullet$\textbf{ Bounds on scalar singlet transforming under $U(1)_{B-L}$:}
In the extended scalar sector, the mixing angle ($\sin\beta$) and the mass of the extra physical state ($m_{h_2}$) faces the following constraints:
i) From $W$ mass corrections at Next to Leading Order (NLO)~\cite{Lopez-Val:2014jva}: 
For $ 250 ~{\rm GeV} \leq m_{h_2} \leq 850~{\rm GeV}$, one has $0.2 \leq \sin\beta \leq 0.3 $. 
ii) For the requirement of perturbative unitarity~\cite{Robens:2016xkb}: $\sin\beta \leq 0.2$ for $m_{h_2} \geq 850~{\rm GeV}$.
iii) Direct search measurement of Higgs signal strength at LHC provides an upper limit on mixing angle $|\sin\beta |<0.36$~\cite{Robens:2016xkb}.

\subsection{Relic abundance of dark matter}\label{blrelics}
\begin{figure}[!ht]
    \centering
    \subfloat{{\includegraphics[width=4cm]{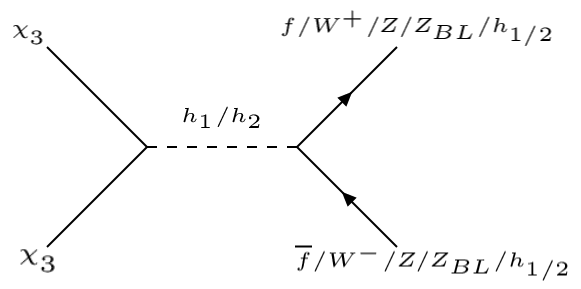} }}%
    \qquad
    \subfloat{{\includegraphics[width=3cm]{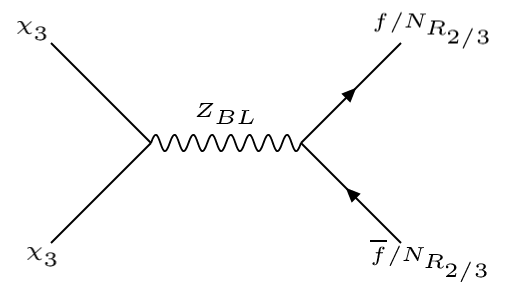} }}%
    \qquad\
    \subfloat{{\includegraphics[width=3.2cm]{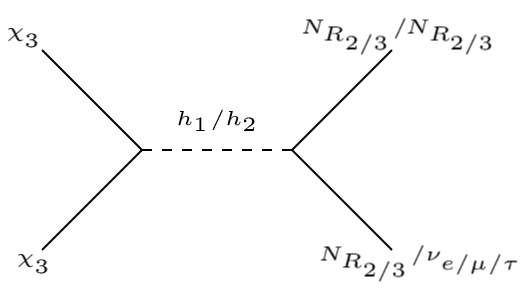} }}%
    \qquad\\
    \subfloat{{\includegraphics[width=3.5cm]{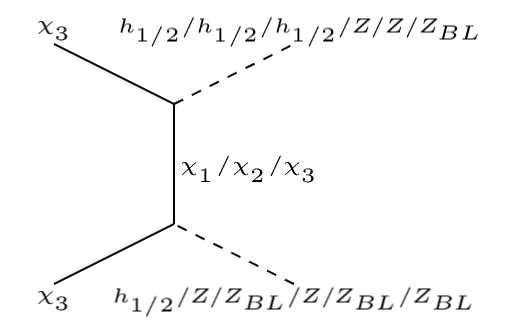} }}%
    \qquad
    \subfloat{{\includegraphics[width=3.5cm]{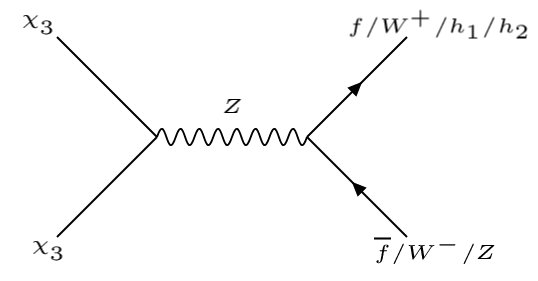} }}%
    \caption{\footnotesize{Additional annihilation channels of the DM ($\chi_{_3}$) to SM particles in $U(1)_{B-L}$ model.}}%
    \label{annihilation_BL}%
\end{figure} 

\begin{figure}[!ht]
    \centering
    \subfloat{{\includegraphics[width=4cm]{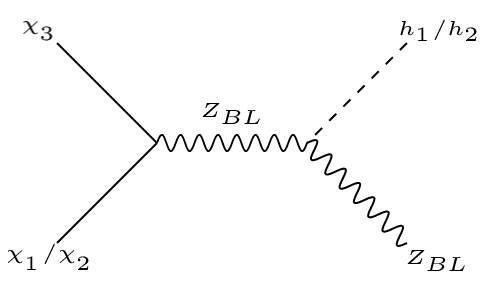} }}%
    \qquad
    \subfloat{{\includegraphics[width=4.5cm]{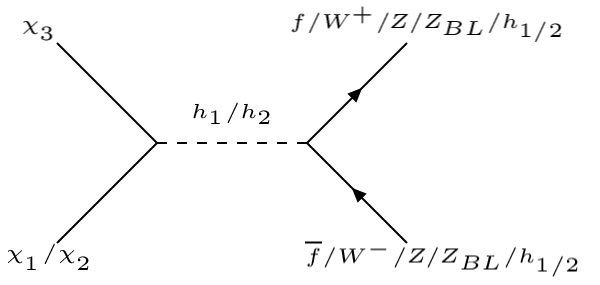} }}%
    \qquad
    \subfloat{{\includegraphics[width=4cm]{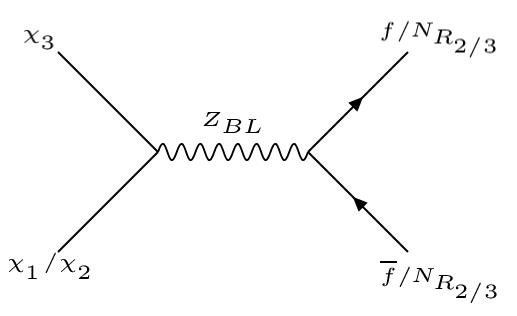} }}%
    \qquad
    \subfloat{{\includegraphics[width=4.3cm]{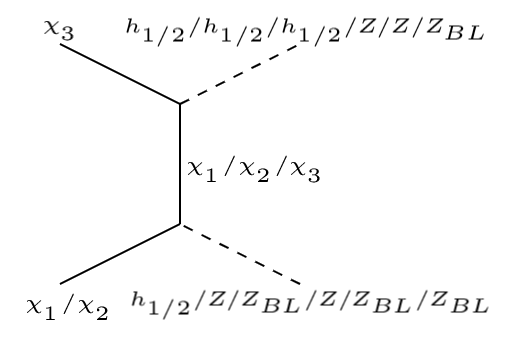} }}%
     \qquad
    \subfloat{{\includegraphics[width=4cm]{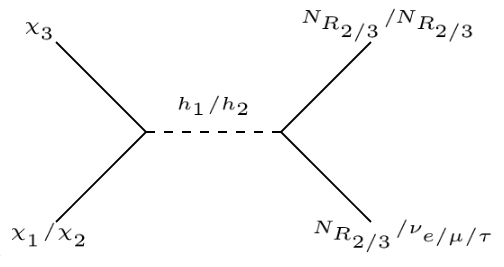} }}%
    \caption{\footnotesize{Additional coannihilation channels of DM ($\chi_{_3}$) with $\chi_{_1}$, $\chi_{_2}$ and $\psi^\pm$ in the $U(1)_{B-L}$ model. }}%
    \label{coanni_BL}
\end{figure}

\begin{figure}[!ht]
    \centering
    \subfloat{{\includegraphics[width=3cm]{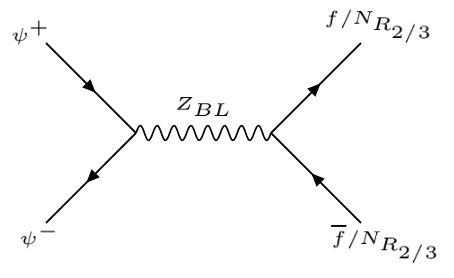} }}%
    \qquad
    \subfloat{{\includegraphics[width=3cm]{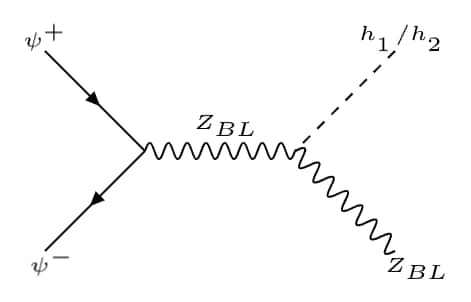} }}
    \qquad
    \subfloat{{\includegraphics[width=3cm]{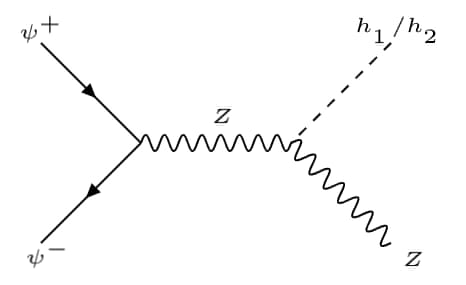} }}%
    \qquad
    \subfloat{{\includegraphics[width=1.8cm]{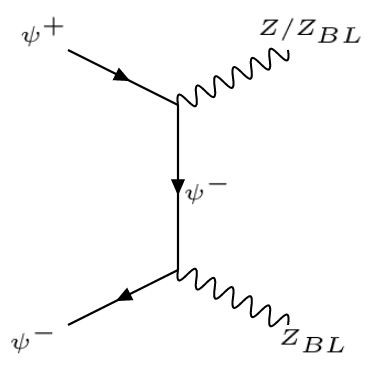} }}%
    \caption{\footnotesize{Additional coannihilation channels of $\psi^+$ and $\psi^-$ that contribute to relic density of 
DM ($\chi_{_3}$) in the $U(1)_{B-L}$ model.}}%
    \label{chargedcoanni_BL}
\end{figure}

The DM-SM interaction terms which deplete the number density of dark sector particles in the gauged $U(1)_{B-L}$ 
case has been discussed in Appendix~\ref{bl_int}. The additional relevant Feynman diagrams of annihilation and 
coannihilation processes over and above those already present in section \ref{relics} are shown in Fig.~\ref{annihilation_BL}, Fig.~\ref{coanni_BL} and Fig.~\ref{chargedcoanni_BL}.

\begin{figure}[htb!]
$$
 \includegraphics[height=5.0cm]{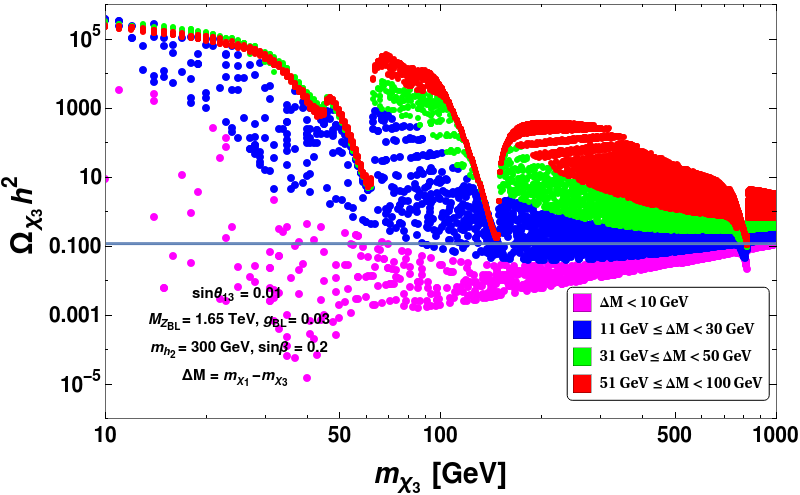} 
 \includegraphics[height=5.0cm]{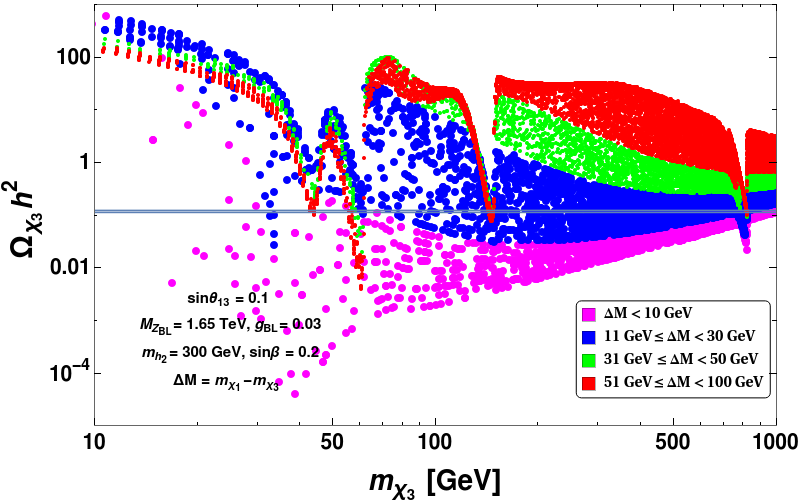}
 $$
 $$
 \includegraphics[height=5.0cm]{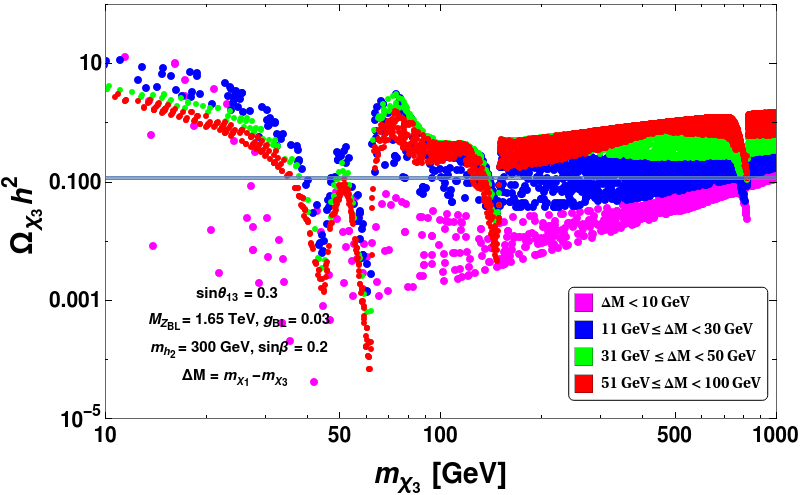} 
 \includegraphics[height=5.0cm]{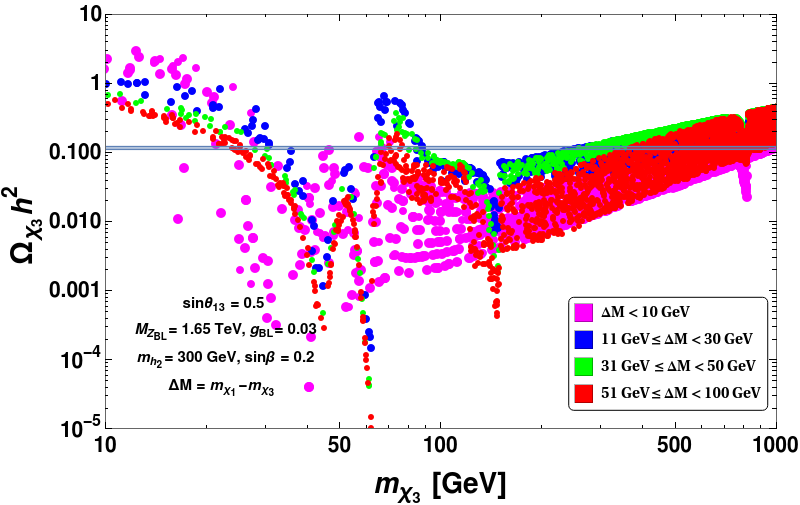}
$$ 
 \caption{\footnotesize{DM relic density as a function of DM mass ($m_{\chi_{3}}$) for different mass splitting $\Delta M$ between the DM and the NLSP (shown by different coloured 
patches as indicated in figure inset) for fixed values of $\sin\theta_{13}=0.01$ (top left panel), $\sin\theta_{13}=0.1$ (top right panel), 
$\sin\theta_{13}=0.3$ (bottom left panel) and  $\sin\theta_{13}=0.5$ (bottom right panel). Correct relic abundance from PLANCK data 
($0.1166\leq\Omega h^2 \leq 0.1206$) is shown by the thick horizontal silver line. The other parameters 
kept fixed are: $M_{Z_{BL}}=1.65 ~{\rm TeV}, g_{BL}=0.03, m_{h_2}=300 ~{\rm GeV}, \sin\beta=0.2$. }}
\label{Relic_BL}
\end{figure}

Again we use {\tt MicrOmegas} to calculate the relic density of DM. The plots for DM relic density $\Omega h^2$ as a function 
of DM mass $m_{DM}=m_{\chi_{_3}}$ are shown in Fig.~\ref{Relic_BL} for different mass splitting $\Delta M$ between the DM and 
the NLSP and for a chosen mixing angle $\sin\theta_{13}$. The main difference in this $B-L$ extended case compared to 
section \ref{relics} is the presence of new resonances at $m_{\chi_3}=m_{h_1}/2$ and $m_{\chi_3}=m_{h_2}/2$. These resonances  
get prominent only when the mass difference $\Delta M$ is sufficiently large such that the coannihilation processes are practically negligible. 
As we can see from Fig.~\ref{Relic_BL}, for small $\Delta M$, the coannihilation through off-diagonal $Z$ and $W^\pm$ 
mediated interactions dominate. Apart from that in Fig.~\ref{Relic_BL} we also see new resonances (in comparison to Fig.~\ref{fig:relicsplot1}) 
occur at $m_{\chi_3}=m_{Z}/2$ and $ m_{\chi_3}=M_{Z_{BL}}/2$. Note that the resonance at $m_{\chi_3}=m_{Z}/2$ is proportional to $\sin \theta_{13}$. As a result 
in the limit $\sin \theta_{13} \to 0$ and new particles, say $h_2$ and $Z_{BL}$ heavy enough we get back to the same situation as in 
Fig.~\ref{fig:relicsplot1}. 
\begin{figure}[!ht]
\centering
\includegraphics[width = 99 mm]{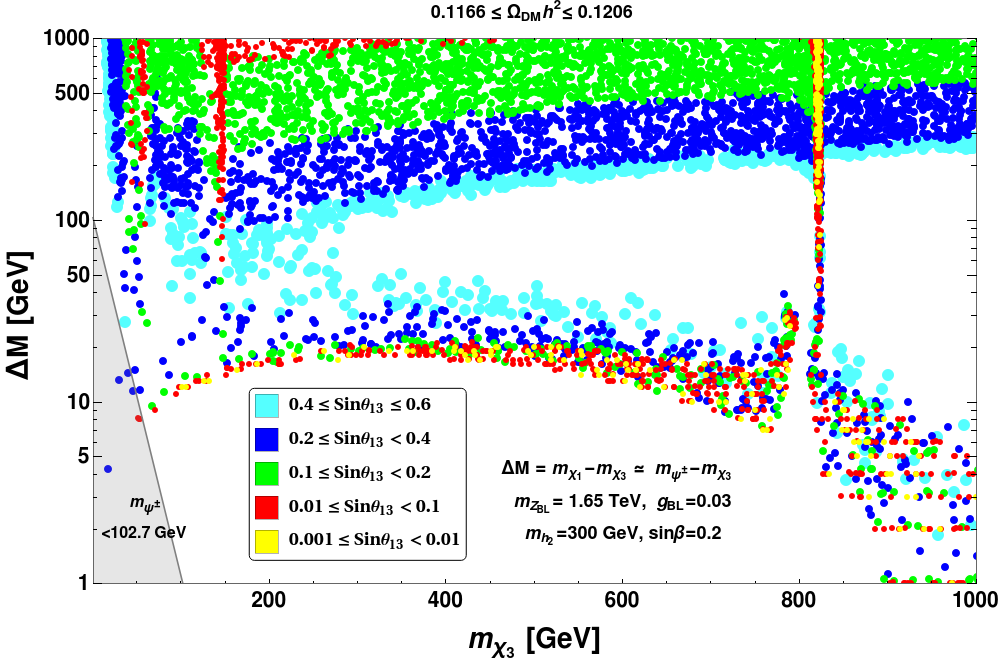}
 \caption{\footnotesize{DM relic density ($0.1166\leq\Omega_{DM} h^2\leq 0.1206$) allowed parameter space shown in $\Delta M-m_{\chi_3}$ plane for the 
 $U(1)_{B-L}$ model. Different coloured points indicate different ranges of $\sin\theta_{13}$ as specified in the figure inset. 
 The parameters kept fixed for the scan are $M_{Z_{BL}}=1.65 ~{\rm TeV}, g_{BL}=0.03, m_{h_2}=300 ~{\rm GeV}, \sin\beta=0.2$. 
 The shaded region in the bottom left corner is ruled out by LEP exclusion bound on charged fermion mass, $m_{\psi^\pm}=M >102.7$ GeV.}}
\label{BLrelic2}
\end{figure}

In Fig.~\ref{Relic_BL}, we have chosen $M_{Z_{BL}}=1.65$ TeV, $g_{BL}=0.03$, $m_{h_{2}}=300$ GeV and 
the mixing parameter of SM Higgs with the new B-L Higgs as $\sin\beta=0.2$, consistent with the available constraints. Also the masses of the two $\mathcal{Z}_2$ even right handed neutrinos are kept fixed as $M_{R_{2/3}}=500$ GeV.  
As $\sin\theta_{13}$ increases, the Yukawa coupling between the doublet and the singlet increases, and hence the $h_1$ (SM-like Higgs) 
mediated interactions become more and more dominant. It is also clear from Fig.~\ref{Relic_BL} that, irrespective of the mass difference 
$\Delta M$, with increasing $\sin\theta_{13}$ the annihilation rates increase making deeper resonance drops. Due to the presence of off-diagonal 
interactions in all cases, all resonance drops has been somewhat broadened up compared to the case of pure diagonal interactions. 

In Fig.~\ref{BLrelic2}, the correct relic abundance is plotted in the plane of $\Delta M$ vs $m_{\chi_{_3}}$, where 
$\Delta M=(m_{\chi_{_1}}- m_{\chi_{_3}})$. Again, the main outcome remain almost similar as before, excepting the presence of additional peak at $m_{\chi_{_3}}=\frac{M_{Z_{BL}}}{2}$ GeV due to $Z_{BL}$ resonance. Other resonances at $m_Z/2,m_{h_1}/2,m_{h_2}/2$ are also visible. Just before the $Z_{BL}$ resonance, we can see the effect of off-diagonal $Z_{BL}$ mediated interactions (see \ref{dm-sm-bl}).

We note here that in the limit $\sin\theta_{23}\to 0$ (alongwith $g_{BL}\to 0, \sin\beta\to 0$ and for very heavy $Z_{BL}$ and $h_2$), Fig.~\ref{BLrelic2} reduces to Fig.~\ref{relic3}, i.e. $U(1)_{B-L}$ extension boils down to the one without it.

\subsection{Direct Detection prospects}\label{bldirect}
\begin{figure}[!ht]
\centering
\includegraphics[width = 40 mm]{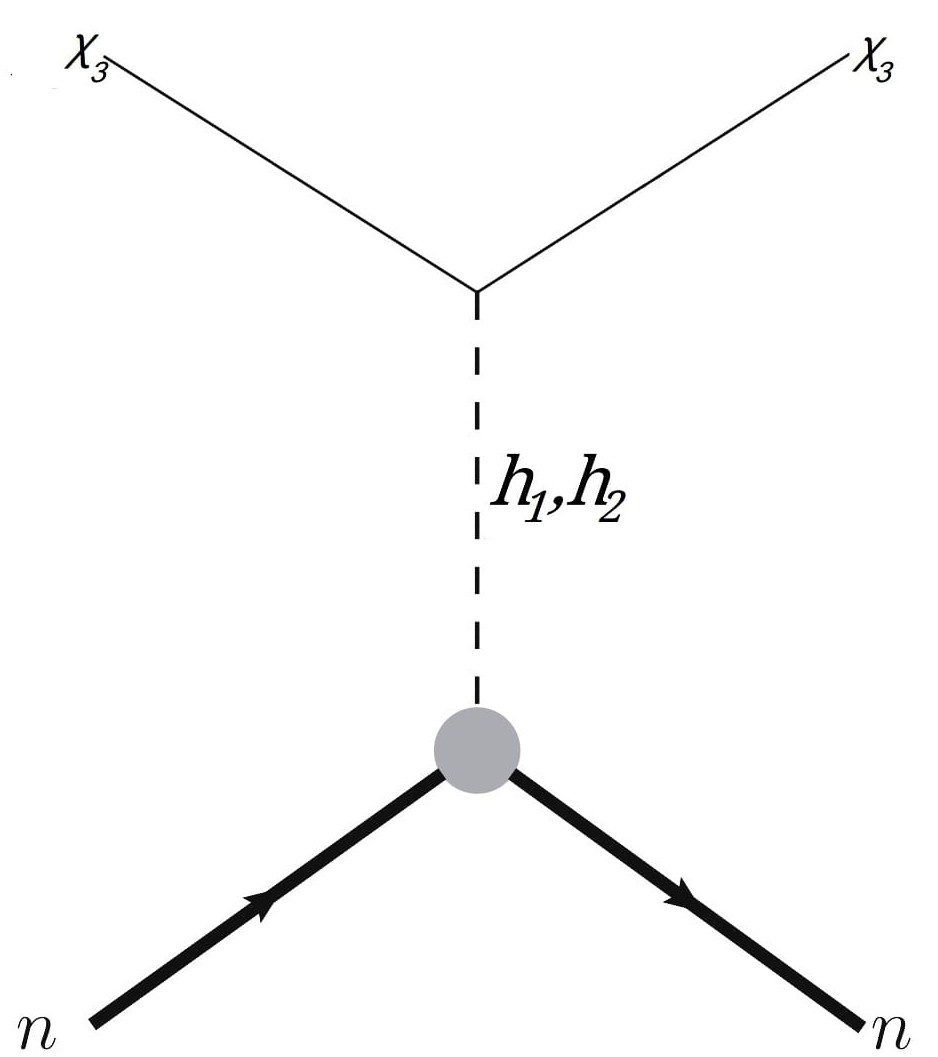}
\caption{\footnotesize{Feynman Diagram for elastic scattering of DM off nuclei at terrestrial laboratory in the $U(1)_{B-L}$ extended model.}}
\label{DD_n3n3}
\end{figure}

The DM candidate ($\chi_3$) in this model is a Majorana fermion, hence the Z and $Z_{BL}$-mediated vector current interaction vanishes. 
Although there is a possibility of spin dependent scattering through axial vector interaction mediated by the 
vector bosons, the sensitivity and bounds are extremely weak. Therefore the prominent channel for direct detection of $\chi_3$ is through $H-\Phi_{BL}$ 
mixing, which results in spin-independent scattering of DM off nuclei. The Feynman diagram for such interaction is shown in Fig.~\ref{DD_n3n3}. 
The spin-independent DM-nucleon elastic scattering cross-section is again given by Eqn.-\ref{da}. However, 
in contrast to the previous case, here there are two propagators ($h_1$ and $h_2$) that can mediate the DM pair production and 
hence direct detection is through the interference of two diagrams. So in this case the effective coupling strength 
$\alpha_q$ is given by:
\begin{equation}\label{BLdda2}
\alpha_q =\frac{m_q}{v}\Big(\frac{\lambda_a \cos\beta}{m^2_{h_1}}-\frac{\lambda_b \sin\beta}{m^2_{h_2}}\Big)\,,
\end{equation}
where 
\begin{equation}
\begin{aligned}
\lambda_a = \frac{Y_1}{2}(s_{13} + s_{23}c_{13})c_{13}c_{23}\cos\beta - \frac{y'_1}{2\sqrt{2}} c^2_{13}c^2_{23} \sin\beta\,,\\
\lambda_b = - \frac{Y_1}{2}(s_{13} + s_{23}c_{13})c_{13}c_{23}\sin\beta -\frac{y'_1}{2\sqrt{2}} c^2_{13}c^2_{23} \cos\beta\,.\\
\end{aligned}
\label{bl_direct_yukawa}
\end{equation}
In the numerical calculation we use the Yukawa coupling $Y_1\approx \Delta M \sin 2\theta_{13}/v$ as given by Eqn.~\ref{theta2} and 
$y'_1=\sqrt{2}M_{R_1}/v_{BL}= 2\sqrt{2}M_{R_1} g_{BL}/M_{Z_{BL}}$. So the direct search cross-section indirectly depends on $\Delta M$, $g_{BL}$ and $M_{Z_{BL}}$ as well. 


The relative minus sign between the two propagators comes from the orthogonal mixing matrix in Eqn.~\ref{higgs_mixing}. From Eqs.~\ref{da}, 
\ref{dd}, \ref{dda} and \ref{BLdda2}, the spin-independent scattering cross-section is given by,
\begin{equation}
\begin{aligned}
\sigma^{SI} &= \frac{\mu^2_r}{\pi A^2}\Big(\frac{\lambda_a \cos\beta}{m^2_{h_1}}-\frac{\lambda_b \sin\beta}{m^2_{h_2}}\Big)^2\Big[Z\frac{m_p}{v}\Big(f^{p}_{Tu} + f^{p}_{Td} + f^{p}_{Ts} + \frac{2}{9}f^{p}_{TG}\\
 &+(A-Z)\frac{m_n}{v}\Big(f^{n}_{Tu} + f^{n}_{Td} + f^{n}_{Ts} + \frac{2}{9}f^{n}_{TG}\Big)\Big]^2
\end{aligned}
\end{equation}

\begin{figure}[ht]
$$
\includegraphics[height=5cm]{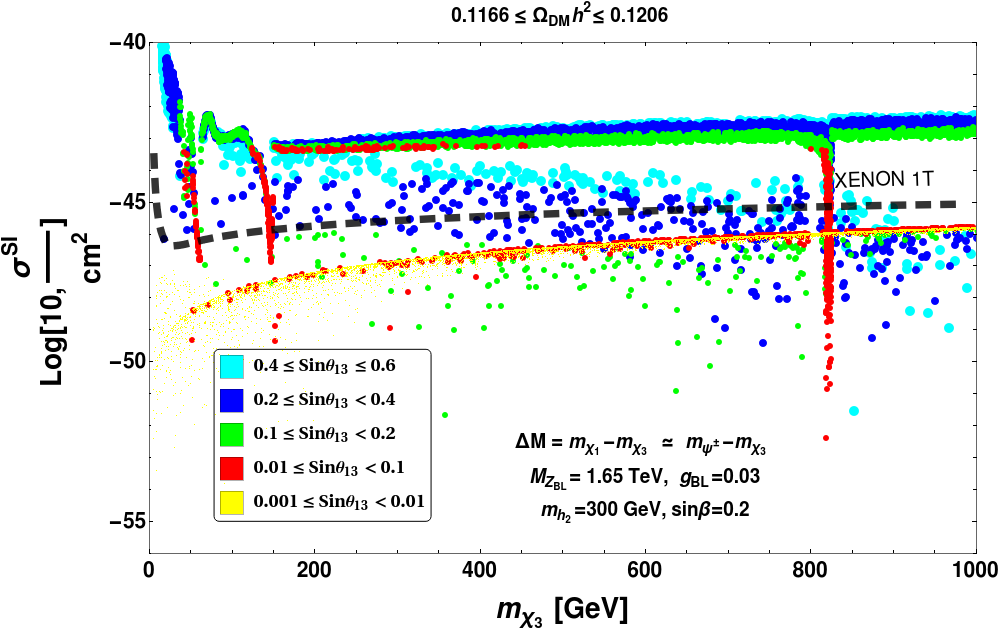}~
 \includegraphics[height=5cm]{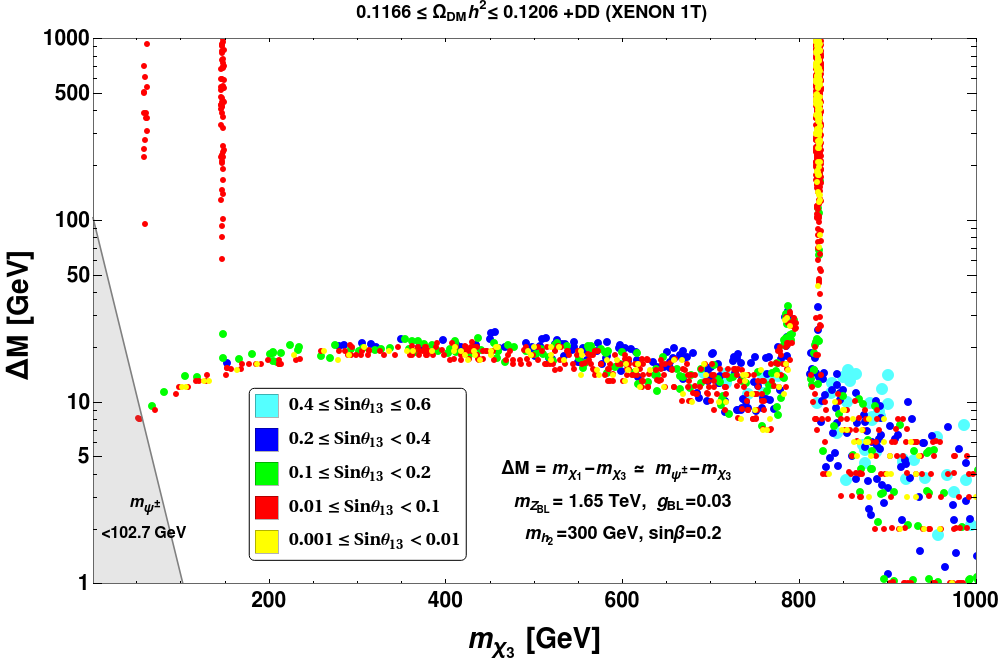}
$$ 
 \caption{\footnotesize{[Left]: Spin-independent direct detection cross section of DM ($\chi_3$) with nucleon as function of DM mass (in GeV) for $U(1)_{B-L}$ model confronted with XENON-1T data  
 over and above relic density constraint from PLANCK;  [Right]: Correct DM relic density allowed parameter space of the model in $\Delta M-m_{\chi_{_3}}$ plane 
 constrained by XENON-1T bound. Different coloured points indicate different ranges of $\sin\theta_{13}$ as mentioned in the figure inset. 
 The parameters kept fixed for the scan are $M_{Z_{BL}}=1.65 ~{\rm TeV}, g_{BL}=0.03, m_{h_2}=300 ~{\rm GeV}, \sin\beta=0.2$. 
 The shaded region in the bottom left corner of right hand plot is ruled out by LEP exclusion bound on charged fermion mass, $m_{\psi^\pm}=M >102.7$ GeV.}}
 \label{directdetectionBL}
\end{figure}

Now we turn to the parameter space of the model consistent with direct search constraints. In left panel of Fig.~\ref{directdetectionBL}, we have 
confronted the points satisfying relic density with the spin independent elastic cross section obtained for the model 
as a function of DM mass. The XENON-1T bound is shown by dashed black line. Again, the region below this line satisfy both 
relic density as well as direct detection constraint. These points (satisfying relic density as well as direct detection 
constraint from XENON-1T) are shown in the right panel of Fig.~\ref{BLrelic2} in the $\Delta M-m_{\chi_{_3}}$ plane. Again we 
see that null observation from direct search crucially tames down the relic density allowed parameter space. The available parameter 
space of the $U(1)_{B-L}$ model is very similar to that without the gauge extension, excepting for the resonance regions at $m_{\chi_3}=m_{h_{1/2}}/2$ 
and $m_{\chi_3}=M_{Z_{BL}}/2$, where $\Delta M$ can be uncorrelated to DM mass.

\subsection{ATLAS bound on $g_{BL}-M_{Z_{BL}}$}
\label{gbl_versus_mzbl}
We now turn to find the allowed parameter space in the $\Delta M-m_{\chi_{_3}}$ plane in light of ATLAS bound on $g_{BL}$  versus $M_{Z_{BL}}$. 
In the previous sections we kept $M_{Z_{BL}}$ fixed at 1650 GeV  corresponding 
to $g_{BL}=0.03$ compatible with ATLAS data~\cite{Aaboud:2017buh}. As a result of choosing such a small value of $g_{BL}$, 
the effect of ${Z_{BL}}$ was only evident at resonance when $m_{\chi_3} \sim M_{Z_{BL}}/2$ (see Fig.~\ref{directdetectionBL}). In the following we 
highlight the effect of $Z_{BL}$ mediated diagrams by varying the coupling and mass. We perform 
a scan by varying the model parameters in the following range:
\begin{equation}\label{eq:neu:NH}
~~ 	
		\left\{
		\begin{array}{l}
		
		 1~{\rm GeV}\leq m_{\chi_3}\leq 2000 ~{\rm GeV}\\
		 1 ~{\rm GeV} \leq\Delta M \leq 1000 ~{\rm GeV}\\
		 20 ~{\rm GeV} \leq M_{Z_{BL}} \leq 4000 ~{\rm GeV}\\
		 0.001 \leq \sin\theta_{13} \leq 0.6\\
		 0.001 \leq g_{BL} \leq 0.3 \,.
		\end{array}
		\right.
		\end{equation}
Other parameters kept fixed are: $\sin\beta =0.2$ and $m_{h_2}= 300$ GeV. Also the masses of the two $\mathcal{Z}_2$ even right handed neutrinos are kept fixed as $M_{R_{2/3}}=500$ GeV. 

\begin{figure}[h]
$$
 \includegraphics[height=5.0cm]{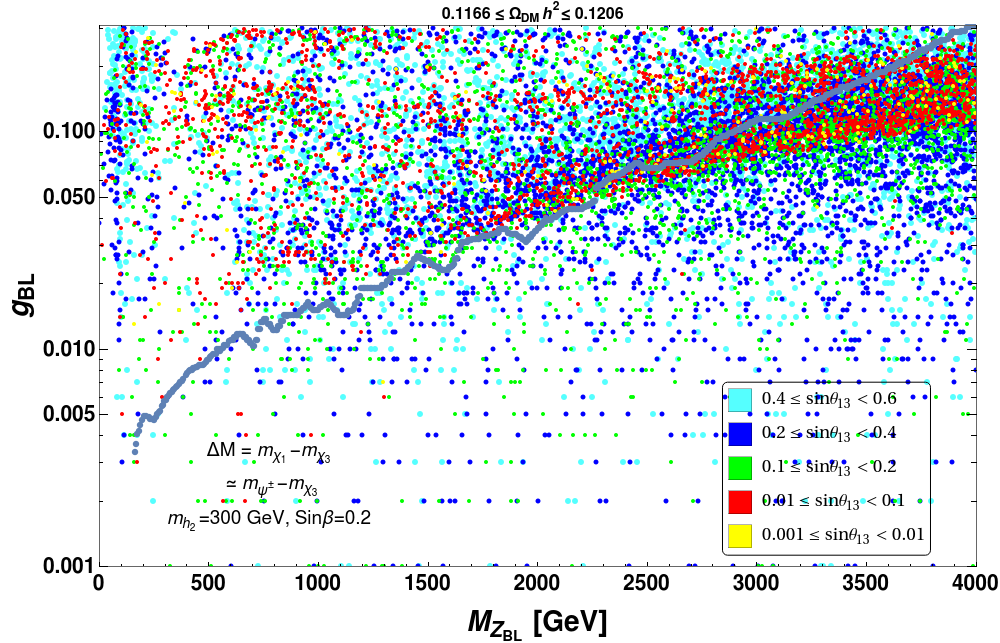} ~
 \includegraphics[height=5.0cm]{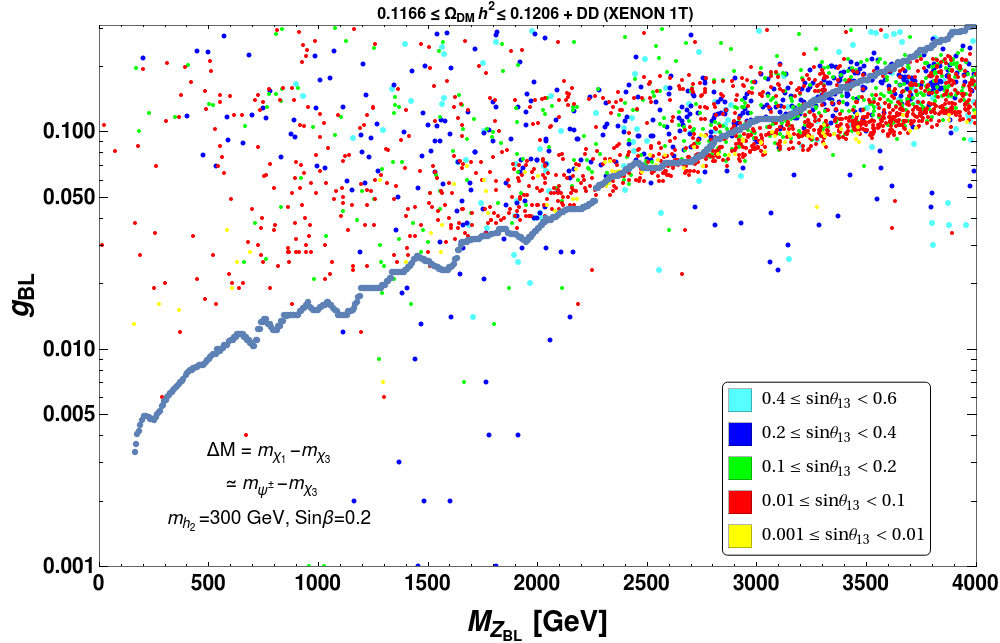}
$$ 
 \caption{\footnotesize{[Left]: Parameter space satisfying relic density constraint from PLANCK ($0.1166\leq\Omega_{DM} h^2\leq 0.1206$) in the plane of $g_{BL}-M_{Z_{BL}}$ for $U(1)_{B-L}$ model; 
 [Right]: Parameter space satisfying both relic density constraint from PLANCK and direct detection bound from XENON-1T in the plane of $g_{BL}-M_{Z_{BL}}$. The thick silver line shows the ATLAS bound on 
 $g_{BL}$ vs $M_{Z_{BL}}$~\cite{Aaboud:2017buh} plane from non-observation of $Z_{BL}$ in collider data.}}
 \label{Atlas-1}
\end{figure}

\begin{figure}[htb!]
$$
 \includegraphics[height=5.1cm]{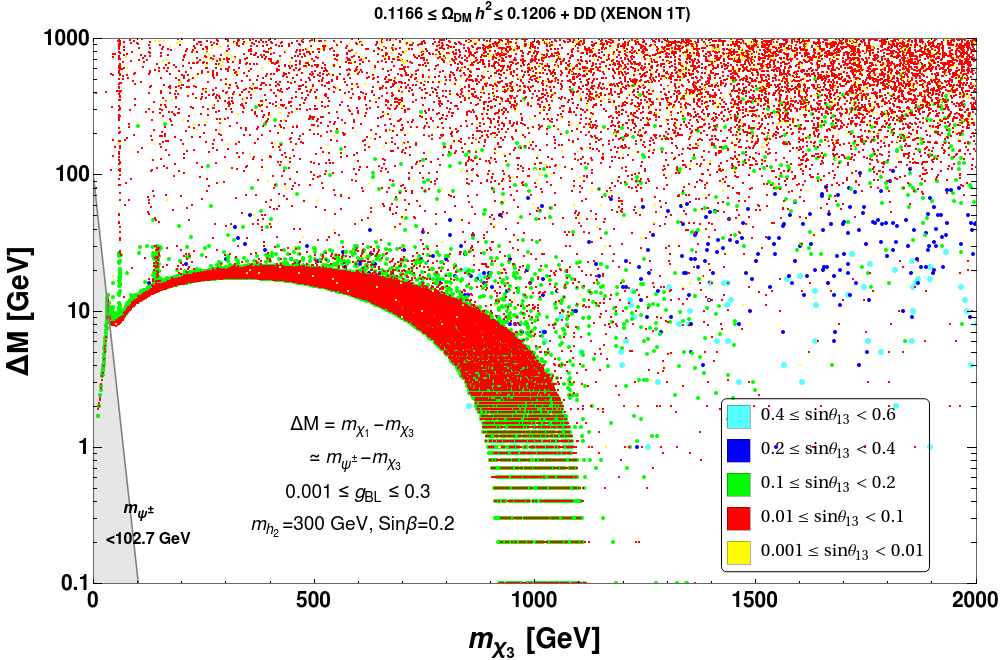} ~
 \includegraphics[height=5.1cm]{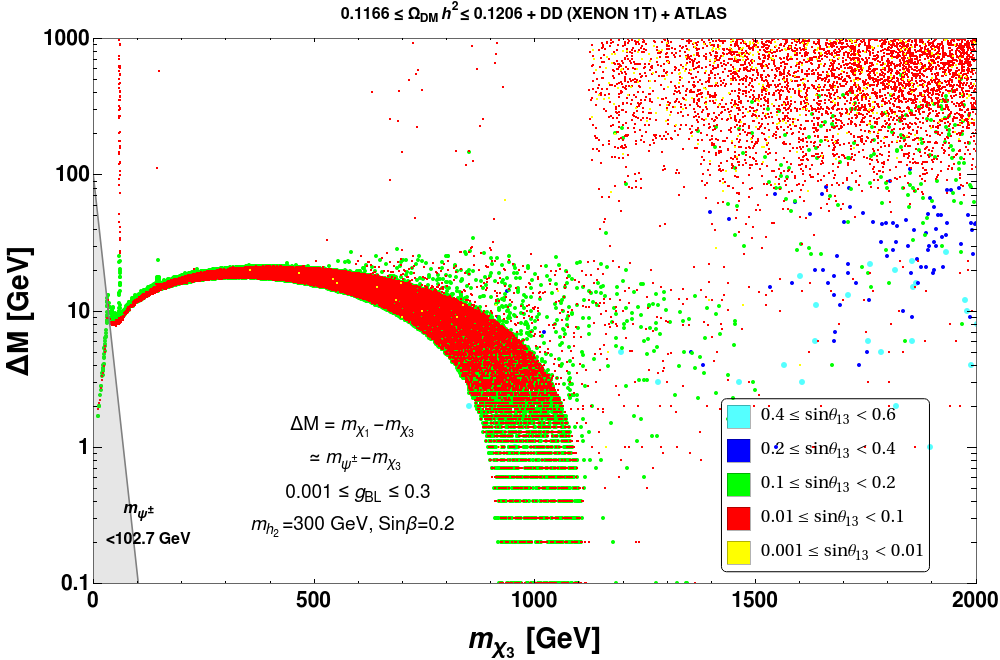}
$$
$$ 
 \includegraphics[height=5.1cm]{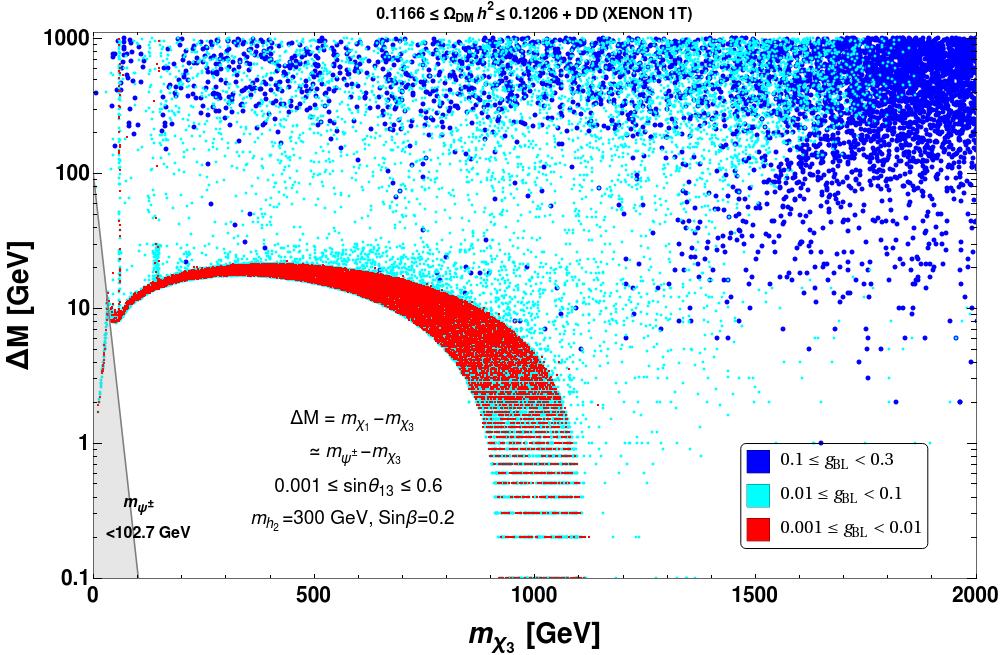} ~
 \includegraphics[height=5.1cm]{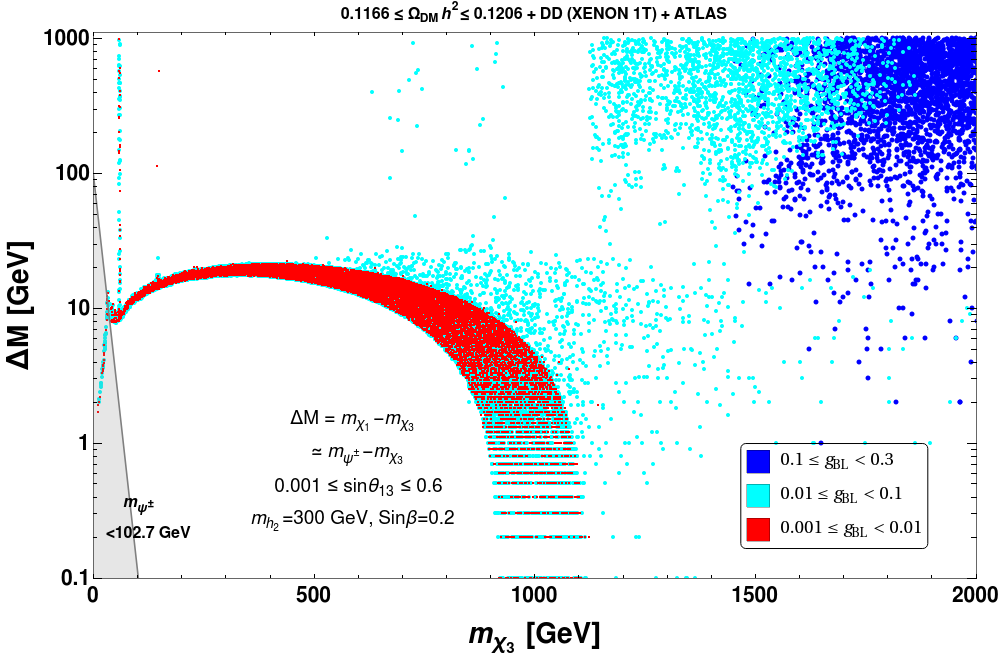}
$$
 \caption{\footnotesize{[Top Left]: Parameter space satisfying relic density (PLANCK) and direct search (XENON-1T) bound in $\Delta M - m_{\chi_3} $ plane, different colours 
 indicate different choices of $\sin\theta_{13}$; [Top Right]: Same as top left but additionally ATLAS bound on $ g_{BL}-M_{Z_{BL}}$~\cite{Aaboud:2017buh} applied;
 [Bottom Panel]: Same as in the top panel, but different coloured points indicate different ranges of $g_{BL}$ coupling as mentioned in figure inset, with left (right) plot without 
 respecting (with) ATLAS bound. The shaded region in the bottom left corner is ruled out by LEP exclusion bound on charged fermion mass, $m_{\psi^\pm}=M >102.7$ GeV. }}
 \label{Atlas-2}
\end{figure}

We first show the constraint coming from non-observation of a new gauge boson ($Z_{BL}$) at LHC coming from ATLAS~\cite{Aaboud:2017buh} analysis 
on $g_{BL}$ for corresponding values of $M_{Z_{BL}}$ shown by the silver thick line in Fig.~\ref{Atlas-1}. This indicates that points below the line with smaller $g_{BL}$ is allowed, 
while those above the line are discarded. The left plot shows points which satisfy relic density constraint from PLANCK ($0.1166\leq\Omega_{DM} h^2\leq 0.1206$) data and right 
plot shows the points which satisfy both relic density and direct search bounds from XENON 1T. Different colours indicate ranges of $\sin\theta_{13}$ as mentioned in figure inset. 
We then showcase the fate of the model when the bound from ATLAS is implemented on the parameter space in 
$\Delta M$ vs $m_{\chi_3}$ plane for different $g_{BL}$ values in Fig.~\ref{Atlas-2}. In the top panel we show the available parameter space in terms of different ranges 
of $\sin\theta_{13}$, while the same is shown in bottom panel for different ranges of $g_{BL}$ coupling for relic density and direct search allowed parameter space of the 
$U(1)_{B-L}$ model. For clarity in inferring how much parameter space gets discarded by the ATLAS bound, in the left panel we show relic density and direct search allowed points without 
ATLAS bound, while on the right panel, we show those after incorporating ATLAS bound~\cite{Aaboud:2017buh}.  

We see from Fig.~\ref{Atlas-2} when $\Delta M \lesssim 10$ GeV, 
the contribution to relic density comes from annihilation, coannihilation and $Z_{BL}$ resonance with relatively 
smaller $g_{BL}$. As we go for further larger $\Delta M$, the coannihilation contribution to relic density decreases 
gradually and gets compensated by $Z_{BL}$ exchange diagrams with increasing values of $g_{BL}$. Beyond $m_{\chi_3}
 =1000$ GeV, the correlation between $\Delta M$ and $m_{\chi_3}$ is lost and relic is mostly dominated by Higgs 
and $Z_{BL}$ mediation. In the right panel of Fig.~\ref{Atlas-2}, we impose bound on $g_{BL}-M_{Z_{BL}}$ from ATLAS data. 
The upper bound on $g_{BL}$ by ATLAS data for lighter $Z_{BL}$ is extremely small (for eg., $M_{Z_{BL}}\sim$ O(1TeV), 
upper bound on $g_{BL} \sim$ 0.009~\cite{Bhattacharya:2019tqq}). 
 Consequently if we have to satisfy ATLAS bound, then all those resonance points with large $g_{BL}$ in the left panel 
of Fig.~\ref{Atlas-2} upto $m_{\chi_3} \sim$ 500 GeV are no longer there in the right panel of Fig.~\ref{Atlas-2}. It is 
only when $M_{Z_{BL}}$ becomes sufficiently large, so that $g_{BL}$ can take somewhat moderate values, we can see the $Z_{BL}$ 
resonance affects. That is why in the right panel of Fig.~\ref{Atlas-2}, such resonance points survive for 
$m_{\chi_3}\geq 500$ GeV. For $m_{\chi_3}> 1000$ GeV, the points which survive the ATLAS bound are mostly due to $Z_{BL}$ resonances 
with relatively large $g_{BL}$. Note that the direct detection cross section has mild dependency on these 
resonance points. Therefore, these resonance points for $m_{\chi_3}> 1000$ GeV also easily survive from XENON-1T bound.
\section{Collider Signatures}
\label{collider}

Both the model frameworks studied here, have attractive signatures at the Large Hadron Collider (LHC) due to the presence of SM isodoublet. 
There exists different types of production processes and decay final states which can be categorized broadly into leptonic and hadronic final states. 
Leptonic final states are favoured over hadronic states for less SM contamination. All the heavier dark sector particles
finally decay into the DM ($\chi_3$), which is missed in the detector and necessarily associate each final state with missing transverse energy ($\slashed{E_T}$) defined 
as: 
\bea
 \slashed{E}_T = -\sqrt{\left(\sum_{\ell,j,unc} p_x\right)^2+\left(\sum_{\ell,j,unc} p_y\right)^2},
 \label{eq:met}
 \eea
where the sum runs over all visible objects that include leptons ($\ell=e,\mu$) and jets, and unclustered components. 
Here we list some of the most important leptonic final states that the models offer. We will refer to the model without $U(1)_{B-L}$ as model I and 
the one with $U(1)_{B-L}$ extension as model II.\\

$\bullet$ \underline{Opposite sign dilepton $(\ell^+ \ell^- +\slashed{E_T})$:}\\

\begin{figure}[htb!]
$$
 \includegraphics[height=4.5cm]{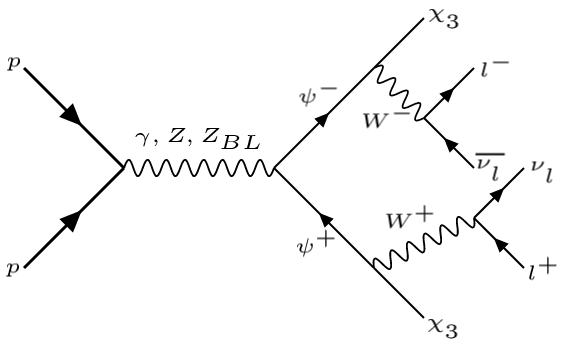} ~
 \includegraphics[height=4.5cm]{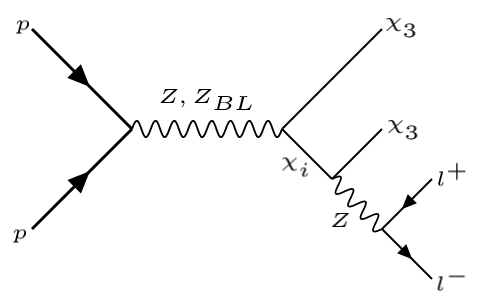}
$$ 
 \caption{\footnotesize{OSD + $\slashed{E_T}$ signal at LHC due to: (i) (Left) $\psi^+ ~ \psi^-$ production and (ii) (Right) $\chi_3 ~ \chi_i$ production. For model II $i=1,2$, for model I, $i=2$.}}
 \label{coll_Feyn1}
\end{figure}

The heavy charged component of $SU(2)_L$ doublet, $\psi^\pm$ (NLSP) can be produced via $(Z,\gamma)$ mediation in model I and 
$(Z,\gamma,Z_{BL})$ mediation in model II. Further they decay to leptonic final state via on-shell or off-shell $W^\pm$ mediator 
(depending on mass splitting $m_{\psi^\pm}-m_{\chi_{_3}}$) and stable DM ($\chi_{_3}$). As a result the process yields 
hadronically quiet opposite sign dilepton (OSD) plus missing energy ($\ell^+ \ell^- + \slashed{E_T}$) signature at collider as shown in 
the Feynman graph in the left panel of Fig.~\ref{coll_Feyn1}: 
\bea
OSD + \slashed{E_T}:~~ p~p \rightarrow \psi^+ ~ \psi^- , ~(\psi^- \rightarrow ~\ell^- ~\overline{\nu_\ell} ~ \chi_{_3}),~~(\psi^+ \rightarrow \ell^+ ~{\nu_\ell} ~\chi_{_3}); ~~\ell = \{e ,\mu\} ~.\nonumber
\eea

Also the production of $\chi_{i}~ \chi_3$ pair via $Z$ propagator in model I and $Z,~ Z_{BL}$ propagator in model II gives rise to OSD final state as shown in right panel of 
Fig.~\ref{coll_Feyn1}:
\bea
OSD + \slashed{E_T}:~~ p~p \rightarrow \chi_{i}~\chi_3 ~(\chi_{1,2} \rightarrow ~\ell^-~\ell^+ ~ \chi_{_3}); ~~\ell = \{e ,\mu\}~~i = \{1 ,2 \} ~.\nonumber
\eea
It is important to note that in model I, $i=2$ is the only possibility [see appendix \ref{dm_sm_int} and Eq.~\ref{dmint} in particular].
Also note that the production of the heavy neutral components as above are proportional to the mixing angle ($\sin\theta$), 
which is small (to respect direct search constraints). Therefore  $\chi_{i}~\chi_3$ production is suppressed than the $\psi^+ ~ \psi^-$ production process, 
which is independent of mixing angle ($\sin\theta$). 

It is worth mentioning that similar process have been studied widely in context of supersymmetric theories by chargino pair production at LHC~\cite{ATLAS:2018diz,Calibbi_2014, Abdughani:2017dqs, Xiang:2016ndq, Belanger:2013pna, Choudhury:2016lku, Cao:2015efs, Aad:2014vma, ATLAS:2018gfq, ATL-PHYS-PUB-2018-048}.  
Non observation of any excess in OSD signal events at LHC results in a bound on the charged fermion mass. The bound(s) obtained for charginos are often specific 
to supersymmetric model given so many additional parameters that the theory inherits and may not be applicable (fully) to our case. Recasting the full analysis in our case is also out of the scope of this draft and 
will be taken up elsewhere. We will however provide a short account of the event simulation procedure and hint towards some broad conclusions. 
Note however that a model independent bound was found in context of LEP experiment as $m_{\psi^\pm} \gtrsim 102.7$ GeV~\citep{2003}. 
One may also look into \cite{Bhattacharya:2017sml, Bhattacharya:2018fus} for event level analysis at LHC without $U(1)_{B-L}$ case and 
in \cite{Barman:2019aku} for $U(1)_{B-L}$ case. \\

$\bullet$ \underline{Three leptons $(\ell \ell \ell +\slashed{E_T})$:}\\

\begin{figure}[htb!]
$$
 \includegraphics[height=4.5cm]{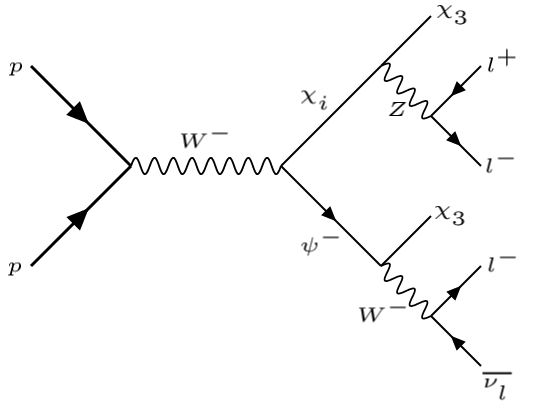} ~~
  \includegraphics[height=4.5cm]{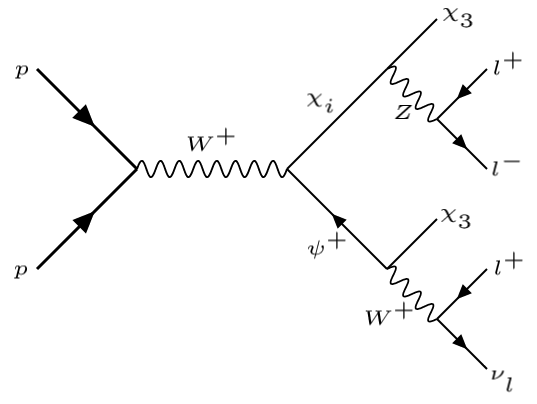} 
$$ 
 \caption{\footnotesize{$\ell\ell\ell+ \slashed{E_T}$ signal at LHC. For model II, $i=\{1,2\}$, while for model II, $i=2$. }}
 \label{coll_Feyn2}
\end{figure}

Hadronically quiet trilepton plus missing energy signature can be obtained from the production of heavy neutral, $\chi_{1,2}$ and charged fermions states, 
$\psi^\pm$ via $W^\pm$ mediator as shown in Fig.~\ref{coll_Feyn2}: 
\bea
3 \ell + \slashed{E_T}:~~ p~p \rightarrow \psi^\pm ~ \chi_{i} , ~(\psi^\pm \rightarrow ~\ell^\pm ~\nu_\ell (\overline{\nu_\ell}) ~ \chi_{_3}),~~(\chi_{i} \rightarrow \ell ^-\ell^+ ~\chi_{_3}); ~~\ell = \{e ,\mu\}~~i = \{1 ,2\} ~.\nonumber 
\eea
Again, it is worth noting that although the production process is same in both model I and model II, subsequent decay of $\chi_i \to \chi_3 Z^*$ is only allowed for $\chi_2$ in model I and provides 
a way of distinguishing the two cases.
The fact that no significant excess in hadronically quiet trilepton events are observed at LHC and the result agrees to SM contribution to a great extent puts a bound 
on the relevant parametrs. From ATLAS data, following constraints
can be obtained: $m_{\chi_{1,2}}, m_{\psi^\pm} < 270$ GeV, $m_{\chi_3} \lesssim 70$ GeV with ${\rm BR}~(\chi_{1,2} \rightarrow Z \chi_3) \gtrsim 60 \%$~\cite{Aad_2014}.
We may note that similar trilepton signature can also arise from Higsino-Bino production in supersymmetric models, which have been studied in context of LHC data~\cite{Calibbi_2014}. \\

$\bullet$ \underline{Four leptons $(\ell \ell \ell \ell +\slashed{E_T})$:}\\

\begin{figure}[htb!]
\centering
 \includegraphics[height=4.5cm]{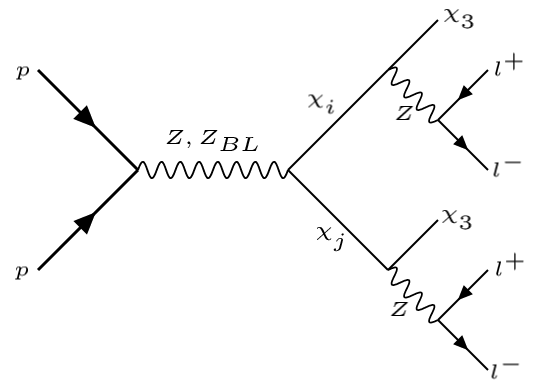}
 \caption{\footnotesize{$\ell\ell\ell\ell+ \slashed{E_T}$ signal in model II at LHC ($i,j=\{1,2\}, ~i\neq j$).}}
 \label{coll_Feyn3}
\end{figure}

The heavy neutral fermionic DM states, $\chi_{1,2}$ (NLSP) can be produced at LHC via $Z$ propagator in model I and $Z, Z_{BL}$ propagator in model II. 
The heavy states, $\chi_{1,2}$ further decay to leptonic final states via $Z$ and produce four leptons plus missing energy signature as shown in Fig.~\ref{coll_Feyn3}:   
\bea
 \ell\ell\ell\ell + \slashed{E_T}:~~ p~p \rightarrow \chi_i ~ \chi_{j} , ~~(\chi_{i,j} \rightarrow \ell ^-\ell^+ ~\chi_{_3}); ~~\ell = \{e ,\mu\}; ~~i,j = \{1,2\} ~.\nonumber 
\eea

We should note here that there are two main issues of producing four lepton states: (i) We need to produce $\chi_1\chi_2$ pair, (ii) then $\chi_{1,2}$ both needs to decay via $Z$ to $\chi_3$. Now from 
interaction vertex in appendix \ref{dm_sm_int} and Eq.~\ref{dmint}, we see that the decay of $\chi_1$ can't occur to $\chi_3 Z^{*}$ unless the model is extended by $U(1)_{B-L}$ [see appendix \ref{bl_int} and Eq.~\ref{eq:intBL}], thus making the 
signal exclusive for the $U(1)_{B-L}$ extension. Apart, one may also have hadronically quiet six lepton states arising from the decay of $\chi_1 \to \chi_2Z^{*}, \chi_2\to\chi_3Z^{*}$, followed by leptonic decays of the off-shell $Z$ from the 
same production process for both models with $U(1)_{B-L}$ extension and without that.  \\

$\bullet$ \underline{Single lepton with jets $(\ell^\pm + j j+ \slashed{E_T})$:}\\

The leptons in the final state arise out of $W$ and $Z$ boson decays (see Figs.~\ref{coll_Feyn1}, \ref{coll_Feyn2}, \ref {coll_Feyn3}), which anyway could also decay to quark antiquark pair to yield jets. 
Therefore, apart from purely leptonic signatures, one may also have hadrons or jet-rich final states. For example, the charged fermion pair production can lead to single lepton with two 
jets plus missing energy signature when one off shell $W$ decays to hadronic final state (see Fig.~\ref{coll_Feyn1}). Obviously when both $W$ decays hadronically, one ends up with four (or more) jets. 
LHC being a QCD machine, hadronic final states are prone to huge SM QCD background and therefore disfavoured. In event analysis, segregating signal from SM background 
is an important task. Missing energy variable as introduced in Eq.~\ref{eq:met} play a crucial role, as in SM contributions to $\slashed{E_T}$ mainly 
arise from neutrinos and mistagging.  \\

%
%
%
%
%
%
 
$\bullet$ \underline{Displaced vertex signature of $\psi^{\pm}$:}\\

We already observed that a large region of available parameter space of the model relies on small $\Delta M$ 
(for example, see in the right panel of Fig. \ref{directdetection}). The decay of $\psi^\pm$ is then phase space suppressed and 
can produce a displaced vertex, which can serve as a very crucial signature of the model. The decay length in its rest frame (following 
from Eq.~\ref{eq:decay}) is given by,
\begin{equation*}
L_0=\frac{1.9\times 10^{-2} ~~{\rm cm}}{\big(\frac{\Delta M}{{\rm GeV}}\big)^5 \sin\theta}. 
\end{equation*}

\begin{figure}[ht]
\centering
\includegraphics[height=5.0cm]{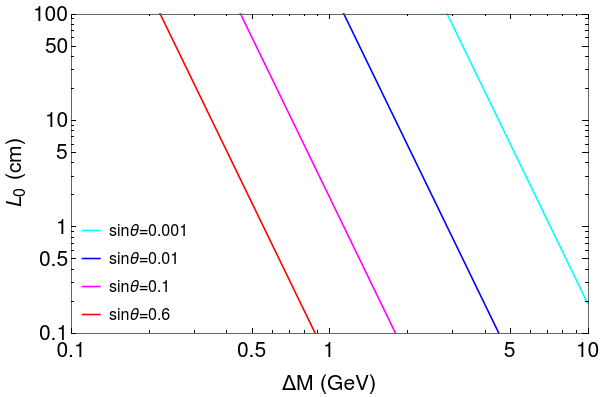}
\caption{\footnotesize{The decay length of $\psi^\pm$ as a function of mass difference $\Delta M$ for fixed $\sin\theta$ values.}}
\label{DV_log}
\end{figure}

In Fig.~\ref{DV_log}, we show the decay length of $\psi^\pm$ as a function of $\Delta M$ for fixed $\sin\theta$ values depicted in different colours. We 
see that for $\Delta M < 10$ GeV,  the displaced vertex of $\psi^\pm$ can be significantly large to be detected at the collider. On the other hand, non-observation 
of a displaced vertex or a charge track will result in a bound on $\Delta M-\sin\theta$ plane.  

$\bullet$ \underline{Effect of $B-L$ gauge extension in $\psi^+ \psi^-$ pair production:}\\
\begin{figure}[ht]
$$
 \includegraphics[height=5.0cm]{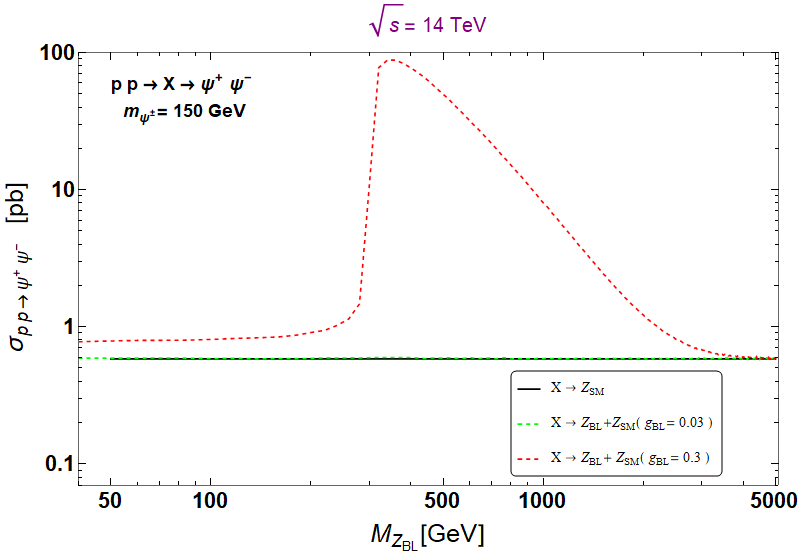} ~
 \includegraphics[height=5.0cm]{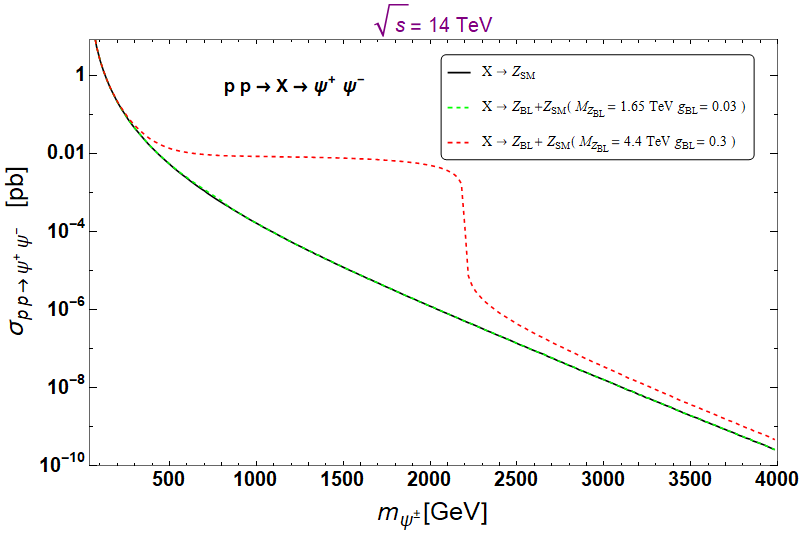}
$$ 
 \caption{\footnotesize{[Left] The production cross-section of $\psi^+ \psi^-$ pairs at collider is shown as a function of $U(1)_{B-L}$ gauge boson mass, 
 $M_{Z_{BL}}$ for fixed $m_{\psi^\pm} = 150$ GeV. Different coloured lines depict different cases: SM production cross-section is shown by black solid line; 
 $U(1)_{B-L}$ case is shown for $g_{BL}=0.03$ (green dashed line) and $g_{BL}=0.3$ (red dashed line). [Right] The production cross-section of $\psi^+ \psi^-$ pairs at collider is shown as a function of $m_{\psi^\pm}$ with $M_{Z_{BL}}=1.65 ~{\rm TeV},  g_{BL}=0.03$ (green dashed line) and $M_{Z_{BL}}=4.4 ~{\rm TeV},  g_{BL}=0.3$ (red dashed line). Pure SM gauge boson mediated production cross-section (Model I) is also shown in black solid line.}}
 \label{coll_prdX}
\end{figure} 

The effect of $U(1)_{B-L}$ gauge boson ($Z_{BL}$) 
mediation in $ p~p \to \psi^+ \psi^-$ production ~\cite{Majee:2010ar} is an important question and we discuss the main features here. We summarise our observations in Fig.~\ref{coll_prdX}. In the left panel of Fig.~\ref{coll_prdX}, we 
have shown the production cross-section of $\psi^+ \psi^-$ pair at LHC as a function of $M_{Z_{BL}}$ for fixed $m_{\psi^\pm} 
= 150$ GeV. On the right panel, we plot the production cross-section of $\psi^+ \psi^-$ as function of $m_{\psi^\pm}$, 
for two different combinations of $Z_{BL}$ parameters: \{$M_{Z_{BL}}=1.65 ~{\rm TeV},  g_{BL}=0.03$\} (green dashed line) and \{$M_{Z_{BL}}=4.4 ~{\rm TeV},  
g_{BL}=0.3$\} (red dashed line), which agree to the current ATLAS bound. Only SM contribution with $Z_{SM}: (Z, \gamma)$ mediation is also shown by black solid line for comparison.   
It is evident from the Fig.~\ref{coll_prdX}, that for the smaller value $g_{BL}=0.03$, the contribution from $Z_{BL}$ mediated production is negligible 
compared to SM and consequently green dashed and black lines fall on top of each other. 
However, with a moderate value of $g_{BL}=0.3$, the production cross-section significantly improves with $Z_{BL}$ mediation, 
which is seen in red dashed line clearly separated from the other two. In the left plot we see that the effect of s-channel resonance in amplitude 
$\sim \frac{1}{{\hat{s}}-M^2_{Z_{BL}}}$ showing up at $M_{Z_{BL}}=2m_{\psi^\pm}=300$ GeV as the minimum subprocess center-of-mass energy required 
for this process to occur is $\sqrt{\hat{s}}=300$ GeV with $m_{\psi^\pm}=150$ GeV. The resonance is extended to account its finite decay width $\sim \frac{1}{{\hat{s}}-M^2_{Z_{BL}}+iM_{Z_{BL}}\Gamma_{Z_{BL}}}$. 
The same effect is seen on the right panel plot where the resonance rise is visible at $m_{\psi^\pm}=\frac{M_{Z_{BL}}}{2}\sim$ 2 TeV for the red dashed curve ($g_{BL}=0.3$). 
To summarise, the effect of $Z_{BL}$ mediation for the production of $\psi^\pm$ pair, 
which contributes to opposite sign dilepton (OSD) plus missing energy signal, can only be realised at relatively larger values of 
gauge coupling ($g_{BL}$) and on-shell $Z_{BL}$ production whenever possible, albeit that current experimental bound requires a higher $Z_{BL}$ mass with larger $g_{BL}$ coupling (see Fig.~\ref{Atlas-1}).\\

\newpage

$\bullet$ \underline{Hadronically quiet OSD events at LHC:}\\

We shall now briefly discuss the event level simulation for the OSD signal ($\ell^+ \ell^- +\slashed{E_T}$) and estimate SM background contamination for the same final state. Our elaboration will be more indicative than exhaustive. For that, 
we refer to two different benchmark points with $\Delta M = 15$ GeV and $300$ GeV keeping DM mass fixed at $m_{\chi_3}=150$ GeV; important to note here that the first case applies to the model I without $B-L$ extension 
where the second possibility with larger $\Delta M$ is only allowed in model II with $B-L$ extension (compare Fig.~\ref{directdetection} to Fig.~\ref{Atlas-2}). 
For the analysis we generate the {\tt {lhe}} file from the model implementation in {\tt FeynRule} \cite{Alloul:2013bka} and run it in 
{\tt Madgraph} \cite{Alwall:2011uj} to generate events and finally pass onto {\tt Pythia} \cite{Sjostrand:2006za} for analysis. Following basic techniques are used in {\tt Pythia} to mimic the actual collider environment: \\
$\bullet$ {\bf{Lepton isolation}}: To identify a lepton $(\ell=e,\mu)$ in the detector, one requires a minimum transverse momentum, which we keep as $p_T > 20$ GeV. We also require the pseudorapidity within $|\eta| < 2.5$, which ensures that 
leptons ejected centrally can only be observed in the detector. Separation of leptons from each other requires
$(\Delta R)_{\ell\ell} \geq 0.2$ in $\eta-\phi$ plane (where $\Delta R= \sqrt{(\Delta \eta)^2+(\Delta \phi)^2}$). We further imposed $(\Delta R)_{\ell j} \geq 0.4$ to separate leptons from jets.\\
 $\bullet$ {\bf{Jet identification}}: Defining a jet ($j$) is an important issue at LHC environment. In the numerical simulation performed here, jets are formed in {\tt Pythia} using cone algorithm inbuilt in {\tt PYCELL}. 
 A jet is then identified with all parton within a cone of $\Delta R \leq 0.4$ around a jet initiator with $p_T > 20$ GeV. We will finally require zero jet veto to ensure hadronically quiet final state.\\
 $\bullet$ {\bf{Unclustered objects}}:  The unclustered objects consist of those objets, which neither qualify as jets nor identified as isolated leptons (following our previous definitions) and only contribute to 
 missing energy. All final state objects with smaller transverse momentum $0.5<p_T <20$ GeV and larger pseudorapidity $2.5< |\eta| < 5$ are therefore identified as unclustered objects. 

Three kinematic variables play a key role in the analysis: Missing Energy ($\slashed{E_T}$), Transverse Mass ($H_T$) and Invariant mass ($m_{\ell\ell}$); where the signal and background show different sensitivity. 
Missing energy has already been defined (Eq.~\ref{eq:met}), the other two are: 
\begin{itemize}
 \item  {\underline{\it Transverse Mass ($H_T$)}:} Transverse mass of an event is identified to:
 \bea
 {H_T} = \sum_{\ell,j} \sqrt{(p_x)^2+ (p_y)^2}= \sum_{\ell,j} p_T,
 \eea
 where the {\it scalar sum} of transverse momentum runs over reconstructed objects like leptons $(\ell)$ and jets ($j$).
 
 \item {\underline{\it Invariant mass ($m_{\ell\ell}$)}:}  Invariant mass of opposite sign dilepton is defined by
  \bea
 {m_{\ell^+\ell^-}} = \sqrt{ (\sum_{\ell^+\ell^-} p_x)^2+ (\sum_{\ell^+\ell^-} p_y)^2+(\sum_{\ell^+\ell^-} p_z)^2}.
 \eea
\end{itemize}

The normalised event distribution for OSD signal events $\ell^+\ell^{-} +({\slashed E}_T)$ at the two benchmark points 
with dominant SM background events are shown in Fig.~\ref{fig:sigbkgdist} with missing energy ($\slashed{E}_T$) in the left panel and transverse mass ($H_T$) on the right panel. 
In both graphs, we note that the peak for $\Delta M=15$ GeV appear on the left side of SM background, while the one for $\Delta M = 300$ GeV is flatter and shifted towards high $\slashed{E}_T/H_T$ value. 
It is then quite apparent, that segregating these two signals from SM background requires different selection cuts on $\slashed{E}_T$, $H_T$ and $m_{\ell\ell}$, which are chosen as follows:

\begin{figure}[htb!]
$$
 \includegraphics[height=4.5cm]{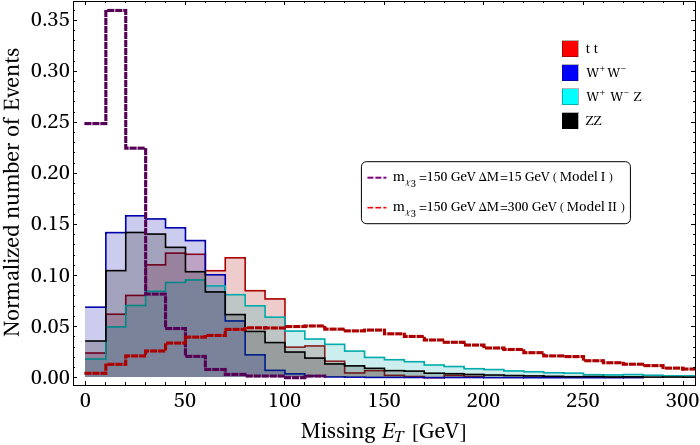}~
 \includegraphics[height=4.5cm]{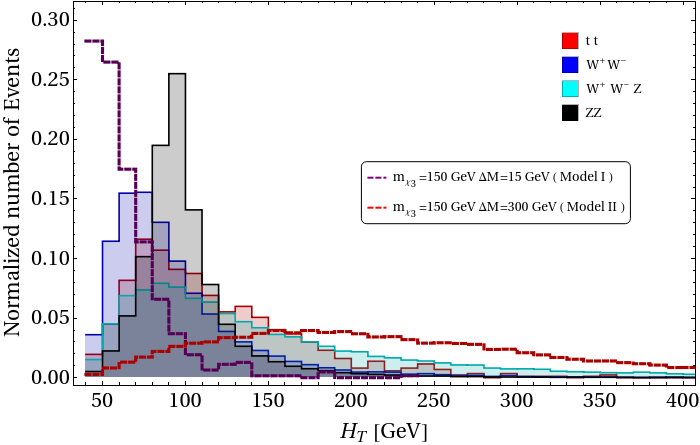}
 $$
  \caption{\footnotesize{ Distribution of missing energy ($\slashed{E}_T$) and transverse mass ($H_T$) for signal events and 
 dominant SM background events at LHC with $\sqrt{s} = 14 \rm ~TeV~$. }}
 \label{fig:sigbkgdist}
\end{figure}

\begin{itemize}
 \item \underline{Invariant mass ($m_{\ell\ell}$) cut}: 
 $m_{\ell\ell} < (m_Z -15)$ GeV and $m_{\ell\ell} > (m_Z +15)$ GeV is imposed to get rid of SM $Z$ boson contribution to OSD final state.

 \item \underline{ $\slashed{E}_T$ and $H_{T}$ cuts}: 
 \begin{itemize}
  \item $\slashed{E}_T < {30}$ GeV, $H_T < {70}$ for  $\Delta M = 15 {\rm GeV} < m_{W^\pm}$.
  \item $\slashed{E}_T > {100}$ GeV, $H_T > {150}$ for  $\Delta M = 300 {\rm GeV} > m_{W^\pm}$ .
 \end{itemize}
\end{itemize}

\begin{table}[htb!]
\begin{center}
\resizebox{\textwidth}{!}{
\begin{tabular}{|c|c|c|c|c|c|c|c|c|c|}
\hline
Model& $m_{\chi_3}$ (GeV)& $\Delta M$ (GeV) & $\sigma^{\ell^+\ell^-X}$ (fb) & $\slashed{E}_T$ (GeV)  & $H_T$ (GeV) & $\sigma^{\ell^+\ell^-X}_{\text{eff}}$(fb) & $N^{\ell^+\ell^-X}_{\text{eff}} (@  \mathcal{L} =10^2 \textrm{fb}^{-1})$ \\
\hline\hline 
\textrm{Model I}& 150 & 15 & 392.37 &  $< 30$  &  $< 70 $ & $ 1.48 $ & $148$\\
\hline 
$\textrm{Model II}$&150 &300 & $9.48$  & $>$100 & $>$150  & $ 1.83$ &  $ 183$ \\
\hline
\end{tabular}
}
\end{center}
\caption {\footnotesize{Signal ($\ell^+\ell^{-} +({\slashed E}_T)$) cross-section for the chosen benchmark points for $\sqrt{s}$ = 14 TeV at LHC with luminosity $\mathcal{L} = 100~fb^{-1}$ 
in Model I (without $B-L$) and Model II (with $B-L$) after the selection cuts employed.}} 
\label{tab:signal1}
\end{table}

\begin{table}[htb!]
\begin{center}
\resizebox{\textwidth}{!}{
\begin{tabular}{|c|c|c|c|c|c|c|c|}
\hline
SM Bkg. & $\sigma^{\ell^+\ell^-X}$ (fb) & $\slashed{E}_T$ (GeV)  & $H_T$ (GeV) & $\sigma^{\ell^+\ell^-X}_{\text{eff}}$(fb) & $N^{\ell^+\ell^-X}_{\text{eff}} (@  \mathcal{L} =10^2 \textrm{fb}^{-1}) $ \\
\hline\hline 
& &  $< 30$  &  $< 70 $ & 6.23 & 623\\  
\cline{3-6}
$t~\bar {t}$ & $36.69 \times 10^3$& $>$100 & $>$150  & 10.64 & 1064 \\
\hline 
& &  $< 30$  &  $< 70 $ & 131.18 & 13118 \\  
\cline{3-6}
$W^+~W^-$ & $4.74 \times 10^3$& $>$100 & $>$150  &  7.72 &  772 \\
\cline{3-6}
\hline 
& &  $< 30$  &  $< 70 $ & 0.53 & 53 \\  
\cline{3-6}
$Z~Z$ & $0.25 \times 10^3$& $>$100 & $>$150  &0.18  &  18 \\
\cline{3-6}
\hline
& &  $< 30$  &  $< 70 $ & 0.01 & 1 \\  
\cline{3-6}
$W^+~W^- Z$ & $1.00$& $>$100 & $>$150  & 0.06 & 6  \\
\cline{3-6}
\hline 
\hline
\end{tabular}
}
\end{center}
 \caption {\footnotesize{Dominant SM background contribution to $\ell^+\ell^{-} +({\slashed E}_T)$ signal events for $\sqrt{s}$ = 14 TeV at LHC for luminosity $\mathcal{L} = 100~fb^{-1}$.The
SM background cross-section are quoted with next-to-leading order (NLO) level with appropriate K-factors~\cite{Alwall_2014}.}} 
\label{tab:background}
\end{table}  

\begin{figure}[htb!]
 $$
 \includegraphics[height=5.0cm]{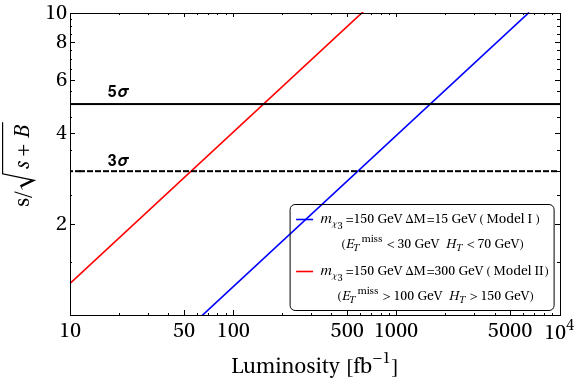}
 $$
 \caption{\footnotesize{Signal significance $\sigma=\frac{S}{\sqrt{S+B}}$ of the benchmark points characteristic to model I (in blue) and model II (in red) at LHC with $\sqrt s=14$ TeV, in terms of luminosity (fb$^{-1}$), subject to the 
 selection criteria imposed in this analysis. 3$\sigma$ and 5$\sigma$ reach }}
 \label{fig:significance1}
\end{figure}

After imposing above cut-flow we list the signal and dominant SM background events in Table
\ref{tab:signal1} and Table \ref{tab:background} respectively for luminosity 100 fb$^{-1}$. We see that $W^+W^-$ production provides the most significant background for OSD at LHC, which couldn't be tamed by the cuts used.
This is surely the key reason for not being able to observe any signal excess over the huge SM background at LHC. The numbers of signal and SM background events thus obtained can provide the discovery reach of the signal for two 
benchmark points in terms of significance defined as $\sigma=\frac{S}{\sqrt{S+B}}$, where $S$ denotes signal events and $B$ denotes 
SM background events, shown as a function of luminosity in Fig.~\ref{fig:significance1}. It shows that $5\sigma$ discovery reach is difficult to achieve for the model I without $U(1)_{B-L}$ characterised by 
low $\Delta M$ ($\mathcal{L}\sim 1500 ~\rm {fb}^{-1}$), while the case with large $\Delta M$ in model II with $U(1)_{B-L}$ extension can be probed in near future with $\mathcal{L}\sim 150 ~\rm {fb}^{-1}$.

\section{Non-zero masses and mixing of light neutrinos}
\label{neutrino}
The very construct of this model is motivated by the fact that we wish to have phenomenologically viable WIMP like DM 
and non-zero masses and mixing of light neutrinos in a minimal extension of the SM. This could be achieved by the 
presence of three RH neutrinos. While, one of them constitute the dark sector being odd under a stabilising $\mathcal{Z}_2$ 
symmetry, the other two can contribute to neutrino sector. In this model, a tiny yet non-zero neutrino mass can be generated via 
Type I seesaw from the following terms in the Lagrangian~\ref{model_Lagrangian},

\begin{equation}
   - \mathcal{L}^{\nu}_{mass} \supset \Big(Y_{j \alpha }\overline{N_{R_j}} \Tilde{H^\dagger} L_{\alpha} + h.c.\Big) + \Big( \frac{1}{2} M_{R_j}\overline{N}_{R_j}(N_{R_j})^c + h.c.\Big)~;
\end{equation}
where $\alpha = e,\mu, \tau$ and $j = 2,3$. After EW symmetry breaking, the SM Higgs acquires a vev to generate the Dirac mass 
terms for the neutrinos. In the gauged B-L scenario, the mass of all three right handed neutrinos are generated through the vev 
of the scalar $\Phi_{BL}$. So for simplicity we can consider the mass of two $\mathcal{Z}_{2}$ even right handed neutrinos that take part 
in the seesaw to be quasi-degenerate and of the same mass scale as that of the $\mathcal{Z}_2$ odd right handed neutrino taking part in 
the dark sector phenomenology. Without loss of generality, we assume the heavy Majorana mass matrix that take part part in 
seesaw to be diagonal, {\it i.e.,} $M_{R} = Diag(0, M_{R_2}, M_{R_3})$. In this basis, the light neutrino mass matrix obtained 
through Type-I seesaw is given as,
\begin{equation}
m_\nu  = - m_D M^{-1}_R m^T_D
\end{equation}
\noindent which is a complex $3\times 3$ matrix and can be diagonalized by the PMNS matrix~\cite{Valle:2006vb} as,
\begin{equation}
(m_\nu)^{{\it diag}}= U^T~ m_\nu ~ U
\end{equation}
where $(m_\nu)^{{\it diag}}=Diag(m_1, m_2, m_3)$ contains at least one zero eigenvalue.

\noindent The PMNS matrix U is given by:
\begin{equation}
U=\left(
	\begin{array}{ccc}
	c_{12}c_{13}                                & s_{12}c_{13}                                & s_{13}e^{-i\delta} \\
	-s_{12}c_{23}-c_{12}s_{23}s_{13}e^{i\delta} & c_{12}c_{23}-s_{12}s_{23}s_{13}e^{i\delta}  & s_{23}c_{13}       \\
	s_{12}s_{23}-c_{12}c_{23}s_{13}e^{i\delta}  & -c_{12}s_{23}-s_{12}c_{23}s_{13}e^{i\delta} & c_{23}c_{13}       \\
	\end{array}
	\right) U_{ph}
	\label{UPMNS}   
\end{equation}
where  $c_{ij}$ and $s_{ij}$ stand for $\cos\theta_{ij}$ and $\sin\theta_{ij}$ respectively and $U_{ph}$ is given by:

\begin{equation}
U_{ph} = Diag(1, e^{-i\alpha/2}, 1) 
\label{phase_matrix}      
\end{equation}
\noindent  where  $\alpha$ is the CP-violating Majorana phase.

Using Casas-Ibarra parameterization~\cite{Casas:2001sr}, the Dirac mass matrix $m_D$ can be parametrized as,
\begin{equation}
({m_D})_{j \alpha }=\sqrt{M_{R_j}}R_{ji}\sqrt{m_i}U^\dagger_{i \alpha  }
\label{eq_y1}
\end{equation}

where $m_i$ are the eigenvalues of the light neutrino mass matrix $m_\nu$ and R is in general a $3\times3$ complex orthogonal matrix. Since in our case, $N_{R_1}$ is decoupled from the spectrum, the corresponding Yukawa coupling $Y_{1 \alpha }$ for a particluar flavour $\alpha$ in the Dirac mass matrix given by Eqn.~\ref{eq_y1} is zero, {\it i.e.},
\begin{equation}
\begin{aligned}
Y_{1 \alpha }
            &=\frac{1}{v}(\sqrt{M_{R_1}}R_{1i}\sqrt{m_i} U^\dagger_{i \alpha} ) \\
            &=\frac{1}{v}(\sqrt{M_{R_1}}R_{11}\sqrt{m_1} U^\dagger_{1 \alpha}+\sqrt{M_{R_1}}R_{12}\sqrt{m_2} U^\dagger_{2 \alpha}+\sqrt{M_{R_1}}R_{13}\sqrt{m_3} U^\dagger_{3 \alpha} )=0
\end{aligned}
\label{eq:y11}
\end{equation}

At present, the oscillation experiments measure
two mass square differences: namely solar ($\Delta m^2_{\odot }$) and atmospheric ($\Delta m^2_{atm}$) along with three mixing angles
$\theta_{23}$, $\theta_{12}$ and $\theta_{13}$. Data indicates that $|\Delta m^2_{atm}|>>\Delta m^2_{\odot }$, but depending on the sign of $\Delta m^2_{atm}$, two cases can arise.\\

\noindent $\bullet$\textbf{ Normal Hierarchy (NH):} 
		\begin{equation}\label{eq:neu:NH1}
~~ 	
		\left\{
		\begin{array}{l}
		
		m_1=0\\
		m_2=\sqrt{\Delta m^2_\odot}\ll m_3 = \sqrt{\Delta m^2_\mathrm{atm}}
		\end{array}
		\right.
		\end{equation}

In Normal Hierarchy(NH), the lightest mass eigenstate $m_1=0$. So, in order for LHS of Eqn~\ref{eq:y11} {\it i.e.,} $Y_{1 \alpha}$ to vanish, $R_{12}$ and $R_{13}$ must be zero, since $m_2$ and $m_3$ are non zero. The orthogonality of R then implies that $R_{11} = 1$ and $R_{21} = 0 = R_{31}$. The four remaining elements of R {\it viz.}, $R_{22}$, $R_{23}$, $R_{32}$ and $R_{33}$, form a $2\times2$ complex orthogonal matrix, defined by one complex angle z~\cite{Narendra:2020hoz}. Thus the structure of R matrix in case of NH is reduced to the simple form:
\begin{equation}
R=\left(
	\begin{array}{ccc}
	1       & 0           & 0       \\
    0       & \cos z      & -\sin z   \\
	0       & \sin z     & \cos z    \\
	\end{array}
	\right) 
\end{equation}

\noindent The neutrino Dirac mass matrix obtained has the form :
\begin{equation}
m_D=v\left(
	\begin{array}{ccc}
	0      &  0     & 0  \\
    Y_{2e }        & Y_{2 \mu}      & Y_{2 \tau}   \\
	Y_{3 e}       & Y_{3 \mu}     &  Y_{3 \tau}    \\
	\end{array}
	\right) 
	\label{neu_dirac}
\end{equation}

\noindent where each element $Y_{\alpha j}$ of $m_D$ is given by Eqn.~\ref{eq_y1}.\\


\noindent $\bullet$\textbf{ Inverted Hierarchy (IH):} 
		\begin{equation}\label{eq:neu:NH2}
		~~
		\left\{
		\begin{array}{l}
		\
		
		m_3=0\\
		
		m_1=\sqrt{\Delta m^2_\mathrm{atm}} \,,~~m_2 = \sqrt{\Delta m^2_\mathrm{atm}+ \Delta m^2_\odot}\\

		\end{array}
		\right.
		\end{equation}
In the case of Inverted Hierarchy(IH), we need to set $m_3 = 0$. So in order for LHS of Eqn.~\ref{eq:y11} {\it i.e.,} $Y_{1 \alpha}$ to vanish, $R_{11}$ and $R_{12}$ must be zero. Again, orthogonality of R demands $R_{13} = 1$  making the first row and the third column of R trivial. The four remaining elements of R {\it viz.}, $R_{21}$, $R_{22}$, $R_{31}$ and $R_{32}$ then form a $2\times2$ complex orthogonal matrix, defined by one complex angle z. Thus the structure of R matrix in case of IH is given by:
\begin{equation}
R=\left(
	\begin{array}{ccc}
	0       & 0          & 1       \\
    \cos z        &-\sin z      & 0   \\
	\sin z        & \cos z      & 0    \\
	\end{array}
	\right) 
\end{equation}

Again we a get a Dirac mass matrix of the same structure as that of NH case, with each element given by Eqn.~\ref{eq_y1}. 

Now, we turn to comment on the charged lepton flavour violation under this parametrization. In particular we study the  the process  $\mu  \to e \gamma$. The branching ratio of this process is given by~\cite{Alonso:2012ji,Marcano:2017ucg,Ilakovac:1994kj,Deppisch:2004fa,Ilakovac:1999md},

\begin{equation}
Br(\mu \to e \gamma)=\frac{\alpha^3_w s^2_w}{256 \pi^2}\frac{m^4_\mu}{M^4_W}\frac{m_\mu}{\Gamma_\mu}|G^{\mu e}_\gamma|^2
\label{br}
\end{equation}

where $\alpha_w$ is the weak coupling strength, $s_w$ is the sin of Weinberg's angle, $m_\mu$ is the muon mass, $M_W$ is the mass of W boson and $\Gamma_\mu \approx 2.996 \times 10^{-19}$GeV denotes the total decay width of muon. The factor $G^{\mu e}_\gamma$ si given by,

\begin{equation}
G^{\mu e}_\gamma=\sum_{i}U_{e i}U^*_{\mu i}G_\gamma(x_i)=\sum_{j}U_{e N_j}U^*_{\mu N_j}G_\gamma(x_{N_j})
\end{equation}

where, $x_i=\frac{m^2_{\nu_i}}{M^2_W}$ and $x_{N_j}=\frac{M^2_{N_j}}{M^2_W}$.
where i (j) runs over total number of light (heavy) physical neutrino states. $m_\nu (M_N)$ denotes the mass of light (heavy) physical neutrinos and $U_{e i} (U_{e N_i})$ represents the mixing matrix elements of light (heavy) neutrinos. The loop integration factor $G_\gamma(x)$ is given by,

\begin{equation}
G_\gamma(x)=-\frac{x(2x^2+5x-1)}{4(1-x^3)}-\frac{2 x^3}{2(1-x^4)}ln(x)
\end{equation}

\begin{figure}[h]
$$
 \includegraphics[height=6.5cm]{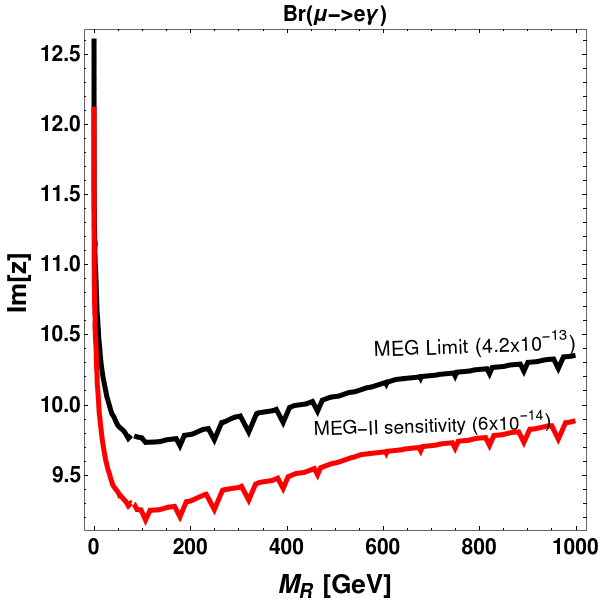} ~
 \includegraphics[height=5.5cm]{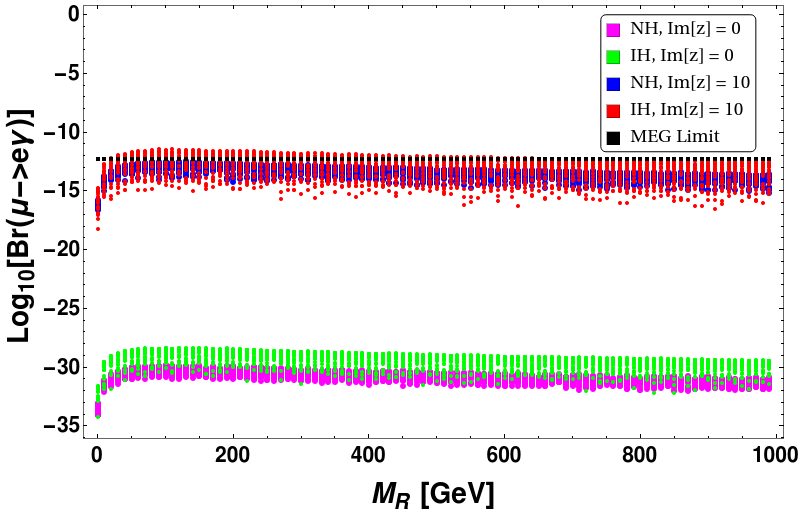}
$$ 
 \caption{\footnotesize{[Left]: $Br(\mu \to e \gamma)$ in $M_R-Im[z]$ plane; [Right]: $Log[Br(\mu \to e \gamma)]$ for $Im[z]=0,10$ for both NH and IH. The black dashed line represents the MEG limit.}}
 \label{CLFV}
\end{figure}

To study the dependence of this branching ratio on the right handed mass scale in light of Casas-Ibarra parameterization, we derive from \ref{br} the following equation,

\begin{equation}
Br(\mu \to e \gamma)=\frac{\alpha^3_w s^2_w}{256 \pi^2}\frac{m^4_\mu}{M^4_W}\frac{m_\mu}{\Gamma_\mu}\frac{4}{M^4_R}G^2_\gamma(x_N)|(m^\dagger_D m_D)_{e \mu}|^2
\label{brfinal}
\end{equation} 

where $M_R$ denotes the mass of the right handed neutrino states. For simplicity we assume the two right handed neutrinos to be degenerate and $M_N = M_R$. The matrix element $(m^\dagger_D m_D)_{e \mu}$ for NH and IH respectively can be written using Eqn.~\ref{eq_y1} as,

\begin{equation}
\begin{aligned}
(m^\dagger_D m_D)_{e\mu}\Big\vert_{NH} &= M_R \big[(m_2 U_{e2} U^*_{\mu2} + m_3 U_{e3} U^*_{\mu3}) \cosh(2 Im[z]) 
\\&+ i  \sqrt{m_2} \sqrt{m_3}(U_{e3} U^*_{\mu2} - U_{e2}U^*_{\mu3}) \sinh(2 Im[z])\big]
\end{aligned}
\label{m_NH}
\end{equation}

\begin{equation}
\begin{aligned}
(m^\dagger_D m_D)_{e\mu}\Big\vert_{IH} &= M_R \big[(m_1 U_{e1} U^*_{\mu1} + m_2 U_{e2} U^*_{\mu2}) \cosh(2 Im[z]) 
\\&+ i  \sqrt{m_1} \sqrt{m_2}(U_{e2} U^*_{\mu1} - U_{e1}U^*_{\mu2}) \sinh(2 Im[z])\big]
\end{aligned}
\label{m_IH}
\end{equation}	

where $U_{\alpha i}$ are the PMNS matrix elements parametrized as in Eqn~\ref{UPMNS}. In Eqn.~\ref{m_NH} and \ref{m_IH}, there are three free parameters namely $M_R$, $Im[z]$ and $\alpha$  all other quantites being measured by oscillation experiments within a range. In left panel of Fig.~\ref{CLFV}, we have shown the $Br(\mu \to e \gamma)$ as a function of heavy neutrino mass $M_R$ and $Im[z]$ taking all oscillation parameters within their $3\sigma$ range as given in~\cite{Capozzi:2018ubv,Tanabashi:2018oca} in case of NH. The Majorana phase $\alpha$ is varied between 0 to $2\pi$. and the amplitude $|(m^\dagger_D m_D)_{e\mu}|$ is almost independent of phase $\alpha$. We confronted our result with current most stringent bound from MEG experiment $Br(\mu \to e \gamma)\leq 4.2 \times 10^{-13}$~\cite{Mori:2016vwi},  represented by the contour in black colour in left panel of Fig.~\ref{CLFV}. The red contour shows the projected MEG-II sensitivity of $Br(\mu \to e\gamma)\sim 6\times10^{-14}$. The region above the black contour is ruled out by MEG experiment while the region below this contour provides us a wide allowed parameter space for $Br(\mu\to e \gamma)$ in the $M_R-Im[z]$ plane simultaneously satisfying MEG limit and low scale neutrino phenomenology. Similar result has been obtained for IH as well. In the right panel of Fig.~\ref{CLFV}, we have shown $log[Br(\mu \to e \gamma)]$ for two particular values of $Im[z]$, $Im[z]=0$ and 10. 

In the simplest scenario $Im[z] = 0$, the branching ratio is very very less than the current sensitivity of worlds leading experiments like MEG for both NH and IH. 
For $Im[z]=10$, the branching ratio is near to the current sensitivity. For intermediate values of $Im[z]$, $Br(\mu \to e \gamma)$ 
is below the current bound by MEG experiment, while for $Im[z]>10$, $Br(\mu \to e\gamma)$ is above the MEG limit for almost all mass range of 
$M_R$ upto 1000 GeV. As it can also be seen from the left panel of Fig.~\ref{CLFV} that only for $M_R\leq 10$ GeV, $Im[z]$ can take values upto 14. 
Naturalness and vacuum stability bounds can also be applied in principle as done in~\cite{Bambhaniya:2016rbb}, but these bounds are extremely 
weaker for $M_R$ upto TeV scale. 


\section{Conclusion}
\label{conclusion}

In this paper, we have studied a minimal extension of the SM by adding a vector like fermion doublet $\Psi$ and three singlet right handed neutrinos 
$N_{R_i}$ to simultaneously address non-zero masses and mixing of light neutrinos as well as a phenomenologically viable dark matter component of the universe. 
An additional $\mathcal{Z}_2$ symmetry is required on top of the SM gauge symmetry to ensure the stability of the DM. Now, 
the $\mathcal{Z}_2$ symmetry crucially distinguishes the added fermions; for example, the vector-like fermion doublet and one of the three right handed 
neutrinos are assumed odd, while the rest are even. As a result the dark matter emerges as the lightest Majorana fermion from the mixture of the neutral component of 
the doublet $\Psi$ and the singlet, which is odd under the same $\mathcal{Z}_2$. The other two right handed neutrinos being even under the 
$\mathcal{Z}_2$ symmetry couple to SM Higgs and generate non-zero masses for light neutrinos via type-I seesaw. The absence of either the 
doublet or the singlet (odd under $\mathcal{Z}_2$), make the DM absurdly constrained from relic density and direct search prospects. Therefore, 
one can simply see that the model under study is possibly the most economical one to simultaneously address neutrino mass and a phenomenologically 
viable DM candidate of the universe. 

We studied the allowed parameter space of the model taking into account all annihilation and co-annihilation channels for 
DM mass ranging from 1 GeV to 1 TeV. The allowed parameter space is shown in the $\Delta M \sim m_{\chi_{_3}}$ plane, 
where $m_{\chi_{_3}}$ is the mass of the dark matter and $\Delta M$ is its mass difference with next to lightest dark 
sector particle. We confronted our results with recent data from both PLANCK and XENON-1T to obtain the correct parameter 
space satisfying both relic density and direct detection constraints. Since the DM is Majorana in nature, it escapes from 
the strong $Z$-mediated direct detection constraint. As a result we end up with relatively large singlet-doublet mixing. 
In particular, for DM mass of 1TeV, the allowed singlet-doublet mixing can be as large as $\sin \theta \sim 0.6$. This crucially 
distinguishes the Majorana singlet-doublet DM from a vector like singlet-doublet DM. This feature also hasn't been highlighted 
in earlier analysis of a similar model framework.   


Since with three right handed neutrinos, the model is qualified for a anomaly-free gauged B-L extension, we studied how 
our results change in light of $U(1)_{B-L}$ gauge extension. Clearly, the model requires an additional complex scalar singlet to break the gauge group 
and the massive gauge boson $Z_{BL}$ further enhances the DM-SM coupling. The relic density allowed parameter space additionally 
enhances due to $Z_{BL}$ resonance in regions where $m_{DM}<M_{Z_{BL}}$. Also, the scalar sector mixes the SM doublet and 
additional singlet to produce two neutral scalar fields to mediate DM-SM interactions and enhance direct search possibility. The 
constraint on $g_{BL}-M_{Z_{BL}}$ from current LHC data is significant enough to ensure the coupling to be minuscule for relatively 
smaller $M_{Z_{BL}} \sim {\rm TeV}$, so that the DM signal at LHC doesn't have any additional contribution from $Z_{BL}$ mediation to 
$\psi^\pm$ pair production. However with larger $M_{Z_{BL}}\sim 4 ~{\rm TeV}$, the coupling ($g_{BL}$) can be large enough to show up additional 
signal strength at LHC, which can be probed in its high luminosity run.  
It is worthy to mention that both the models offer variety of leptonic signatures 
like hadronically quiet opposite sign dilepton (OSD), trilepton or four lepton in association with missing energy. In a toy simulation for OSD events at LHC, we showed that 
it is easier to probe large $\Delta M$ regions of the model, characteristic to the $U(1)_{B-L}$ scenario than the small $\Delta M$ regions characteristic to the framework without 
$U(1)_{B-L}$. The model may also offer displaced vertex or stable charge track whenever the mass splitting $\Delta M$ between the charged companion and DM becomes very small.

Neutrino mass generation although fused naturally in this model, do not have direct influence on the dark sector. However, the RH neutrino 
mass turns out crucial for the neutrino sector, constrained from flavour changing decays like ($\mu \to e \gamma$). On the other hand, in the small mixing scenario, 
DM mass is dominantly controlled by the RH neutrino odd under $\mathcal{Z}_2$ symmetry {\it i.e.} $m_{DM}\sim M_{R_1}$. 
Since in the context of $U(1)_{B-L}$ model, the Majorana masses of all the three RH neutrinos (including the one in the dark sector) 
are generated uniformly from the same symmetry breaking scale, we can treat them as a common parameter of the framework constrained by 
both dark sector and neutrino sector as a bridging ligand of the model.

\paragraph*{Acknowledgments\,:} 
MD acknowledges Department of Science and Technology (DST), Govt. of India for providing the financial assistance for the research under the grant 
DST/INSPIRE/03/ 2017/000032. MD also acknowledges Anirban Karan for his help in learning some computational techniques. PG would like to acknowledge the support from DAE, India for the Regional Centre for Accelerator based Particle Physics(RECAPP), Harish Chandra Research Institute. SB would like to acknowledge to DST-SERB 
grant CRG/2019/004078 at IIT Guwahati.
\appendix
\section*{Appendix}
\vspace{0.2cm}
\section{DM-SM Interaction in model I}
\label{dm_sm_int}

Expanding the covariant derivative of the Lagrangian given by Eq.\ref{model_Lagrangian}, we get the interaction term of $\psi^0$ and $\psi^\pm$ with the SM gauge bosons as follows:
\begin{equation}
\begin{aligned}
 \mathcal{L}_{int} &=\overline{\Psi}i\gamma^\mu(-i\frac{g}{2}\tau.W_\mu - ig'\frac{Y}{2}B_\mu )\Psi
 \\ 
   &= \Big(\frac{e}{2\sin\theta_W \cos\theta_W}\Big)\overline{\psi^0}\gamma^\mu Z_\mu \psi^0
   \\
   & +\frac{e}{\sqrt{2}\sin\theta_W}(\overline{\psi^0}\gamma^\mu W^+_\mu \psi^- + \psi^+\gamma^\mu W^-_\mu \psi^0) \\
   & - e \psi^+\gamma^\mu A_\mu \psi^-\\ 
    & -\Big(\frac{e \cos2\theta_W}{2\sin\theta_W \cos\theta_W}\Big)~ \psi^+\gamma^\mu Z_\mu \psi^- .
\end{aligned}
\end{equation}
where $g = \frac{e}{\sin\theta_W}$ and $g'= \frac{e}{\cos\theta_W}$ with $e$ being the electromagnetic coupling constant and $\theta_W$ being the Weinberg angle.\\

These interactions, when written in terms of the physical states becomes:
\begin{equation}
\label{dmint}
\begin{aligned}
\mathcal{L}_{int} &=\Big(\frac{e}{2\sin\theta_W \cos\theta_W}\Big)(-\cos\theta\overline{\chi_{_{1L}}}i\gamma^\mu Z_{\mu}\chi_{_{2L}}-\sin\theta\overline{\chi_{_{2L}}}i\gamma^\mu Z_{\mu}\chi_{_{3L}} + h.c.)\\
 & +\frac{e}{\sqrt{2}\sin\theta_W}(\cos\theta \overline{\chi_{_1}}\gamma^\mu W^+_\mu \psi^-+ \overline{\chi_{_2}}i\gamma^\mu W^+_\mu \psi^- -\sin\theta \overline{\chi_{_3}}\gamma^\mu W^+_\mu \psi^-) \\
 & +\frac{e}{\sqrt{2}\sin\theta_W}(\cos\theta \psi^+\gamma^\mu W^-_\mu \chi_{_1}-\psi^+i\gamma^\mu W^-_\mu \chi_{_2}-\sin\theta\psi^+ \gamma^\mu W^-_\mu\chi_{_3})\\
 & - e ~\psi^+\gamma^\mu A_\mu \psi^-\\ 
 & -(\frac{e\cos2\theta_W}{2\sin\theta_W \cos\theta_W})~ \psi^+\gamma^\mu Z_\mu \psi^-.\\
\end{aligned}
\end{equation}\\

Another possibility of interaction between DM sector and the visible sector arises from the Yukawa interaction term $\frac{Y_1}{\sqrt{2}}\overline{\Psi}\Tilde{H}(N_{R_1}+(N_{R_1})^c)$ and its hermitian conjugate by expanding the SM Higgs $H$ around its vev.
 Writing in terms of physical bases,\begin{equation}
\label{DMHiggs}
\begin{aligned}
-\mathcal{L}_{DM-Higgs} ={} & \frac{Y_1}{\sqrt{2}}\Big[\sin 2\theta(\overline{\chi_{_1}}h\chi_{_1} - \overline{\chi_{_3}}h\chi_{_3}) + \cos2\theta (\overline{\chi_{_1}}h\chi_{_3} + \overline{\chi_{_3}}h\chi_{_1})\Big].\\
\end{aligned}
\end{equation}Additionally, dark sector particles can annihilate into $\mathcal{Z}_2$ even right handed neutrino $N_{R_2/3}$ and SM neutrinos via the Yukawa term $\left(Y_{j \alpha }\overline{N_{R_j}} \Tilde{H^\dagger} L_{\alpha} + h.c.\right)$ present in Eqn.~\ref{eq:yukawa}. As it has been stated already, the lightest stable particle $\chi_{_3}$ serves as the DM. The relic abundance of $\chi_{_3}$ can be obtained through its annihilations to as well as through coannihilations with $\chi_{_1}$, $\chi_{_2}$ and $\psi^\pm$ to SM particles. The main processes which contribute to the relic abundance of DM are noted below:\\
\begin{shaded}
\noindent $\overline{\chi_{_1}}\chi_{_1} \rightarrow hh, W^+W^-, ZZ, f\Bar{f}, N_{R_{2/3}}\Bar{\nu}_{e/\mu/\tau}$\\
$\overline{\chi_{_1}}\chi_{_2} \rightarrow hh, Zh, W^+W^-, ZZ, f \Bar{f}$\\
$\overline{\chi_{_1}}\chi_{_3} \rightarrow hh, W^+W^-, ZZ, f \Bar{f}, N_{R_{2/3}}\Bar{\nu}_{e/\mu/\tau}$\\
$\overline{\chi_{_2}}\chi_{_1} \rightarrow hh, Zh, W^+W^-, ZZ, f \Bar{f}$\\
$\overline{\chi_{_2}}\chi_{_3} \rightarrow hh, Zh, W^+W^-, ZZ, f \Bar{f}$\\
$\overline{\chi_{_3}}\chi_{_1} \rightarrow hh, W^+W^-, ZZ, f \Bar{f}, N_{R_{2/3}}\Bar{\nu}_{e/\mu/\tau}$\\
$\overline{\chi_{_3}}\chi_{_2} \rightarrow hh, Zh, W^+W^-, ZZ, f \Bar{f}$\\
$\overline{\chi_{_3}}\chi_{_3} \rightarrow hh, W^+W^-, ZZ, f \Bar{f}, N_{R_{2/3}}\Bar{\nu}_{e/\mu/\tau}$\\
$\chi_{_1}\psi^\pm \rightarrow W^\pm \gamma, W^\pm h, W^\pm Z, f'f $\\
$\chi_{_2}\psi^\pm \rightarrow W^\pm \gamma, W^\pm h, W^\pm Z, f'f $\\
$\chi_{_3}\psi^\pm \rightarrow W^\pm \gamma, W^\pm h, W^\pm Z, f'f $\\
$\psi^\pm \psi^\mp \rightarrow W^\pm W^\mp, Zh, \gamma Z, \gamma \gamma, ZZ, f \Bar{f}$\\
\end{shaded}
\section{DM-SM Interaction in model II with $U(1)_{B-L}$ extension}
\label{bl_int}
The interaction terms of the dark and visible sector particles in the gauged $U(1)_{B-L}$ scenario can be obtained by expanding the kinetic terms of $\Psi$ and $N_{R_1}$ given in Eq.-\ref{mdl_Lag_BL} as the following,
\begin{equation}
\begin{aligned}
 \mathcal{L}_{int} &= \overline{\Psi}i\gamma^\mu[-i\frac{g}{2}\tau.W_\mu - ig'\frac{Y}{2}B_\mu - i g_{BL}  Y_{BL}(Z_{BL})_\mu]\Psi \\
 &+ \overline{N_{R_1}}i\gamma^\mu(-i g_{BL} Y_{BL}(Z_{BL})_\mu) N_{R_1} 
 \\ 
   &= \Big(\frac{e}{2\sin\theta_W \cos\theta_W}\Big)\overline{\psi^0}\gamma^\mu Z_\mu \psi^0
   \\
   & +\frac{e}{\sqrt{2}\sin\theta_W}(\overline{\psi^0}\gamma^\mu W^+_\mu \psi^- + \psi^+\gamma^\mu W^-_\mu \psi^0) \\
   & - e ~\psi^+\gamma^\mu A_\mu \psi^-\\ 
   & -\Big(\frac{e \cos2\theta_W}{2\sin\theta_W \cos\theta_W}\Big) \psi^+\gamma^\mu Z_\mu \psi^- \\
    & -g_{BL}\Big[\overline{\psi^0}\gamma^\mu (Z_{BL})_\mu \psi^0 + \psi^+ \gamma^\mu (Z_{BL})_\mu \psi^- 
    + \overline{N_{R_1}}\gamma^\mu (Z_{BL})_\mu N_{R_1}\Big].
\end{aligned}
\end{equation}
where $g = \frac{e}{\sin\theta_W}$ and $g'= \frac{e}{\cos\theta_W}$ with $e$ being the electromagnetic coupling constant, $\theta_W$ being the Weinberg angle and $g_{BL}$ is the $U(1)_{B-L}$ coupling constant. The other interaction is through the Yukawa interaction term ${Y_1}\overline{\Psi}\Tilde{H}N_{R_1}$, where we now have to also take into account the mixing between $H$ ans $\Phi_{BL}$. In terms of physical bases $\chi_{_1}, \chi_{_2}$ and $\chi_{_3}$, the interaction terms of 
DM with the SM gauge bosons are given by:

\begin{equation}
\begin{aligned}
\mathcal{L}_{DM-SM}  ={} & \left(\frac{e}{2\sin\theta_W \cos\theta_W}\right)\Big[(2s_{23}s_{13}c_{13})\big(\overline{\chi_{_{3L}}}\gamma^\mu Z_\mu \chi_{_{3L}} - \overline{\chi_{_{1L}}}\gamma^\mu Z_\mu \chi_{_{1L}}\big)\\
      & + \big(c_{23}c_{13}\overline{\chi_{_{1L}}}\gamma^\mu Z_\mu \chi_{_{2L}}  -c2_{13}s_{23}\overline{\chi_{_{1L}}}\gamma^\mu Z_\mu \chi_{_{3L}} - s_{13}c_{23}\overline{\chi_{_{2L}}}\gamma^\mu Z_\mu \chi_{_{3L}} + h.c.\big)\Big] \\      
       & + \frac{e}{\sqrt{2}\sin\theta_W} \Bigg[\frac{1}{\sqrt{2}}\Big((c_{13}- s_{13}s_{23})\overline{\chi_{_{1L}}} + c_{23}\overline{\chi_{_{2L}}} - (s_{13}+s_{23}c_{13})\big) \overline{\chi_{_{3L}}}\Big)\gamma^\mu W^+_\mu \psi^-_L \\
       &+\frac{1}{\sqrt{2}}\Big((c_{13}+ s_{13}s_{23})\overline{\chi_{_{1L}}} - c_{23}\overline{\chi_{_{2L}}} - (s_{13}-s_{23}c_{13}) \overline{\chi_{_{3L}}}\big)\Big)\gamma^\mu W^+_\mu \psi^-_R+ h.c.\Big]\\
      & - e ~\psi^+\gamma^\mu A_\mu \psi^-
    -\Big(\frac{e \cos2\theta_W}{2\sin\theta_W \cos\theta_W}\Big) \psi^+\gamma^\mu Z_\mu \psi^- .
    \end{aligned}
    \label{eq:intBL}
\end{equation}
Additionally we have the interactions of DM with $Z_{BL}$ as follows:
\begin{equation}
\begin{aligned}
\mathcal{L}_{DM-Z_{BL}} &= - g_{BL}\Big[(s_{23}s2_{13}+c^2_{13}c^2_{23})\big(\overline{\chi_{_{3L}}}\gamma^\mu (Z_{BL})_\mu \chi_{_{3L}}\\
&+(s^2_{13}c^2_{23}-s_{23}s2_{13}) \overline{\chi_{_{1L}}}\gamma^\mu (Z_{BL})_\mu \chi_{_{1L}} + s^2_{23}\overline{\chi_{_{2L}}}\gamma^\mu (Z_{BL})_\mu \chi_{_{2L}} \\
      & + (\frac{1}{2}s2_{23}s_{13}+c_{23}c_{13})(\overline{\chi_{_{1L}}}\gamma^\mu (Z_{BL})_\mu \chi_{_{2L}}+h.c) \\
      & +(\frac{1}{2}s2_{13}c^2_{23}-c2_{13}s_{23})(\overline{\chi_{_{1L}}}\gamma^\mu (Z_{BL})_\mu \chi_{_{3L}} + h.c.)\\
      & + ( \frac{1}{2}s2_{23}c_{13}- s_{13}c_{23})\overline{\chi_{_{2L}}}\gamma^\mu (Z_{BL})_\mu \chi_{_{3L}} + h.c.)\Big] \\
 &-g_{BL} \psi^+ \gamma^\mu (Z_{BL})_\mu \psi^-.
\end{aligned}
\label{dm-sm-bl}
\end{equation}

\vspace{0.5cm}

Here, we abbreviated $\sin2\theta_{ij}$ and $\cos2\theta_{ij}$ as $s2_{ij}$ and $c2_{ij}$ respectively. We note that in the limit $\sin\theta_{23}\to 0$ (along with $g_{BL}\to 0$), we get back to the interactions present in~\ref{dmint}.
DM-Scalar interaction also have additional channels from $H$ and $\Phi_{B-L}$ mixing given by,
\begin{equation}
\begin{aligned}
-\mathcal{L}_{DM-Higgs} ={} & \frac{Y_1}{2}(h_1 \cos\beta - h_2 \sin\beta)\Big[\Big((c_{13}-s_{13}s_{23})\overline{\chi_{_{1L}}}+ c_{23}\overline{\chi_{_{2L}}} -(s_{13}+s_{23}c_{13}) \overline{\chi_{_{3L}}}\Big)\\
&\Big( s_{13}c_{23}(\chi_{_{1L}})^c+s_{23}(\chi_{_{2L}})^c+c_{13}c_{23}(\chi_{_{3L}})^c \Big) + h.c.\Big]\\
 & + \frac{y'_1}{2\sqrt{2}}(h_2 \cos\beta + h_1 \sin\beta)\Big[\Big(s_{13}c_{23}\overline{(\chi_{_{1L}})^c}+ s_{23}\overline{(\chi_{_{2L}})^c} + c_{13}c_{23} \overline{(\chi_{_{3L}})^c}\Big)\\
& \Big(s_{13}c_{23}\chi_{_{1L}}+ s_{23}\chi_{_{2L}} + c_{13}c_{23} \chi_{_{3L}}\Big) + h.c.\Big],
\end{aligned}
\end{equation}
where $h_1$, $h_2$ are the two physical scalars of the model and $\beta$ represents $H-\Phi_{B-L}$ mixing angle.
The annihilation channels of dark matter in the $U(1)_{B-L}$ extended case differs from the one without it, by having additional $Z_{BL}$ and an additional scalar present both in mediator as well as in final states. 
The following processes contributes to the relic abundance of the DM particle $\chi_{_3}$ in this model with $U(1)_{B-L}$ extension. \\

\begin{shaded}
\noindent {\footnotesize $\overline{\chi_{_1}}\chi_{_1} \rightarrow h_1h_1, h_2h_2, h_1h_2, W^+W^-, ZZ,Z_{BL} Z_{BL}, Z Z_{BL},  f \Bar{f},  N_{R_{2/3}}N_{R_{2/3}}, N_{R_{2/3}}\Bar{\nu}_{e/\mu/\tau}$\\
$\overline{\chi_{_1}}\chi_{_2}\rightarrow  h_1h_1, h_2h_2,h_1h_2, Z h_1, Z h_2, Z_{BL} h_1, Z_{BL} h_2,  W^+W^-, ZZ,Z_{BL} Z_{BL}, Z Z_{BL}, f \Bar{f}, N_{R_{2/3}}N_{R_{2/3}}, N_{R_{2/3}}\Bar{\nu}_{e/\mu/\tau}$\\
$\overline{\chi_{_1}}\chi_{_3}\rightarrow  h_1h_1, h_2h_2,h_1h_2, W^+W^-, ZZ,Z_{BL} Z_{BL},Z Z_{BL},  f \Bar{f}, N_{R_{2/3}}N_{R_{2/3}}, N_{R_{2/3}}\Bar{\nu}_{e/\mu/\tau}$\\
$\overline{\chi_{_2}}\chi_{_1} \rightarrow  h_1h_1, h_2h_2,h_1h_2, Z h_1, Z h_2, Z_{BL} h_1, Z_{BL} h_2,  W^+W^-, ZZ,Z_{BL} Z_{BL},Z Z_{BL},  f \Bar{f}, N_{R_{2/3}}N_{R_{2/3}},  N_{R_{2/3}}\Bar{\nu}_{e/\mu/\tau}$\\
$\overline{\chi_{_2}}\chi_{_2}\rightarrow  h_1h_1, h_2h_2,h_1h_2, Z h_1, Z h_2, Z_{BL} h_1, Z_{BL} h_2,  W^+W^-, ZZ,Z_{BL} Z_{BL},Z Z_{BL},  f \Bar{f}, N_{R_{2/3}}N_{R_{2/3}}, N_{R_{2/3}}\Bar{\nu}_{e/\mu/\tau}$\\
$\overline{\chi_{_2}}\chi_{_3} \rightarrow  h_1h_1, h_2h_2,h_1h_2, Z h_1, Z h_2, Z_{BL} h_1, Z_{BL} h_2,  W^+W^-, ZZ,Z_{BL} Z_{BL},Z Z_{BL},  f \Bar{f}, N_{R_{2/3}}N_{R_{2/3}}, N_{R_{2/3}}\Bar{\nu}_{e/\mu/\tau}$\\
$\overline{\chi_{_3}}\chi_{_1} \rightarrow  h_1h_1, h_2h_2,h_1h_2, W^+W^-, ZZ,Z_{BL} Z_{BL},Z Z_{BL},  f \Bar{f}, N_{R_{2/3}}N_{R_{2/3}}, N_{R_{2/3}}\Bar{\nu}_{e/\mu/\tau}$\\
$\overline{\chi_{_3}}\chi_{_2} \rightarrow  h_1h_1, h_2h_2,h_1h_2, Z h_1, Z h_2, Z_{BL} h_1, Z_{BL} h_2,  W^+W^-, ZZ,Z_{BL} Z_{BL}, Z Z_{BL}, f \Bar{f}, N_{R_{2/3}}N_{R_{2/3}}, N_{R_{2/3}}\Bar{\nu}_{e/\mu/\tau}$\\
$\overline{\chi_{_3}}\chi_{_3} \rightarrow  h_1h_1, h_2h_2,h_1h_2, W^+W^-, ZZ,Z_{BL} Z_{BL},Z Z_{BL},  f \Bar{f}, N_{R_{2/3}}N_{R_{2/3}}, N_{R_{2/3}}\Bar{\nu}_{e/\mu/\tau}$\\
$\chi_{_1}\psi^\pm \rightarrow W^\pm \gamma, W^\pm h_1, W^\pm h_1, W^\pm Z, W^\pm Z_{BL}, f'f $\\
$\chi_{_2}\psi^\pm \rightarrow  W^\pm \gamma, W^\pm h_1, W^\pm h_1, W^\pm Z, W^\pm Z_{BL}, f'f $\\
$\chi_{_3}\psi^\pm \rightarrow  W^\pm \gamma, W^\pm h_1, W^\pm h_1, W^\pm Z, W^\pm Z_{BL}, f'f $\\
$\psi^\pm \psi^\mp \rightarrow W^\pm W^\mp, Z h_1, Z h_2, Z_{BL} h_1, Z_{BL} h_2, \gamma Z, \gamma \gamma, ZZ, Z_{BL} Z_{BL}, Z Z_{BL}, f \Bar{f}$}
\end{shaded}

\bibliographystyle{JHEP}
\bibliography{ref}


\end{document}